\documentclass[a4paper, american,11pt]{article}
\pdfoutput=1 

\usepackage{jinstpub} 
\usepackage{graphicx}
\usepackage{caption}
\usepackage{subcaption}                      
\usepackage{float}

\title{\boldmath Gain Stabilization of SiPMs with an Adaptive Power Supply}

\author[a,1]{G. Eigen,\note{Corresponding author.}}
\author[b]{J. Cvach,}
\author[b]{J. Kvasnicka,}
\author[b]{I. Polak,}
\author[a]{A. Tr{\ae}et}
\author[a]{and J. Zalieckas}

\affiliation[a]{Department of Physics and Technology, University of Bergen, N-5007 Bergen, Norway}
\affiliation[b]{Institute of Physics of the ASCR, Prague, Czech Republic}

\emailAdd{gerald.eigen@uib.no}

\abstract{The gain of silicon photomultipliers (SiPMs) increases with bias voltage and decreases with temperature. To operate SiPMs at stable gain, the bias voltage can be readjusted to compensate for temperature changes. We have tested this concept with 30 SiPMs from three manufacturers (Hamamatsu, KETEK and CPTA) operating in a climate chamber at CERN by varying temperatures between $1^\circ \rm C$ and $48^\circ \rm C$. We built an adaptive power supply that uses a linear dependence of the bias voltage  on temperature. We stabilized four SiPMs simultaneously with only one compensation parameter for the readjustment of the bias voltage of four SiPMs. For all tested Hamamatsu and CPTA SiPMs 
we achieved our goal of limiting gain changes  to less than $\pm 0.5\%$ in the $20-30^\circ C$ temperature range. }


\arxivnumber{1902.06261}

\collaboration[c]{}
\proceeding{}

\begin{document}
\maketitle
\flushbottom

\section{Introduction}
\label{sec:intro}
\begin{figure}[htbp!]
\centering
\includegraphics[width=75mm]{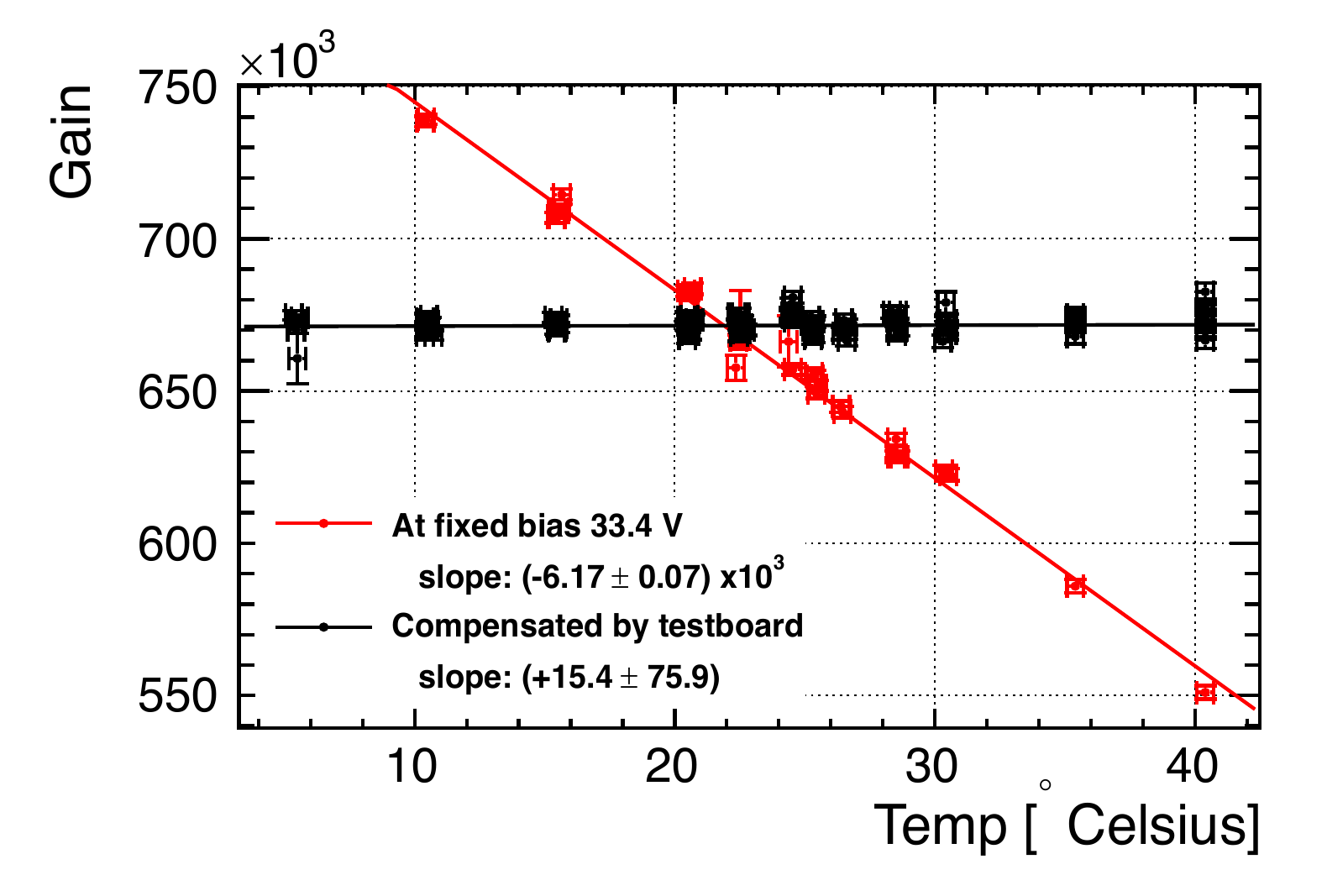}
\caption{The temperature dependence of the gain of SiPMs without $V_{\rm b}$ adjustments (red curve) and with $V_{\rm b}$ adjustments (black curve).}
\label{fig:uncorrected-gain} 
\end{figure}

The gain of silicon photomultipliers (SiPMs)~\cite{Bondarenko, Buzhan2002, Buzhan2003}\footnote{Hamamatsu calls these photosensors MPPCs.} 
depends on temperature ($T$) and on the overvoltage $\Delta V_{\rm b} = V_{\rm b} - V_{\rm break}$ where $V_{\rm b}$ is the bias voltage and $V_{\rm break}$ is the break-down voltage. Typically, the gain
increases nearly linearly with $\Delta V_b$ or  $V_{\rm b}$ and decreases nearly linearly with $T$. For stable operation the gain needs to be kept constant, especially in large detector systems such as an analog hadron calorimeter~\cite{analogCal} planned for an ILC detector, operating with $\mathcal{O}(10^{6})$ SiPMs~\cite{intent}. Variations of the ambient temperature and heat produced by electronics typically induce gain changes in SiPMs. Figure~\ref{fig:uncorrected-gain} illustrates the dependence of the gain on temperature without and with $V_{\rm b}~(\Delta V_{\rm b})$ readjustments.
The method of keeping the gain constant consists of adjusting $V_{\rm b}$ when $T$ changes. This, however, requires knowledge of $dV_{\rm b}/dT$, which we obtain from measurements of gain versus bias voltage ($dG/dV_{\rm b}$)   and gain versus temperature ($dG/dT$). We assume a linear dependence of the gain both on  bias voltage and on temperature. This implies constant values for $dG/dV_{\rm b}$ versus $T$ and $dG/dT$ versus $V_{\rm b}$ and in turn a  constant value for $dV_{\rm b}/dT$. We built a bias voltage regulator that adjusts $V_{\rm b}$ using a linear  temperature dependence.

We present herein a gain stabilization study of 30 SiPMs from three manufacturers (Hamamatsu~\cite{hamamatsu}, KETEK~\cite{ketek} and CPTA~\cite{cpta}) in the temperature range of  $1- 48^{\circ}$C. Our goal consists of keeping gain changes smaller than $ \pm0.5\%$ in the temperature range of $20- 30^{\circ}$C. We accomplish this in two steps. First, we determine $dV_{\rm b}/dT$ from $dG/dV_{\rm b}$ and $dG/dT$ measurements for each SiPM.  Second,  we select a common value of $dV_{\rm b}/dT$ to test gain stabilization of four SiPMs simultaneously. The purpose is to demonstrate that a set of SiPMs can be stabilized with one compensation parameter $dV_{\rm b}/dT$. This is an essential requirement for large arrays of SiPMs.

In section~\ref{sec:setup}, we describe the experimental setup, summarize properties of all tested  SiPMs and discuss the bias voltage regulator. In section ~\ref{sec:gain},  we present two methods for extracting the photoelectron (p.e.) spectra from waveforms and two fit methodologies for extracting the gain from photoelectron spectra. In section~\ref{sec:dvdt}, we present measurements of gain versus bias voltage and gain versus temperature from which we extract $dG/dV_{\rm b}$, $dG/dT$ and in turn $dV_{\rm b}/dT$. In section~\ref{sec:gainstab}, we show the performance of our gain stabilization studies for all 30 SiPMs before we end with our conclusion and outlook  in section~\ref{sec:conclusion}.

\section{Experimental Setup}
\label{sec:setup}

\subsection{Experimental Setup in the Climate Chamber}
\label{sec:exp_setup}
\begin{figure}
\centering 
\includegraphics[width=65 mm]{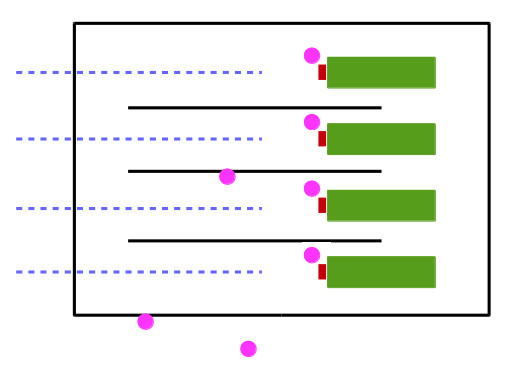}
\hspace{5mm}
\includegraphics[width=75 mm]{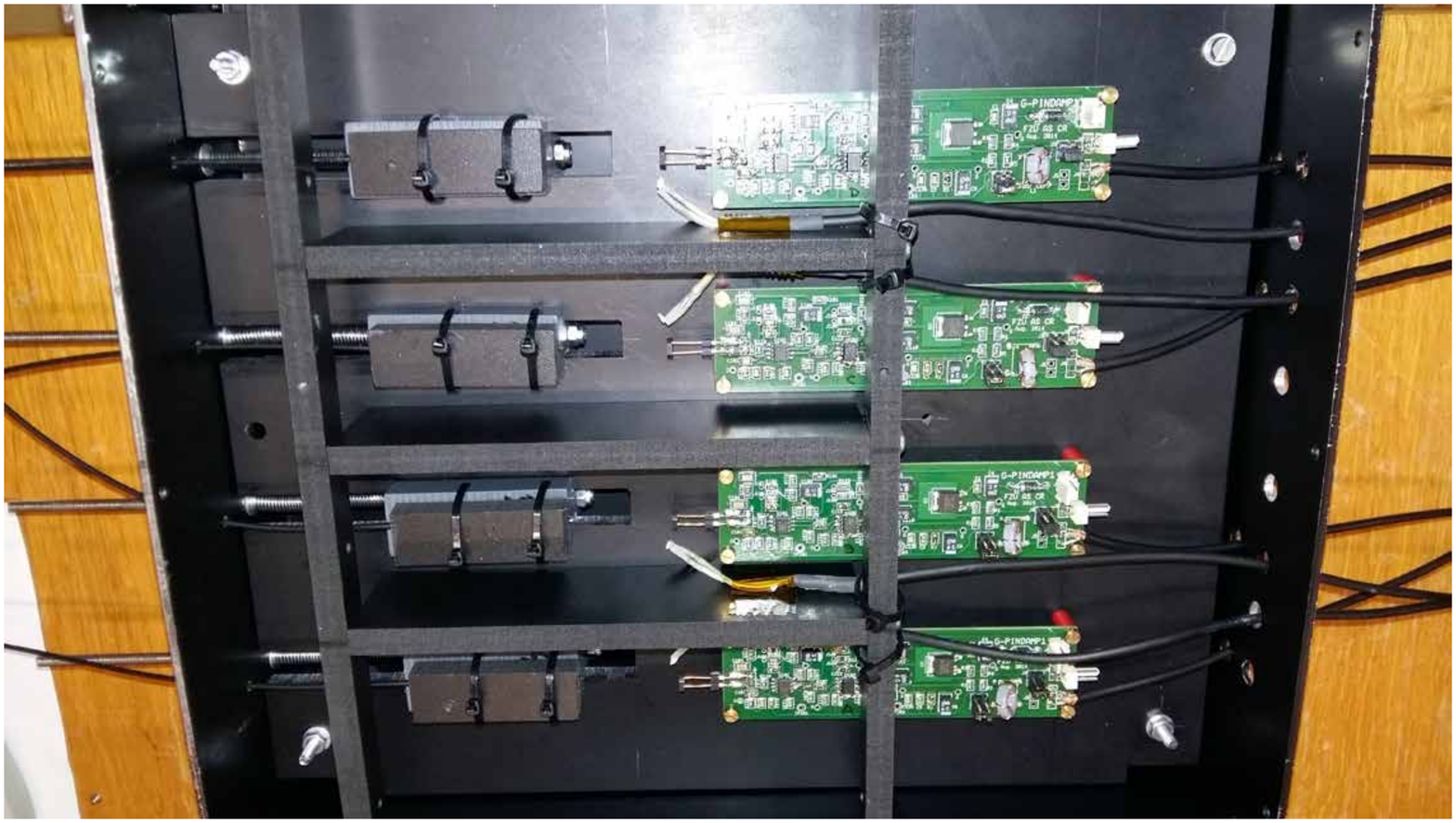}
\caption{\label{fig:exp} Left: schematic setup of the gain stabilization and temperature characterization measurements inside the black box, showing SiPMs (red  rectangles), preamplifiers (green rectangles), temperature sensors (magenta points) and optical fibers (dashed blue lines).
Right: top view of the black box. The green circuit boards each host a preamplifier and the signal readout. On the right-hand side of the board we can see
a black signal cable (lower corner), the  SiPM high-voltage cable (middle) and  white connectors for the preamplifier power  (upper corner).
A PT1000 sensor is positioned near each SiPM (white extensions near a SiPM). The clear fibers transporting the blue LED light run inside black, distance-adjustable foam boxes on the left-hand side, which are  mounted precisely to illuminate the SiPMs uniformly.}
\end{figure}

\begin{figure}[htbp!]
\centering 
\includegraphics[width=70 mm]{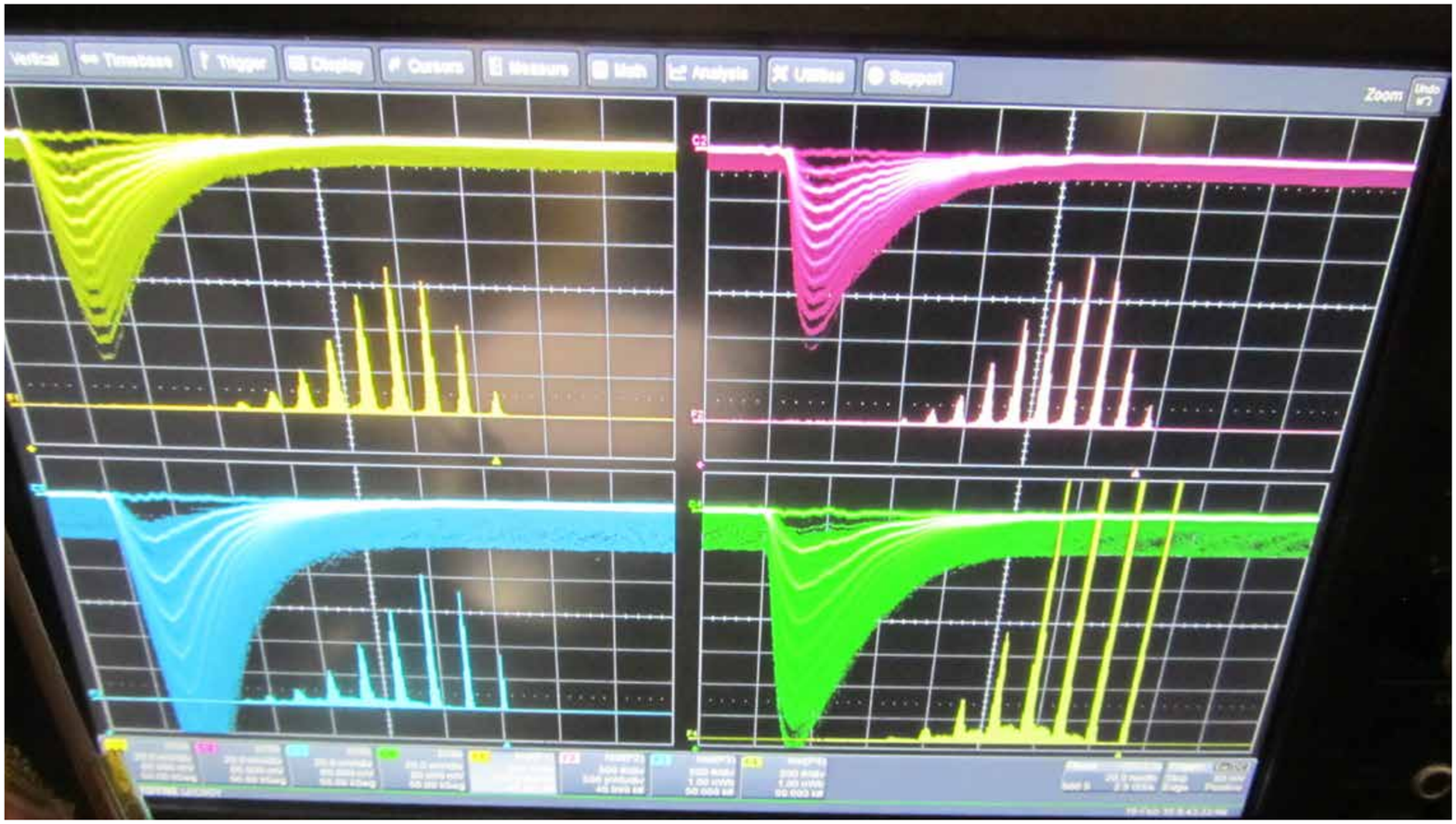}
\caption{Waveform and photoelectron spectra of four Hamamatsu S13360  MPPCs. }
\label{fig:waveform}
\end{figure}

\begin{figure}[htbp!]
\centering 
\includegraphics[width=120 mm]{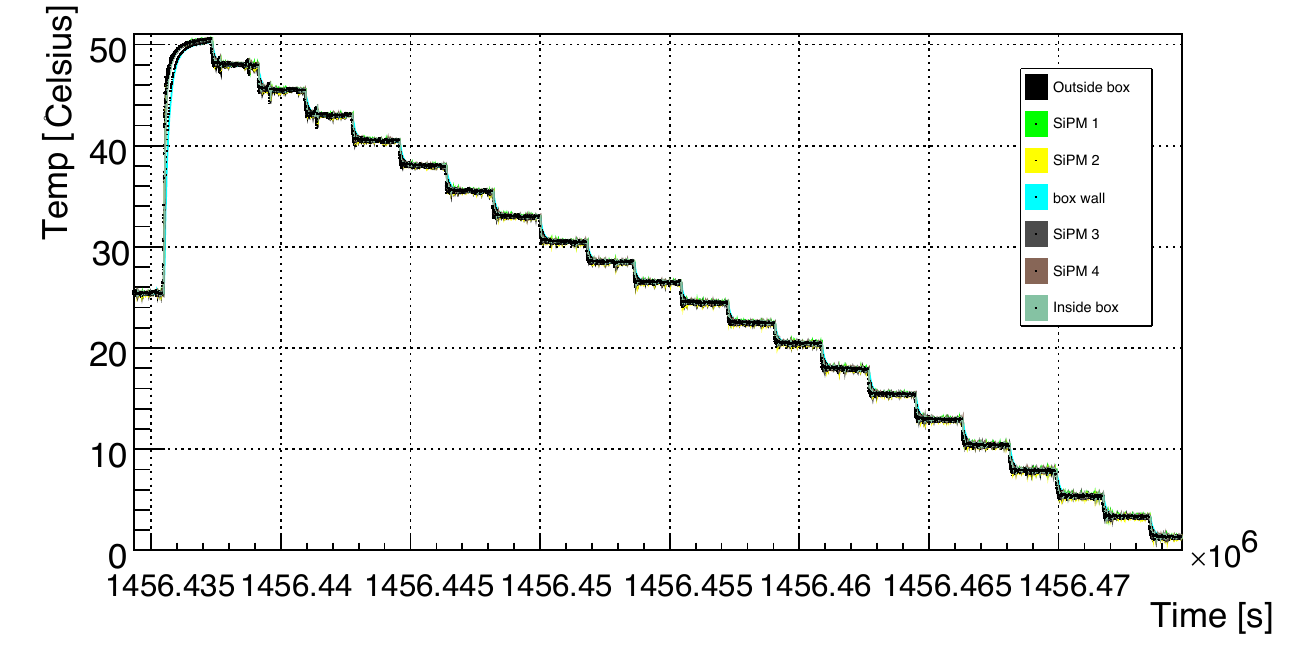}
\caption{Typical temperature profile for an overnight stabilization run.  Green, yellow, navy blue and brown data points show measured temperatures at each of the four SiPMs denoted as SiPM-1, SiPM-2, SiPM-3 and SiPM-4,  respectively.  The dark green, light blue and black points represent temperatures inside the black box, on the box wall and in the climate chamber outside the box, respectively. Stable temperature is reached after $\sim15$ min.}
\label{fig:tprofile}
\end{figure}

Figures~\ref{fig:exp}  (left, right) respectively show a schematic view and a photo of the measurement setup in the climate chamber (Spiral3 from Climats) at CERN.  Four SiPMs are mounted inside a black box separated by black walls to prevent optical cross talk. During measurements the entire climate chamber is kept light tight. Each  SiPM is inserted into a socket soldered onto the circuit board that houses  a voltage-operational two-stage preamplifier operating with time constants of $8~\rm ns$ and $25~\rm ns$, respectively. 
The amplified signals are recorded by four channels of a digital oscilloscope from LeCroy (model 6104), which uses 12 bit ADCs and samples data at a rate of 2.5 GS/s. We illuminate each SiPM with blue LED light. The LED is placed outside the climate chamber to minimize noise pickup. It is attached to four optical  plastic fibers that transport the blue light of similar intensity to each SiPM. We position the fibers such that the surface of each SiPM is uniformly illuminated.  The LED is generated by a  3.4 ns wide light pulser signal that is obtained from a sinusoidal pulse above an adjustable threshold. Both,  repetition rate and light intensity are adjustable.  We operate the light pulser at a rate of 10 kHz and set the LED  light intensity such that several single-photoelectron peaks are visible in addition to the pedestal.  We operate the digital oscilloscope in the mode that records directly SiPM waveforms, which are stored on disk allowing us to keep the entire raw data sample for offline analysis. For example, Fig.~\ref{fig:waveform} shows  50000 recorded waveforms and the resulting photoelectron spectra for Hamamatsu MPPCs with trenches (S13360). Individual photoelectron peaks are clearly separated.  
For data taking, we use a dedicated LabView program that operates the digital oscilloscope, controls the intensity of the light pulser and sets the bias voltages of the SiPMs. The low voltages of the preamplifiers are set manually and the temperature profiles are recorded by a separate dedicated system built by a group from MPI Munich~\cite{temp}. 
To accurately record the temperature  inside the climate chamber, we use seven PT1000 sensors. Four PT1000 sensors record the temperature close to each SiPM, one sensor is placed inside the black box, one sensor is attached to the outside wall of the black box, and the seventh sensor is placed in the climate chamber outside the black box. Figure~\ref{fig:tprofile}  shows a typical temperature profile used in the gain stabilization studies.

\subsection{Properties of the Photodetectors}
\label{sec:sipm}

A silicon photomultiplier is a pixelated avalanche photodiode operated in the Geiger mode~\cite{Bondarenko, Buzhan2002, Buzhan2003}. A photon impinging on the surface triggers a Geiger-M\"uller avalanche with well-defined probability that depends on three factors:  the wavelength-dependent quantum efficiency of the pixel, the geometric efficiency of the SiPM layout and the probability that an absorbed photon triggers a Geiger-M\"uller avalanche. A fired pixel is insensitive to record following photons until the avalanche is broken off via a quenching resistor.  The dynamic range of SiPMs is given by the number of pixels. The bias  voltage lies slightly above the breakdown voltage. For Hamamatsu MPPCs, the nominal bias voltage ranges between 50 V and 70 V while for KETEK and CPTA SiPMs it is around 30 V.

\begin{figure}
\centering
\includegraphics[width=65 mm]{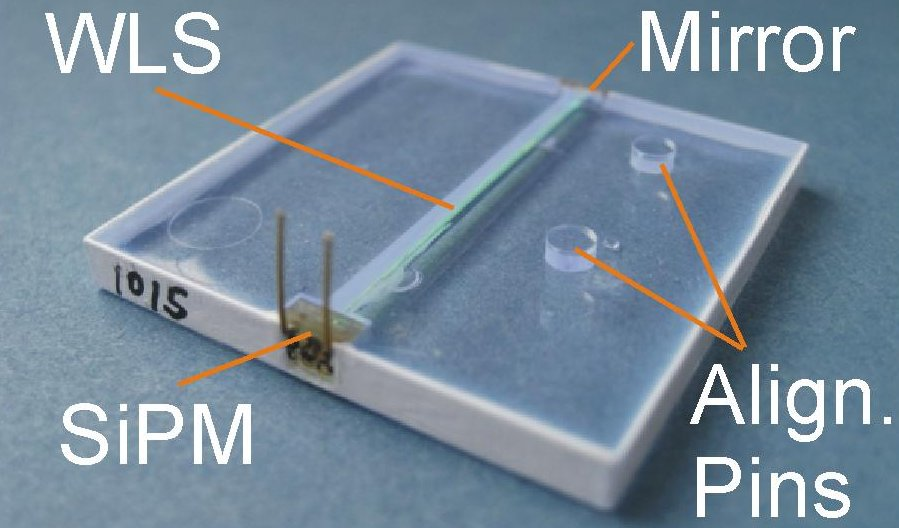}    
\caption{\label{fig:tile} Setup of a scintillator tile that is read out with a CPTA SiPM. The sensor is glued to green wavelength-shifting fiber that is inserted into a groove in the tile.  }
\end{figure}

Table~\ref{tab:PropSiPMs} in appendix~\ref{sec:appSiPM} summarizes the properties of the 30 tested SiPMs, consisting of 18 Hamamatsu MPPCs, eight KETEK SiPMs  and four CPTA SiPMs. The Hamamatsu   $\rm A-$type, $\rm B-$type and S12571 MPPCs are highly-pixelated conventional photodetectors without trenches. The $\rm A-$type and $\rm B-$type  MPPCs come with a pixel pitch of $~20~\mu \rm m$ and $\rm 15~\mu \rm m$ while  the S12571 sensors have a pixel pitch of  $\rm 15~\mu \rm m$  and $ ~10~\mu \rm m$. At the nominal bias voltage the gain is around $2\times 10^5$ except for the  $\rm 10~\mu \rm m$ S12571 sensors that have a gain of  $1.4 \times 10^5$. In addition, we obtained six new MPPCs with trenches, which have reduced pixel cross talk and low noise: four S13360 (two LCT4) MPPCs have a pixel pitch of $\rm 25~(50)~\mu m$ operating at a gain of  $7.0 \times 10^5$ ($1.6 \times 10^6$) at a nominal bias voltage of  57~(51)~V. 
From KETEK we tested two experimental devices (W12) and six conventional SiPMs (PM3350). The W12 SiPMs come with a  pixel pitch of $\rm 20 ~\mu \rm  m$ and operate with a gain of $5.4 \times 10^5$ at the nominal bias voltage of 28~V while the PM3350 SiPMs have a pixel pitch of $\rm 50 ~\mu \rm m$  
yielding a gain of $2 \times 10^6$ at a nominal bias voltage of 29.5~V.
From CPTA we tested four SiPMs that have a pixel pitch of  $\rm 40~\mu \rm m$. The  gain is $7 \times 10^5$ at the nominal bias voltage of  33~V.  However, the SiPMs were glued to a green wavelength-shifting fiber inserted into a groove in a 3 mm thick $3 \times 3~\rm cm^2$ plastic scintillator tile. Figure ~\ref{fig:tile}  depicts the details of  the tile setup. 

Typically, we illuminate the bare SiPM directly except for CPTA sensors where
we illuminate the tile close to the SiPM. The blue light either reaches the fiber directly or via absorption and reemission in the scintillator. The green double-cladded fiber shifts the absorbed scintillation  light to higher wave lengths and transports the re-emitted photons that satisfy the total reflection criteria to the SiPM where they trigger Geiger-M\"uller avalanches. In this setup, the light is recorded with a delay of a few nanoseconds, which is small compared to the collection time. Due to the more complicated illumination procedure it was not obvious if we could carry out the stabilization tests of the CPTA SiPMs successfully. 

The nominal bias voltage for each of the four $\rm A-$type and four $\rm B-$type  MPPCs  is rather similar. So, we selected a single $V_{\rm b}$ for each MPPC type.  For S12571 MPPCs with  $15~\rm \mu m$ pitch and $10~\rm \mu m$ pitch, the nominal bias voltage differs by a factor of 0.974. We, therefore, used
 a voltage divider to correct for this difference.  
 Similarly, the bias voltage between the S13360-25 and the LCT MPPCs differs by a factor of 0.944. To correct for this difference, we
 adjusted the voltage divider appropriately.  Since for KETEK and CPTA SiPMs the nominal bias voltage was similar for each set of four tested SiPMs, we
could select a common bias voltage for each set of four SiPMs. 

\subsection{The Bias Voltage Regulator}

\begin{figure}
    \centering
    \begin{subfigure}[c]{0.8\textwidth}
        \includegraphics[width=\textwidth]{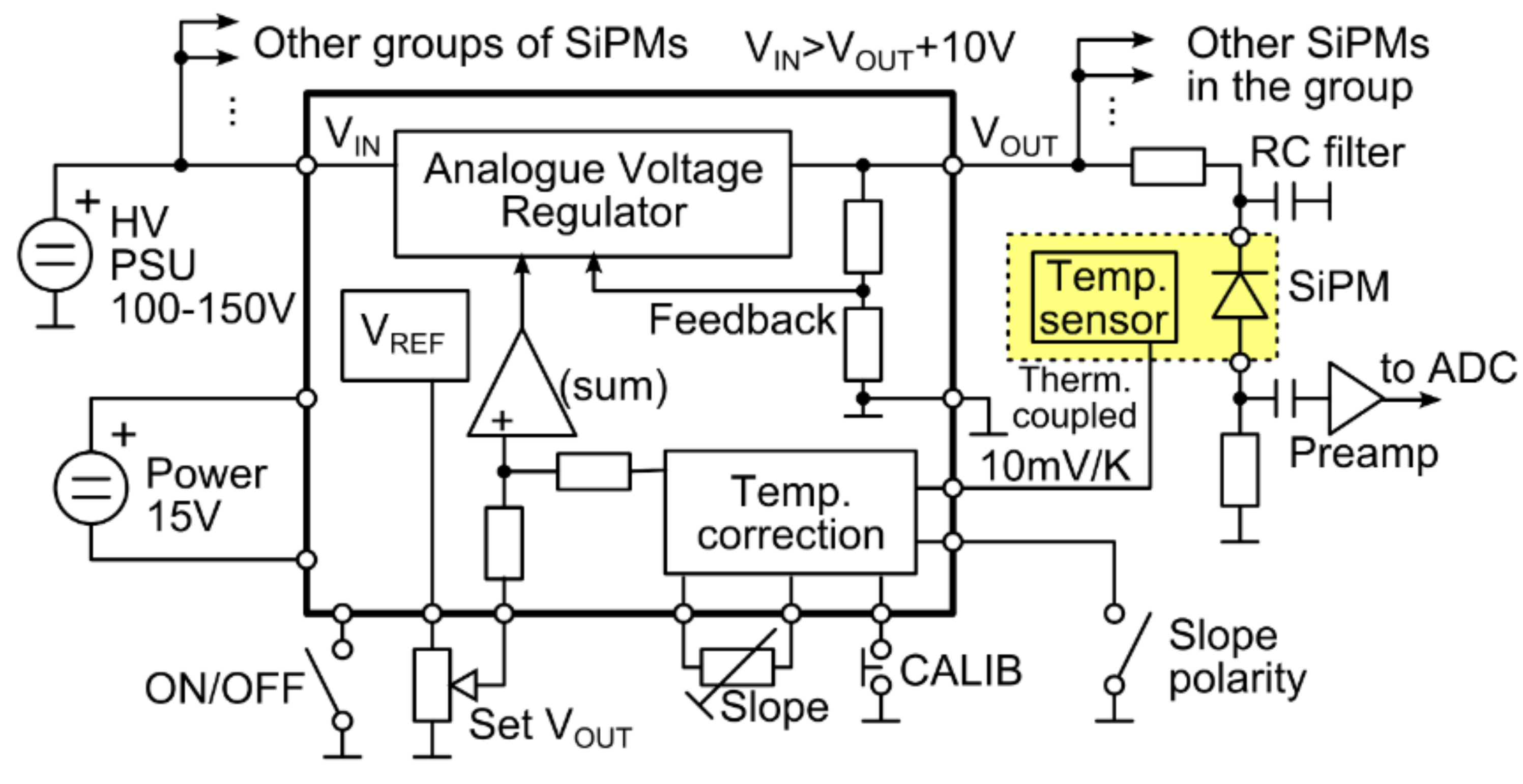}
    \end{subfigure}    
\caption{\label{fig:exp2} Schematic layout of the bias voltage regulator board.}
\end{figure}

In order to keep the gain of SiPMs constant, we developed and built a bias voltage regulator, which automatically adjusts the bias voltage 
 proportional to the
 temperature change. Figure~\ref{fig:exp2} shows a schematic layout of this device, which consists of three blocks: a high-voltage block, a voltage reference block and a temperature correction circuit. The high-voltage block is supplied with a positive nominal high-voltage input of 130~V (range: 100-150 V) and a low voltage of 15 V. The high voltage is  generated from 230~V AC by a custom-made transformer and rectifiers. The generation of the low voltage needs  a linear regulator in addition. On the output side, the high voltage is variable up to 100 V.  The input is decoupled by a radio frequency interference filter to keep the noise as low as possible. On the output side, the SiPM is decoupled by an RC filter and is connected to the bias voltage regulator via a high-voltage  cable. The high-voltage regulation uses two HV MOSFETs in a totem pole configuration. We have implemented a current limitation at 10 mA  using an NPN transistor that checks the output current at the upper MOSFET. The output voltage divided by a factor of ten is used as a feedback to the analog voltage regulator. To reach the appropriate precision, we use an integrated resistor voltage divider with precision resistors that have a very low tracking temperature coefficient of less than 2.5 ppm/$^{\circ}$C. This high-precision divider (CNS471 series by Vishay) is one of the main key elements for guaranteeing voltage stability.

The voltage reference block is based on a stable voltage of 10~V with a precision of 1~mV that corresponds to 100~V at the regulator output with a precision of $10^{-5}$. We use the precision reference LT1021 from Linear Technology, which has a very low drift of  $\sim2$ ppm/$^{\circ}$C, a low noise of $<6~\mu \rm V$  and extremely good long-term stability at 15 ppm/1000 h. The voltage is set by two ten-turns potentiometers (one for coarse and one for fine tuning) in an output range of 15 V to 100 V. To achieve smooth ramp-up and ramp-down of the output voltage, we implemented a soft start circuit, which needs about 15 seconds till the output voltage becomes stable with a precision of 10 mV.

The temperature correction circuit is designed to match the signal from the semiconductor LM35D temperature sensor from Texas Instruments, which
sends a linear analog output in units of 10~mV/$^\circ$C. So, at $25^\circ$C the signal corresponds to 250 mV. The signal is amplified by a factor
 of ten before  it is compared with a reference.
 The ambient temperature on the sensor close to the SiPM is monitored in units of 100~mV/$^\circ$C with a precision of $10^{-4}$. For calibration purpose, the circuit has built in a switch to ground that sets the zero point. Another switch allows to set the +10$^\circ$C point to 100~mV by  dropping 100 $\mu$A  current over a resistor of 1~$\rm k\Omega$. An inverting amplifier 
further amplifies the  signal from the temperature sensor to the required final value. The amplifier is operable with both polarities. However, all SiPMs under study show a positive compensation slope. The signal is added in a summing amplifier  to set the reference voltage. The output of summing amplifier is driving an analog high-voltage regulator.
We use a simple algorithm to set the operation point of the SiPMs with the bias voltage regulator. The operation requires the nominal output voltage at a given temperature and the slope of the correction voltage $dV_{\rm b}/dT$ in mV/$^\circ$C as inputs. The nominal bias voltage at a given temperature is specified by the SiPM manufacturer. The slope is the one we determine in our measurements of gain versus bias voltage and gain versus temperature (see section~\ref{sec:dvdt}). While the feed back-loop has a time constant of $\sim 100$~ms, the timing of the compensation loop is determined by a thermal constant of the material around the thermo-sensor yielding a time constant of the order of 10~s.

We monitor the output voltage with a precision better than $10^{-6}$. To check the short-term stability, we use a freeze spray that cools the board to $-40^\circ$C and a 100 W lamp to heat up the board to about $50^\circ$C. We measure a drift of the output voltage of about 60~mV corresponding to a stability of better than 10 ppm/$^\circ$C.  In order to improve the stability, we need to shield the voltage reference from ambient air turbulence since that produces low-frequency noise ($<$ 1 Hz) due to thermoelectric differences between the integrated circuit package leads and the printed circuit board materials. Similar effects result from the temperature coupling between the SiPM and temperature sensor. Stability may be further improved by using more stable resistors (better than 50 ppm/$^\circ$C) in the signal traces and an output voltage divider with a better long-term stability.

\section{Gain Measurement Methodology}
\label{sec:gain}

\subsection{Definition of the SiPM Gain}
\label{sec:gain_def} 
A photon triggering a Geiger-M\"uller avalanche produces a definite charge 
\begin{equation}
Q(V_{\rm b}, T) =A(V_{\rm b},  T) \cdot e
\end{equation}
at a given bias voltage and temperature, 
where $A(V_{\rm b},  T)$ is the gain and $e$ is the unit charge. Thus, the gain is the difference between the charge of the first photoelectron peak and that of the pedestal divided by $e$, 
\begin{equation}
A(V_{\rm b},  T)=[Q_1(V_{\rm b},  T) - Q_{\rm ped}]/e
\end{equation}
or the charge difference between two adjacent photoelectron peaks $n$ and $n-1$ divided by $e$
\begin{equation}
A(V_{\rm b},  T)=[Q_{\rm n}(V_{\rm b},  T) - Q_{\rm n-1}(V_{\rm b},  T)]/e.
\end{equation}

Though both definitions yield the same gain, we extract the gain from the charge difference between the second and first photoelectron peaks,
\begin{equation}
A(V_{\rm b},  T)=[Q_2(V_{\rm b},  T) - Q_1(V_{\rm b},  T)]/e. 
\end{equation}
In principle,  the charge difference between any two adjacent photoelectron peaks should be the same at fixed $V_{\rm b}$ and $T$. Fits to  photoelectron spectra with up to ten photoelectrons confirm this. Note that the measured gain is a product of the SiPM gain and that of the preamplifier. The latter gain is independent of $V_{\rm b}$ and $T$. For this study, we do not decouple  the intrinsic SiPM gain and use the total gain. 

To extract the SiPM gain we proceed in three steps. First, we record 50000 waveforms and save them on disk. In the second step, we convert them offline to photoelectron ($pe$) spectra. We had to develop two methods for this task: we either integrate the waveform over a selected time window or we determine the minimum value of the waveform. We call these methods waveform integration and waveform minimum, respectively. The waveform integration methods works well for all tested Hamamatsu MPPCs. However, for KETEK and CPTA SiPMs it fails at certain values of $V_{\rm b}$ and $T$. Since it is important to apply the same procedure for a given SiPM over  the entire range of measured  $V_{\rm b}$ and $T$ values, we had to develop a  rather robust second method.
The KETEK SiPMs have much longer decay times than Hamamatsu MPPCs because their capacitance is much larger as their area is typically a factor of nine larger. Thus, integration times became too long for a variable time window whereas for a fixed time window the first and second photoelectron peaks were not well separated at certain values of $V_{\rm b}$ and $T$. For CPTA SiPMs, the photoelectron peaks were not well enough separated at certain values of $V_{\rm b}$ and $T$ in order to use the waveform integration method. In the final step, we extract the gain from a fit to the photoelectron spectrum.

\subsection{Measurement of Photoelectron Spectra}
\label{sec:pe_spec}   

\begin{figure}[htbp!]
\centering 
\includegraphics[width=2.9in]{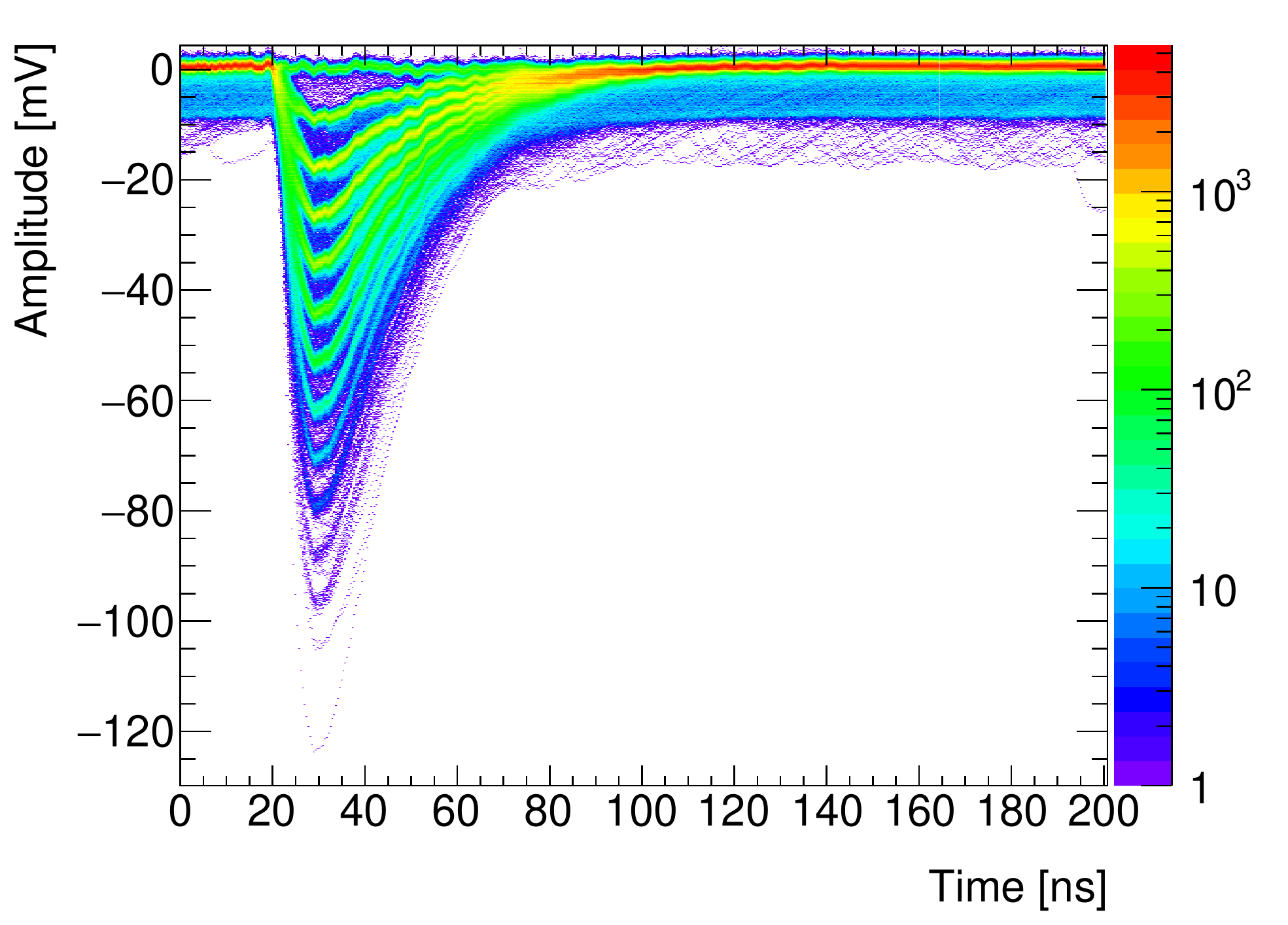}
\includegraphics[width=2.9in]{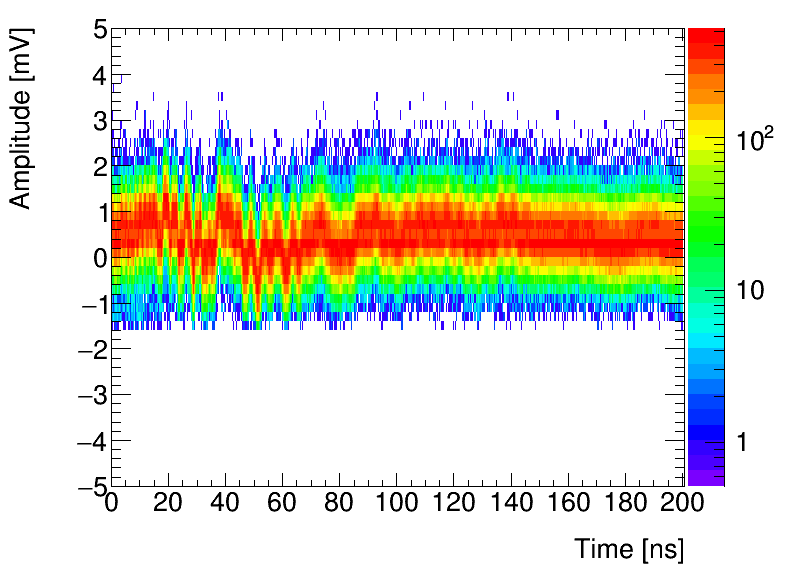}
\includegraphics[width=2.9in]{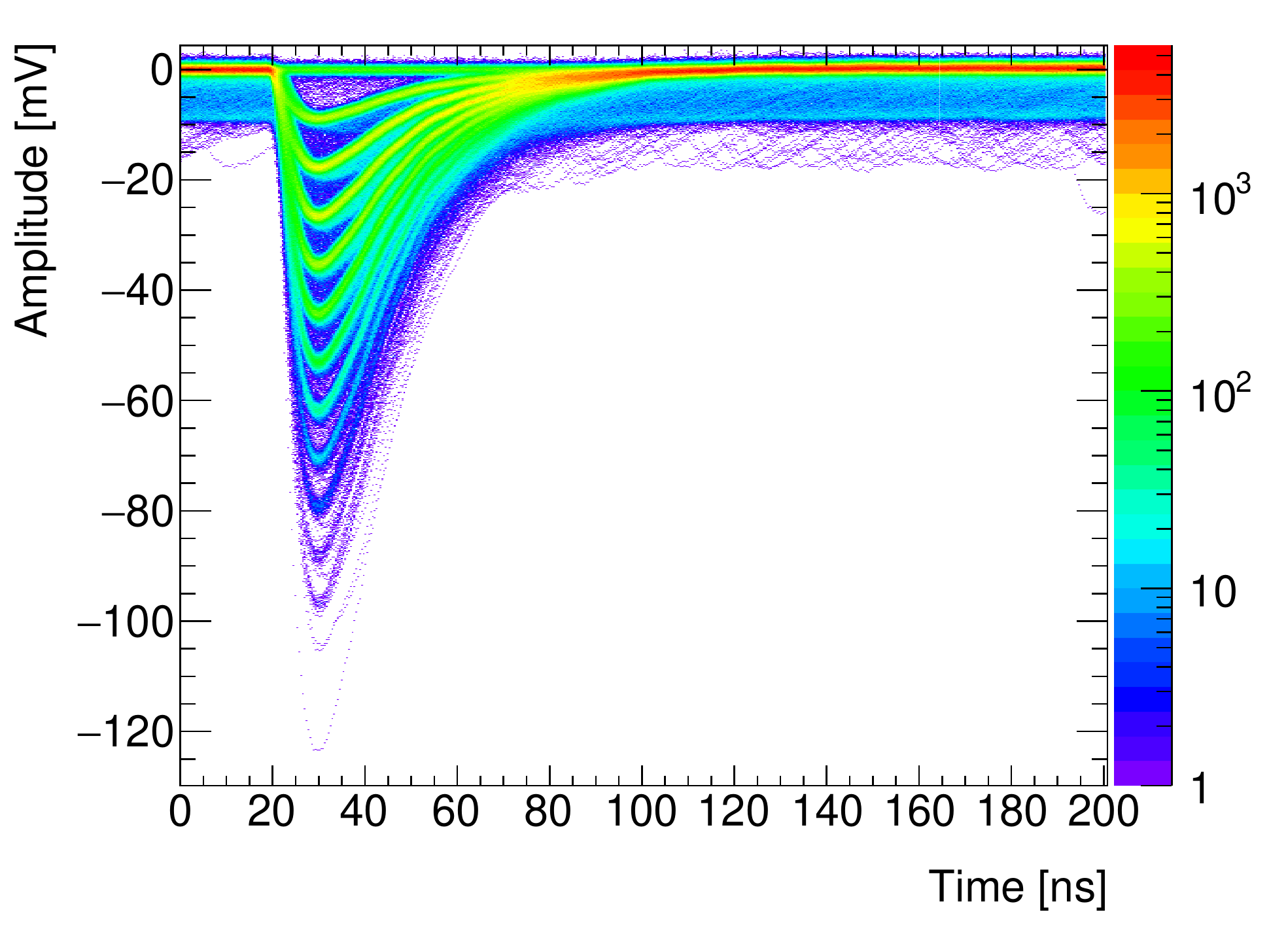}
\includegraphics[width=2.9in]{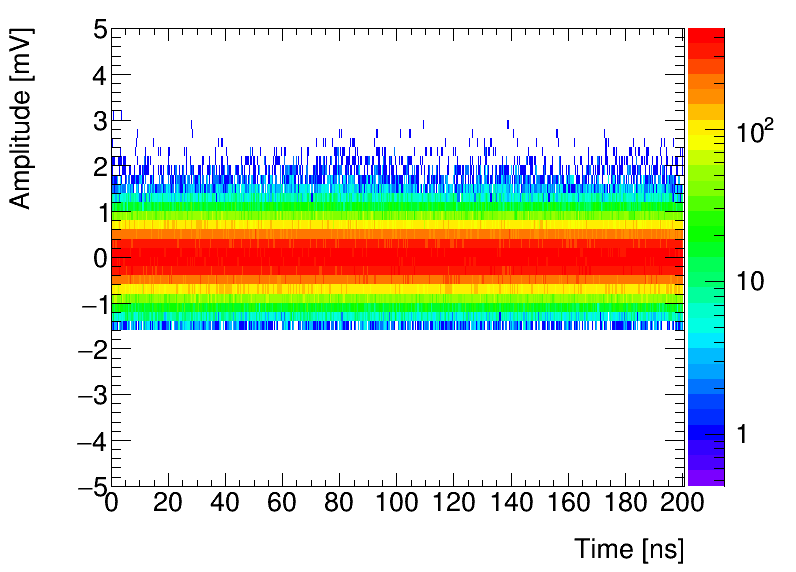}
\caption{\label{fig:WF} Top: superimposed waveforms measured with the Hamamatsu MPPC S13360-1325b (left) and 
pedestal distribution (right) before subtraction of the parasitic pickup signal. Bottom: same superimposed waveforms (left) and pedestal distribution (right) after subtraction of the parasitic pickup signal.}
\end{figure}

Figure~\ref{fig:WF} (top left) depicts raw waveforms measured with the Hamamatsu MPPCs S13360-1325b. The spectrum clearly shows some 
pickup noise. Despite extensive efforts on grounding and shielding, all waveforms show a small parasitic pickup noise signal, which was caused by a defective light pulser cable that we could not replace during the run period at CERN. Thus, we developed a procedure to subtract the pickup noise from each waveform in order to improve the resolution of the photoelectron peak positions. 
First, we sample 21 points before the signal waveform starts corresponding to a time window of 8.4~ns. 
We fit the distribution with a Gaussian function and define a threshold taking the mean value minus three standard deviations. Then, we select all pedestal distributions that lie above the threshold shown in Fig.~\ref{fig:WF} (top right). We determine the average of the pedestal distribution and subtract it from all waveforms. Figures~\ref{fig:WF} (bottom left, right)  respectively show 
the waveform and pedestal distributions after subtraction of the parasitic noise signal. 
The removal of the pick-up noise  obviously provides an improvement.

\begin{figure}[htbp!]
\centering 
\includegraphics[width=2.9in]{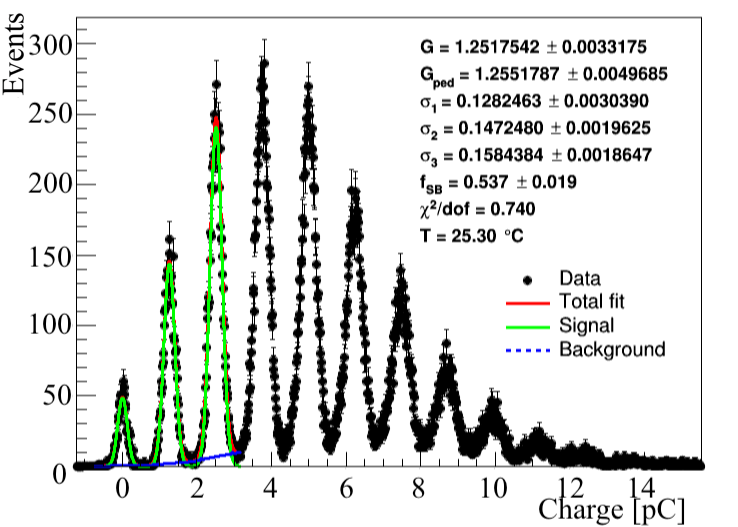}
\includegraphics[width=2.9in]{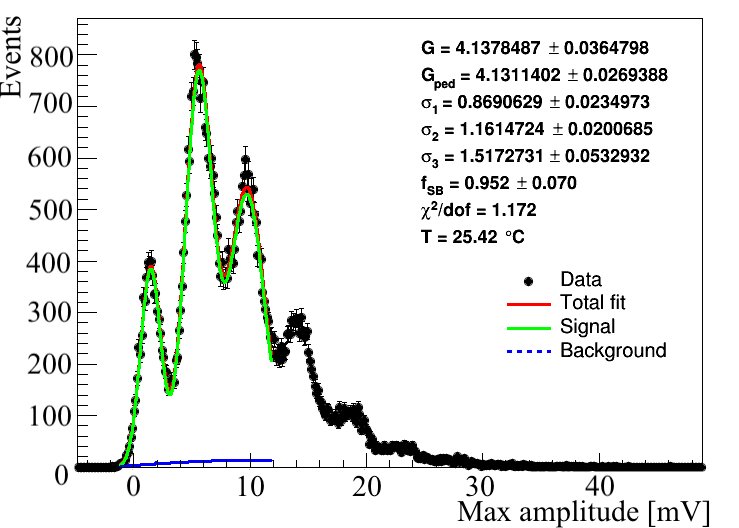}
\caption{\label{fig:typ_pe} Left: measured photoelectron spectrum for the Hamamatsu MPPC B2-20. Right:  measured photoelectron spectrum for the KETEK $\rm SiPM ~PM3350\#1$. In addition to the data (black points with error bars) fits with our first gain fit model (solid green curves) are shown  in which the probability density function consists of  three Gaussian functions (solid red curves) plus a smooth background (dotted blue curves). 
 }
\end{figure}

For all Hamamatsu MPPCs, we perform the waveform integration method by integrating the noise-subtracted waveforms over a time window  $\Delta t = t_2 -t_1$ where $t_1$ is the fixed time of the signal start and $t_2$ is the variable time when the signal reaches the baseline again. We tested different definitions of $t_1$ and $t_2$ and found that the above selection gave the best performance. Since the integration is over the full waveform, we always record entire photoelectrons. Pixel cross talk 
and afterpulsing increase $t_2$ but the additional photoelectron is completely recorded. Pixel cross talk triggers an avalanche in an adjacent pixel while afterpulsing triggers a new avalanche in the same pixel before the initial avalanche is fully recorded. Figure~\ref{fig:typ_pe} (left) shows a typical photoelectron spectrum obtained with the waveform integration method. Individual photoelectron peaks are well separated. The green curve shows a fit with our first gain fit model that is discussed in the next section~\ref{sec:gain_det}.   For all KETEK and CPTA SiPMs, we extract the photoelectron spectra using the waveform minimum method on the noise-subtracted waveforms. 
Figure~\ref{fig:typ_pe} (right) shows a typical photoelectron spectrum for  a KETEK SiPM obtained with the waveform minimum method. Again, the green curve shows a fit with our first gain fit model.

\subsection{Gain Determination}
\label{sec:gain_det}  
We employ two gain fit models to determine the gain, which in our case is the product of the SiPM gain and the gain of the shaping amplifier. In the first gain fit model we perform fits of the pedestal and the first two photoelectron peaks with Gaussian functions on top of a parameterized background. In the second model we fit all visible photoelectron peaks in addition to the pedestal with Gaussian functions without any background. In both models we use
 the  multi-peak finder tool implemented in the  ROOT~\cite{Root} $TSpectrum$ class, which is based on one-dimensional peak search algorithms with advanced spectra processing functions~\cite{TSp1, TSp2, TSp3} for identifying individual photoelectron peaks. The located peak positions are used to define an analytical model of photoelectron spectra. For the first model, we define the  likelihood function

\begin{equation}
\mathcal{L}=\prod_{{\rm i}=1}^{50000}f_{\rm s}F_{\rm sig}(w_{\rm i})+(1-f_{\rm s})F_{\rm bkg}(w_{\rm i})
\end{equation} 

\noindent where $f_s$ is the signal fraction. For the waveform integration method $w_i$ stands for the charge, $w_{\rm i}= Q_{\rm i}$, while for the waveform minimum method it stands for the minimum amplitude in the waveform, $w_{\rm i}=V_{\rm i}$.
In the first model, the signal probability density function (PDF) consists of three Gaussian functions:

\begin{equation}
F_{\rm sig}(w_{\rm i})=f_{\rm ped}G_{\rm ped}(w_{\rm i}) + f_{1}G_1(w_{\rm i}) + (1-f_{1}-f_{\rm ped})G_2(w_{\rm i})
\end{equation}    

\noindent where $f_{\rm ped}$ and $f_{1}$ respectively denote the fractions of the pedestal and first p.e. peak while $G_{\rm ped}(w_{\rm i})$,  $G_1 (w_{\rm i})$ and $ G_2(w_{\rm i})$ respectively model the shape of the pedestal, first p.e. and second p.e. peaks.
The fractions, positions and widths of the three Gaussian functions are not constrained in the fit.

Potential background  may originate from residual noise and dark current contributions or from tails of the third and higher p.e. peaks that are not included in the fit. Crosstalk and afterpulsing yield additional photoelectrons in the waveform integration method. Since both processes are delayed with respect to the original signal, they typically do not affect the photoelectron spectra extracted from the waveform minimum method.
The probability for crosstalk and afterpulsing increases with higher bias voltage~\cite{bkg1}. The dark count rate depends both on temperature and on 
 bias voltage~\cite{bkg2} being different for each phase space $(T, V_{\rm b})$ point. Since it is difficult to model the background PDF $F_{\rm bkg}$ with  an analytic function, we parameterize  it by a sensitive nonlinear iterative peak (SNIP) clipping algorithm that is implemented in the ROOT~\cite{Root} $TSpectrum$ class. We perform binned fits of the spectra, which have at least two visible p.e. peaks plus the pedestal. 
The statistical error on the gain is estimated by combining the uncertainties in the first and second p.e. peak positions, $\sigma_G=\sqrt{\sigma^2_{\rm 1p.e.}+\sigma^2_{\rm 2p.e.}}$.   
 
For low noise and dark currents, the observed photoelectron spectra extracted from both methods should result from  a superposition of individual Gaussian functions without additional background. This assumption provides the basis for  the second gain fit model in which  we fit the pedestal and all visible photoelectron peaks with Gaussian functions ($G_{\rm ped}(w_{\rm i})$ and $G_i (w_{\rm i})$), leaving all widths and fractions as free parameters. Thus, the PDF for the second model is given by 
\begin{equation}
F_{\rm sig}(w_{\rm i})=f_{\rm ped} G_{\rm ped} (w_{\rm i})+ \sum_{\rm {j}=1}^{\rm n-1} f_{\rm j} G_{\rm j} (w_{\rm i})+   G_{\rm n}(w_{\rm n}) \sum_{{\rm j}=1}^{\rm n-1} (1-f_{\rm j}-f_{\rm ped} ).
\end{equation}
The second gain  fit model works well for Hamamatsu MPPCs without trenches as well as for KETEK and CPTA SIPMs.  For Hamamatsu MPPCs with trenches, the second gain fit model sometimes fails because the photoelectron peaks are rather narrow and show non-Gaussian tails on the right-hand side of each peak that become larger for higher photoelectron peaks. Rather than developing a suitable  parameterization, we use the first model for MPPCs with trenches. 
Furthermore, since we used the first gain fit model in the voltage scans to determine  $d V_{\rm b}/dT$ values for compensating temperature changes in the gain stabilization studies, we keep this gain fit model as the default and use the results from the second gain fit model to estimate systematic errors. 
However, for the gain stabilization studies we use the second gain fit model as the default except for Hamamatsu MPPCs with trenches. 
 
\begin{figure}[htbp!]
\centering 
\includegraphics[width=72 mm]{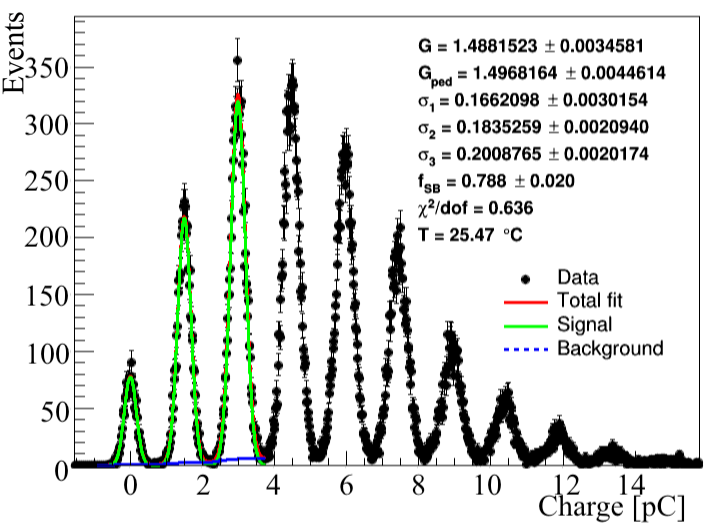}
\includegraphics[width=72 mm]{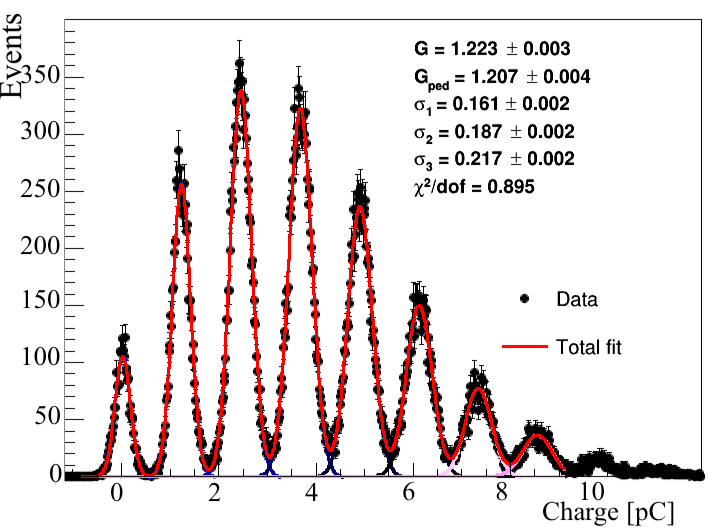}
\caption{Left: measured photoelectron spectrum (black points with error bars) for a Hamamatsu A1-20 MPPC with results of the first gain fit model  overlaid (solid red curve) showing signal (solid green curve) and background (dotted blue curve). Right: measured photoelectron spectrum (black points with error bars)  for a Hamamatsu A1-20 MPPC with results of the second signal model overlaid (solid red curve).  }
\label{fig:model}
\end{figure}

Figure~\ref{fig:model} (left) shows a photoelectron spectrum of the Hamamatsu A1-20  MPPC with the result of the first gain fit model overlaid.   Figure.~\ref{fig:model} (right) shows another photoelectron spectrum the same  MPPC with the result of the second gain fit model overlaid. For Hamamatsu MPPC without trenches both gain fit models yield consistent results. In appendix~\ref{sec:appfits}, we show a direct comparison of the two fitting methodologies for the Hamamatsu MPPC B2-20, the CPTA SiPM 1065 and the KETEK SiPM $\rm PM3350\#1$. To determine the systematic uncertainty due to  the gain fit model we  fit the difference between the  $d V_{\rm b}/dT$ values 
obtained with the two fit methods with a Gaussian function and extract the standard deviation from the fit as the systematic error yielding $\rm 0.3~ mV/^{\circ}$C. 

\begin{figure}[htbp!]
\centering 
\includegraphics[width= 72 mm]{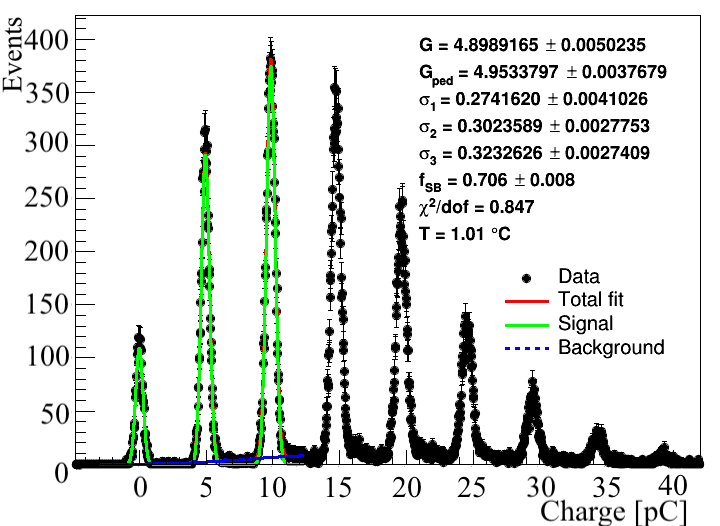}
\includegraphics[width= 72 mm]{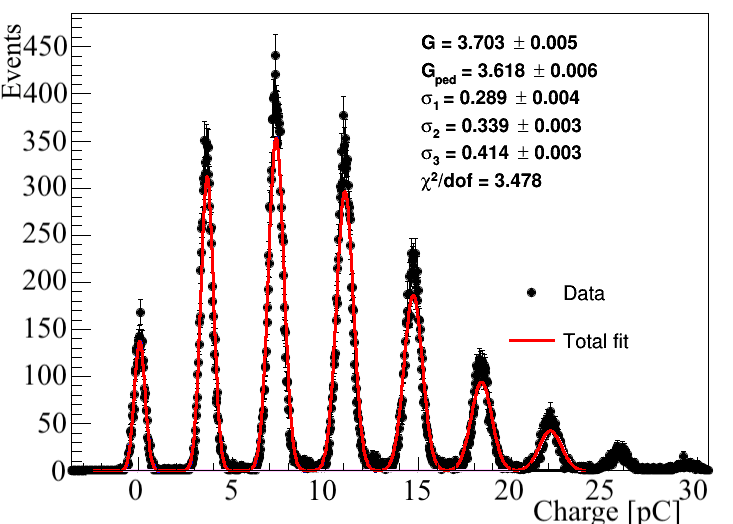}
\caption{Left: measured photoelectron spectrum (black points with error bars) of  the Hamamatsu S13360-1325b MPPC with results of the first gain fit model overlaid (solid red curve) showing signal (solid green curve) and background (dotted blue curve).  Right: measured photoelectron spectrum (black points with error bars) of the Hamamatsu S13360-1325b MPPC with results of the second signal model overlaid (solid red curve).  }
\label{fig:model2}
\end{figure}

Figure~\ref{fig:model2} (left) shows a photoelectron spectrum of the Hamamatsu S13360-1325b MPPC with the result of the first gain fit model overlaid. The fit provides a good representation of the data. The tail of the first photoelectron peak is accounted for  by a small background component while that of the second photoelectron peak has no effect. Figure~\ref{fig:model2} (right) shows a photoelectron spectrum  of the Hamamatsu S13360-1325b MPPC, which is fitted with the second gain fit model. Both the peak and tail regions are not well described by this fit though the peak positions are correctly determined.

\section{Determination of $d V_{\rm b}/dT$}
\label{sec:dvdt}
For the gain stabilization studies we need to know the compensation parameter $dV_{\rm b}/dT$ for each SiPM. We first performed voltage scans at different temperatures in which we typically varied the bias voltage at least by  $\pm 1.5~$V in steps of 0.1~V around the nominal value.  We repeated these measurements at different temperatures in the 5$^{\circ}$C to 45$^{\circ}$C temperature range\footnote{For Hamamatsu $\rm A-$type  MPPCs we used the extended  1$^{\circ}$C to 48$^{\circ}$C temperature range.}. For each temperature, we first plot the gain determined with the first gain fit model as a function of $V_{\rm b}$. Since the gain is  expected to increase linearly with $V_{\rm b}$, we  fit the $V_{\rm b}$ dependence at each temperature point with a first-order polynomial and plot the $dG/dV_{\rm b}$ slopes versus temperature. We expect a uniform distribution since $dG/dV_{\rm b}$ is proportional to the capacitance of the SiPM, which is  expected to be temperature independent.  However, the $dG/dV_{\rm b}$ slopes typically show a small temperature dependence. This was already reported in \cite{SiPMdep}. Though this behavior is not fully understood, an explanation may be that the depletion zone in the high-gain region changes with temperature, which in turn leads to a change of the capacitance. Fits of $dG/dV_{\rm b}$ versus $T$ 
 to a linear function yield deviations from uniformity that are less than $\pm 1\%$ in the entire temperature range. Thus, the average value serves as a good approximation.
The uncertainty on the temperature is dominated by the temperature stability of the climate chamber, which is $\delta T =0.2^{\circ}$C. The error on the $dG/dV_{\rm b}$ is given by the uncertainty of the fit.

Next, we plot  the gain as a function of temperature at each value of $V_{\rm b}$. Since the gain is expected to decrease linearly with temperature, we fit the $T$ dependence at each  $V_{\rm b}$ point with a linear function and extract the slope $dG/dT$. The $dG/dT$  slopes show a small $V_{\rm b}$ dependence that is typically less than  $\pm 1\%$ in the entire $V_{\rm b}$ range. Thus, the average value again provides a good approximation.
   
The value $dV_{\rm b}/dT$ can be obtained from
\begin{equation}
dV_{\rm b}/dT=-\frac{dG/dT}{dG/dV_{\rm b}}.
\label{eq:dvdt}
\end{equation}
Since the errors of  $dG/dV_{\rm b}$  and $dG/dT$ are highly correlated we have to include the covariance term between $V_{\rm b}$ and $T$ in the determination of the statistical error of $dV_{\rm b}/dT$. Therefore, we cannot simply divide  $dG/dT$ by  $dG/dV_{\rm b}$ obtained from the one-dimensional fits. 
Instead, we  perform a two-dimensional fit parameterizing the gain both in terms of  $V_{\rm b}$ and $T$ with the following PDF:
\begin{equation}
G(V_{\rm b},T)= G_0+ \frac{dG(V_{\rm b},T)}{dV_{\rm b}} \big(V_{\rm b} -V_0 \big ) + \frac{dG(V_{\rm b},T)}{dT} \big(T -T_0 \big )  
\end{equation}
where $G_0 = G(V_0,T_0) $, $T_0 =25^\circ \rm C$ and $V_0 =V_{\rm b} (25^\circ \rm C)$.
For all Hamamatsu MPPCs we use all gain measurements  for which the $\chi^2$ value of the fit  with the first gain fit model is less than two. For CPTA and KETEK SiPMs the limit is increased to ten.  We fit for $G_0 $, $dG(V_{\rm b},T)/dV_{\rm b}$ and $dG(V_{\rm b},T)/dT $. The fit correctly determines all the errors on the three fit parameters.  We then use eq.~\ref{eq:dvdt} to determine $dV_{\rm b}/dT$. The error is obtained from the errors of $dG/dV_{\rm b}$  and $dG/dT$ and the covariance term between $V_{\rm b}$ and $T$.
In addition, we scale the final errors of  $dG/dV_{\rm b}$,  $dG/dT$ and $dV_{\rm b}/dT$ with the square root of the $\chi^2/DOF$ value of the two-dimensional gain fit where $DOF$ denotes the degrees of freedom in the fit. The latter step is particularly  important for KETEK SiPMs for which the gain dependence on $V_{\rm b}$ and $T$ is more complicated and for which the $\chi^2$ values of the fit tend to be rather high. We applied this procedure to all 30 SiPMs. Table ~\ref{tab:dvdt-Hamamatsu} summarizes the results for $dG/dV_{\rm b}$, $dG/dT$  and $dV_{\rm b}/dT$ for all Hamamatsu MPPCs obtained with the first gain fit model. 
The systematic error is obtained from the $rms$ of a Gaussian that is used to fit  the difference of $dV_{\rm b}/dT$ values determined with the first and second gain fit models. Please note that we tested the S13360-1325 MPPCs twice, once together with the LCT MPPCs and a second time with the S13360-3025 MPPCs. Table ~\ref{tab:dvdt-Ketek} summarizes the corresponding results for KETEK and CPTA SiPMs.

\begin{table}
\centering
\caption{Temperature characterization measurements for Hamamatsu MPPCs showing the studied temperature range, the measured values for $dG/dV_{\rm b}$, $dG/dT$  and $dV_{\rm b}/dT$ as well as the compensation parameter $(dV_{\rm b}/dT)_s$ used in the stabilization runs. The results come from a two-dimensional fit. Errors on $dG/dV_{\rm b}$ and $dG/dT$ are statistical only. The first error on $dV_{\rm b}/dT$ is statistical and the second error is systematic. }
\label{tab:dvdt-Hamamatsu}
\begin{tabular}{|c|c|c|c|c|c|}
\hline
SiPM  & $T$ range   & $dG/dV_{\rm b}$ & $dG/dT$ & $dV_{\rm b}/dT$  & $(dV_{\rm b}/dT)_s$ \\
&   $[^{\circ} C]$  &$[10^{6}/V]$  & $[10^{5}/^{\circ} C]$ & $[mV/^\circ \rm C]$ & $[mV/^\circ \rm C]$  \\ 
\hline 
A1-20&  2-48 & $4.306\pm0.008$ & $-2.536\pm0.008$ & $59.1\pm0.1\pm 0.3$ & 59.0 \\  
A2-20&  2-48 & $3.600 \pm0.007$ & $-2.135\pm0.007$ & $59.3\pm0.1\pm0.3$ & \\  
A1-15 &  2-48 & $3.427\pm0.006$ & $-2.020\pm0.005$ & $58.9\pm0.1\pm0.3$ &\\  
A2-15 &  2-48 & $3.518\pm0.008$ & $-2.085\pm0.008$ & $59.3\pm0.2\pm0.3$ & \\  
\hline
B1-20 &  5-45 & $4.343\pm0.008$ & $-2.471\pm0.008$ & $56.9\pm0.1\pm0.3$ & 57.8 \\  
B2-20 &  5-45 & $3.640\pm0.008$ & $-2.143\pm0.008$ & $57.8\pm0.2\pm0.3$ & \\  
B1-15 &  5-45 & $3.320\pm0.021$ & $-1.933\pm0.021$ & $58.2\pm0.4\pm0.3$& \\  
B2-15 &  5-45 & $3.329\pm0.008$ & $-1.902\pm0.008$ & $57.1\pm0.2\pm0.3$&  \\  
\hline
S12571-010a &  5-45 & $1.285\pm0.004$ & $-0.824\pm0.004$ & $64.1\pm0.3\pm 0.3$ &64.3 \\  
S12571-010b &  5-45 & $0.987\pm0.003$ & $-0.646\pm0.003$ & $65.5\pm0.3\pm 0.3$ &\\  
S12571-015a &  5-45 & $2.989\pm0.012$ & $-1.848\pm0.012$ & $61.6\pm0.4\pm 0.3$& 63.1 \\  
S12571-015b &  5-45 & $3.060\pm0.014$ & $-1.904\pm0.013$ & $62.2\pm0.4\pm 0.3$ & \\
\hline
S13360-1325a &  5-45 & $6.298\pm0.015$ & $-3.482\pm0.025$ & $55.3\pm0.4\pm 0.3$ & 57.0 \\  
S13360-1325b & 5-45 & $6.287\pm0.010$ & $-3.488\pm0.018$ & $55.5\pm0.3\pm 0.3$  &\\  
LCT4\#6 &   5-45 &$11.085\pm0.020 $ & $-5.930\pm 0.020$ & $53.5\pm  0.1\pm 0.3 $ & 54 \\
LCT4\#9 &  5-45 & $10.96\pm 0.010$ & $-5.820\pm 0.015$&$ 53.1\pm 0.1\pm 0.3$  & \\
\hline
S13360-3025a &  5-45 & $5.521\pm0.012$ & $-3.144\pm0.025$ & $56.9\pm0.3 \pm 0.3$  & 57.2 \\  
S13360-3025b &  5-45 & $4.602\pm0.015$ & $-2.726\pm0.021$ & $59.2\pm0.5 \pm 0.3$ & \\  
S13360-1325a &  5-45 & $5.535\pm0.008$ & $-3.160\pm0.009$ & $57.1\pm0.1\pm 0.3$ & \\  
S13360-1325b &  5-45 & $4.647\pm0.007$ & $-2.745\pm0.008$ & $59.1\pm0.1\pm 0.3$ & \\  

\hline
\end{tabular}
\end{table}

\begin{table}
\centering
\caption{Temperature characterization measurements for KETEK and CPTA SiPMs showing the studied temperature range, the measured values for $dG/dV_{\rm b}$, $dG/dT$  and $dV_{\rm b}/dT$ as well as the compensation parameter used in the stabilization runs. For CPTA 975 the two rows correspond to values before and after the gain change. Errors on $dG/dV_{\rm b}$ and $dG/dT$ are statistical only. The first error on $dV_{\rm b}/dT$ is statistical and the second error is systematic.  }
\label{tab:dvdt-Ketek}
\begin{tabular}{|c|c|c|c|c|c|}
\hline
SiPM &  $T$ range  & $dG/dV_{\rm b}$  & $dG/dT$ & $dV_{\rm b}/dT$  & $(dV_{\rm b}/dT)_s$\\ 
 &    $[^{\circ} C]$  & $[10^{6}/V]$  & $[10^{5}/^{\circ} C]$ &$[mV/^\circ \rm C]$ &$[mV/^\circ \rm C]$  \\ 
\hline 
W12A &  2-40 & $7.29\pm0.02$ & $-1.53\pm0.03$ & $20.9\pm0.4\pm 0.3$& 18.3  \\  
W12B &  2-40 & $6.15\pm0.05$ & $-1.47\pm 0.07$ & $23.8\pm1.1\pm 0.3$ & \\  
\hline
PM3350\#1 & 1-30 & $11.2\pm0.16$ & $-2.84\pm0.22$ & $25.4\pm2.0\pm 0.3$&18.3 \\  
PM3350\#2 & 1-30 & $10.75\pm0.10$ & $-2.20\pm0.11$ & $20.4\pm1.1\pm 0.3$ &\\  
PM3350\#5  &  1-30 & $10.30\pm0.23$ & $-0.92\pm0.13$ & $20.9\pm1.6\pm 0.3$  & \\  
PM3350\#6  &  1-30 & $8.97\pm0.18$ & $-1.90\pm0.25$ & $21.1\pm2.7\pm 0.3$  &\\  
PM3350\#7  &  1-30 & $10.68\pm0.25$ & $-1.65\pm 0.31$ & $15.4\pm2.9\pm 0.3$  & \\  
PM3350\#8  &  1-30 & $10.48\pm$ 0.12 & $-2.31\pm0.15$ & $22.1\pm1.4\pm 0.3$  &\\  
\hline \hline
CPTA ~ 857&  5-40 & $7.32\pm0.33$ & $-1.74\pm0.13$ & $23.8\pm0.1\pm 0.3$ & 21.2\\  
CPTA ~922 &  5-40 & $18.02\pm0.54$ & $-3.71\pm0.32$ & $20.6\pm1.6\pm 0.3$&  \\  
CPTA ~975  &  5-40 & $16.05\pm1.94$ ~& $-4.41\pm1.14$ & $27.6\pm6.8\pm 0.3$ & \\  
CPTA~975  & 5-40 & $8.22\pm0.19$ ~& $-2.13\pm0.08$ & $25.9\pm0.5\pm 0.3$ & \\  
CPTA ~1065  &  5-40 & $20.42\pm0.14$ & $-4.70\pm0.07$ & $23.0\pm0.3\pm 0.3$ &  \\  
\hline
\end{tabular}
\end{table}

\begin{figure}
\centering 
\includegraphics[width=75 mm]{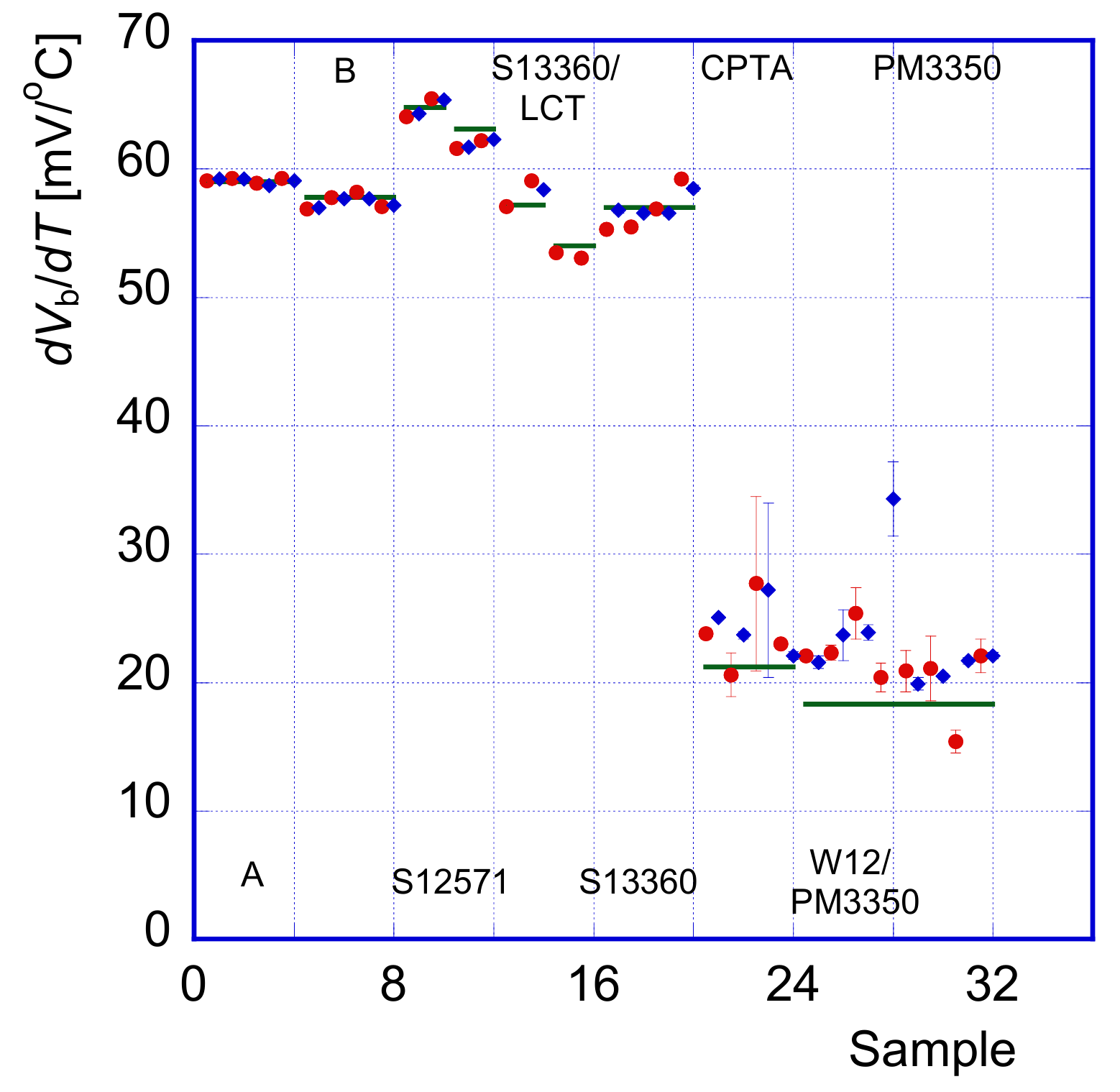}
\includegraphics[width=75 mm]{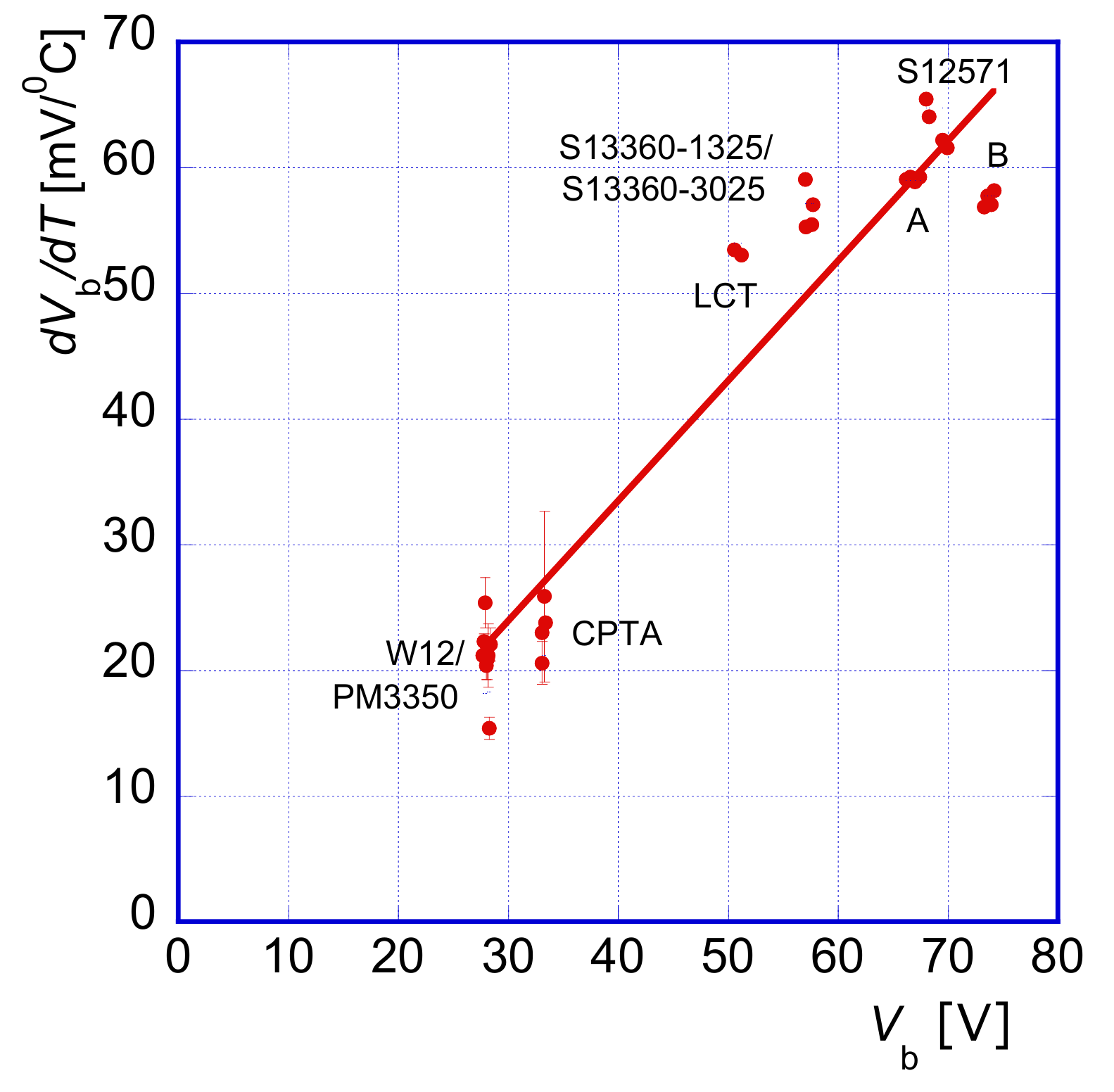}
\caption{Left: Measured $dV_{\rm b}/dT$ values for gains extracted with the first gain fit model (red points with error bars) and those extracted with the second gain fit model (blue diamonds with error bars) in comparison to the selected compensation parameters used in the gain stabilization study (green lines). Right: Correlation of $dV_{\rm b}/dT$ (red points with error bars) versus nominal bias voltage. The gains were extracted with the first gain fit model. The red line shows a fit to the data points.}
\label{fig:dvdt-vbias}
\end{figure}

Figure~\ref{fig:dvdt-vbias} (left) shows the $dV_{\rm b}/dT$ values obtained with the first and second gain fit models for all 30 tested SiPMs in addition to the compensation parameter used in the gain stabilization study. For Hamamatsu MPPCs, the two gain fit models produce $dV_{\rm b}/dT$ values that are in good agreement. The $\chi^2$ values of the two-dimensional fits are acceptable and the compensation parameters, which were selected from the results of 
bias voltage scans at a few temperature points before the overnight gain stabilization run, are consistent with the $dV_{\rm b}/dT$ values obtained from the complete bias voltage scans.
While for CPTA SiPMs the compensation value is consistent with the extracted $dV_{\rm b}/dT$, that for KETEK SiPM is systematically too low. In addition, we observe large $\chi^2$ values in the two-dimensional fits indicating that the $V_{\rm b}$-versus-$T$ dependence 
is more complicated than linear.

Figure~\ref{fig:dvdt-vbias} (right)  shows the measured compensation parameter $dV_{\rm b}/dT$ versus the nominal  bias voltage. For Hamamatsu $\rm A-$type   and S12571 MPPCs and KETEK W12/PM3350 SiPMs the measurements lie on a line. 
For $\rm B-$type  MPPCs and CPTA SiPMs compensation parameters  lie slightly below this line while that for MPPCs with trenches they lie slightly above.  This indicates that the compensation parameters scale approximately with the nominal bias voltage. 
The relative spread in $dV_{\rm b}/dT$ is typically larger for KETEK and CPTA SiPMs  than that for Hamamatsu MPPCs.

\subsection{Measurements of Hamamatsu Detectors}
\label{sec:dvdt-hamamatsu}

\begin{figure}[htbp!]
\centering 
\includegraphics[width=70 mm]{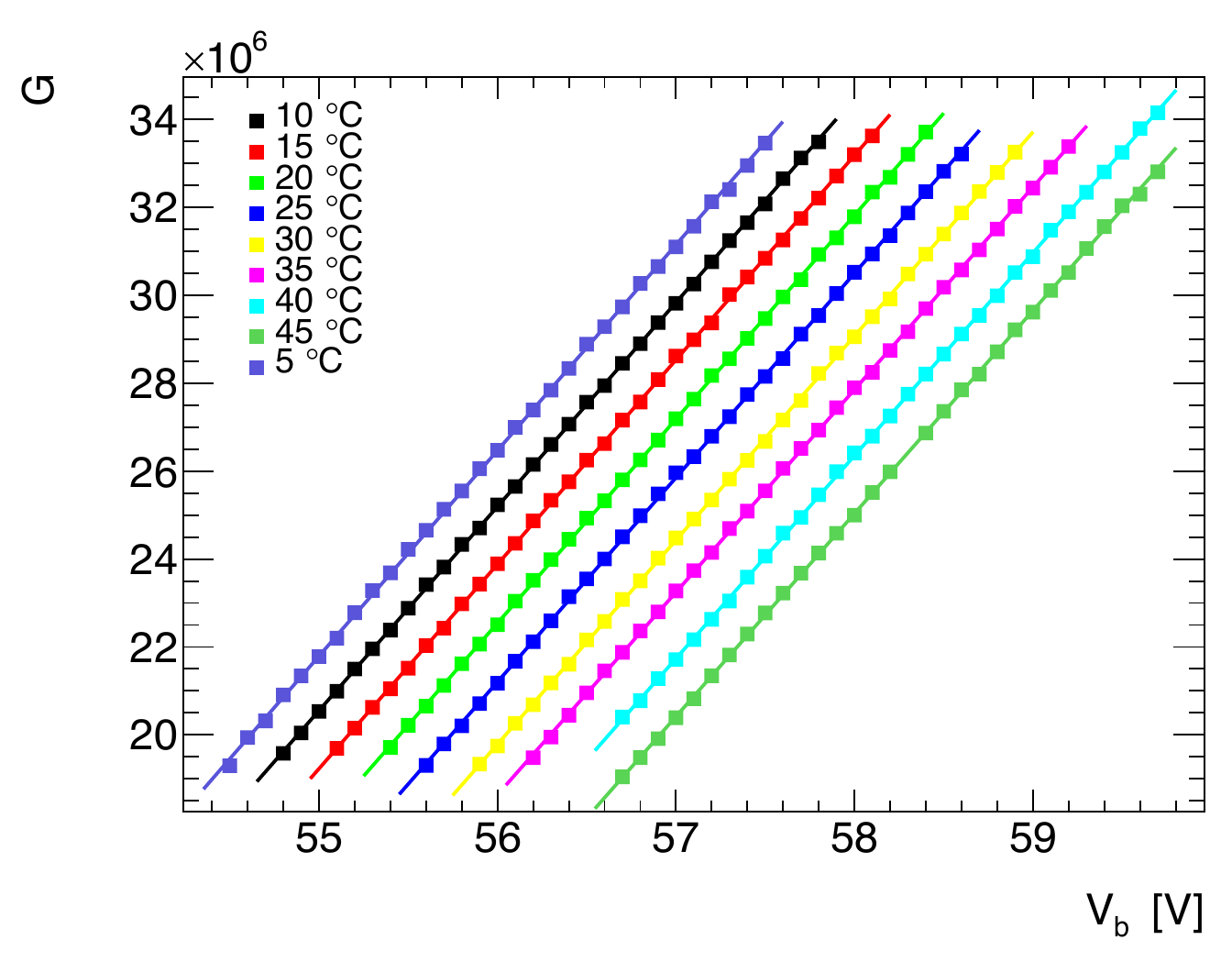}
\includegraphics[width=70 mm]{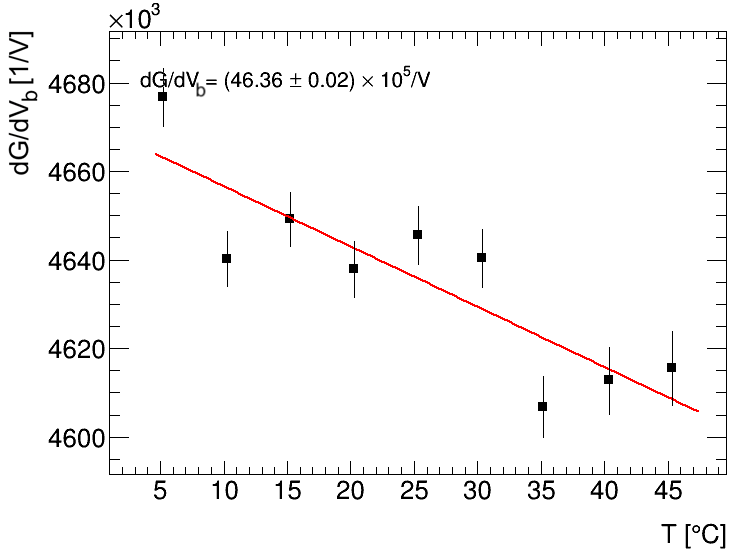}
\caption{Left: measurements  of gain (points) versus bias voltage at different temperatures for a Hamamatsu MPPC S13360-1325b with the results of the one-dimensional fits  overlaid (solid lines). Right:
extracted  $dG/dV_{\rm b}$ slopes (points with error bars) versus temperature with the result of a fit to a first-order polynomial overlaid (solid red line).  }
\label{fig:dgdv}
\end{figure}

\begin{figure}[htbp!]
\centering 
\includegraphics[width=70 mm]{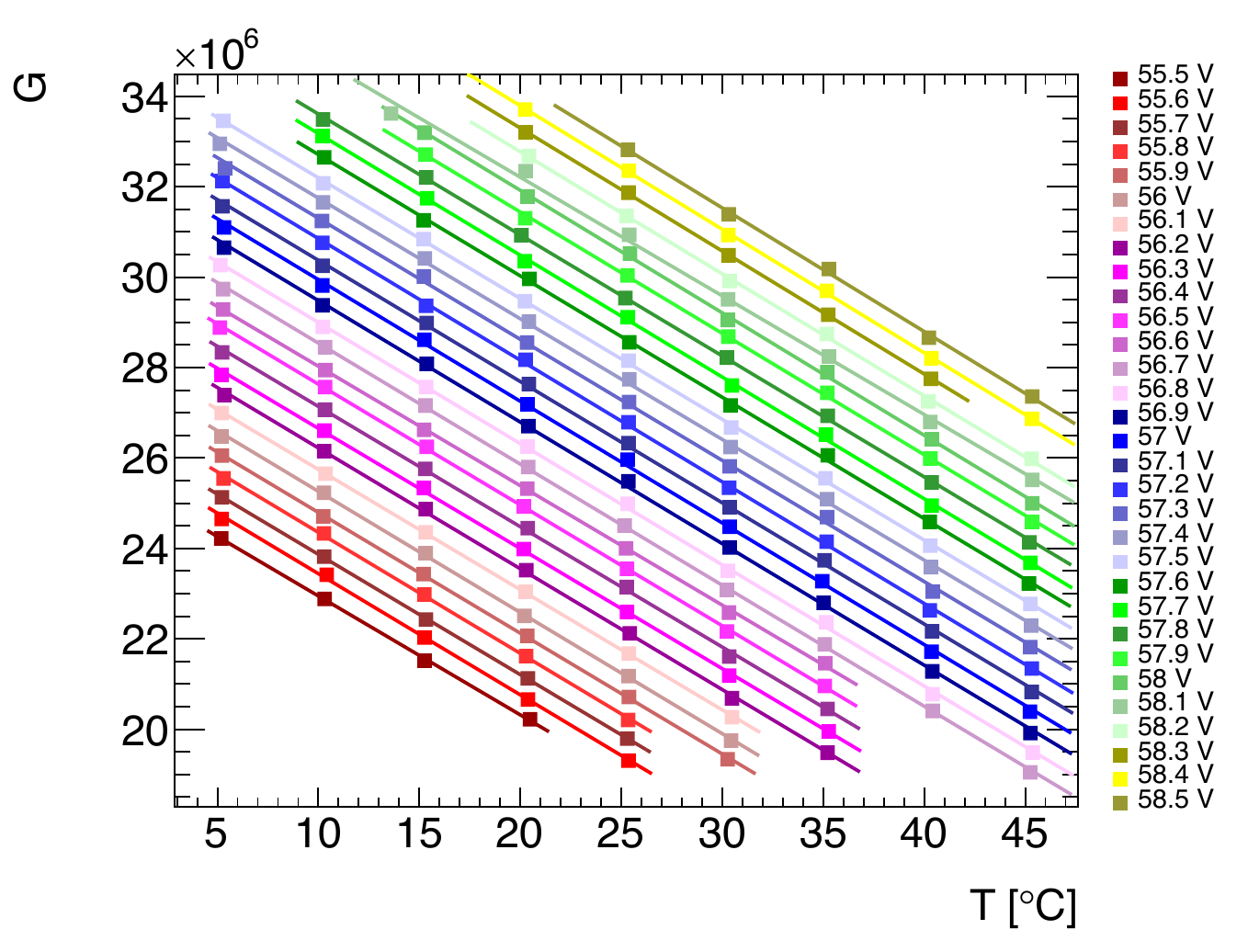}
\includegraphics[width=70 mm]{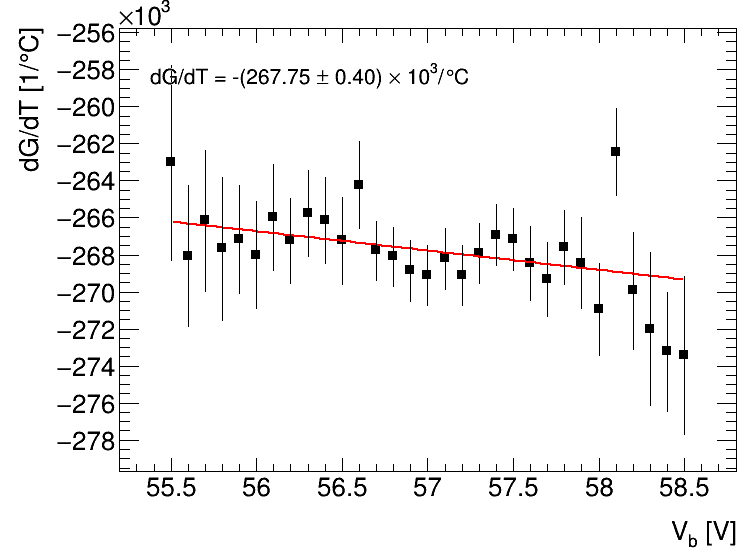}
\caption{Left: measurements of gain (points) versus  temperature for different  bias voltages for a Hamamatsu  MPPC S13360-1325b with the results of the one-dimensional fits overlaid (solid lines). Right: extracted  $dG/dT$ slopes (points with error bars) versus bias voltage with the result of a fit to a first-order polynomial overlaid (solid red line).  
}
\label{fig:dgdt}
\end{figure}

\begin{figure}[htbp!]
\centering 
\includegraphics[width=70 mm]{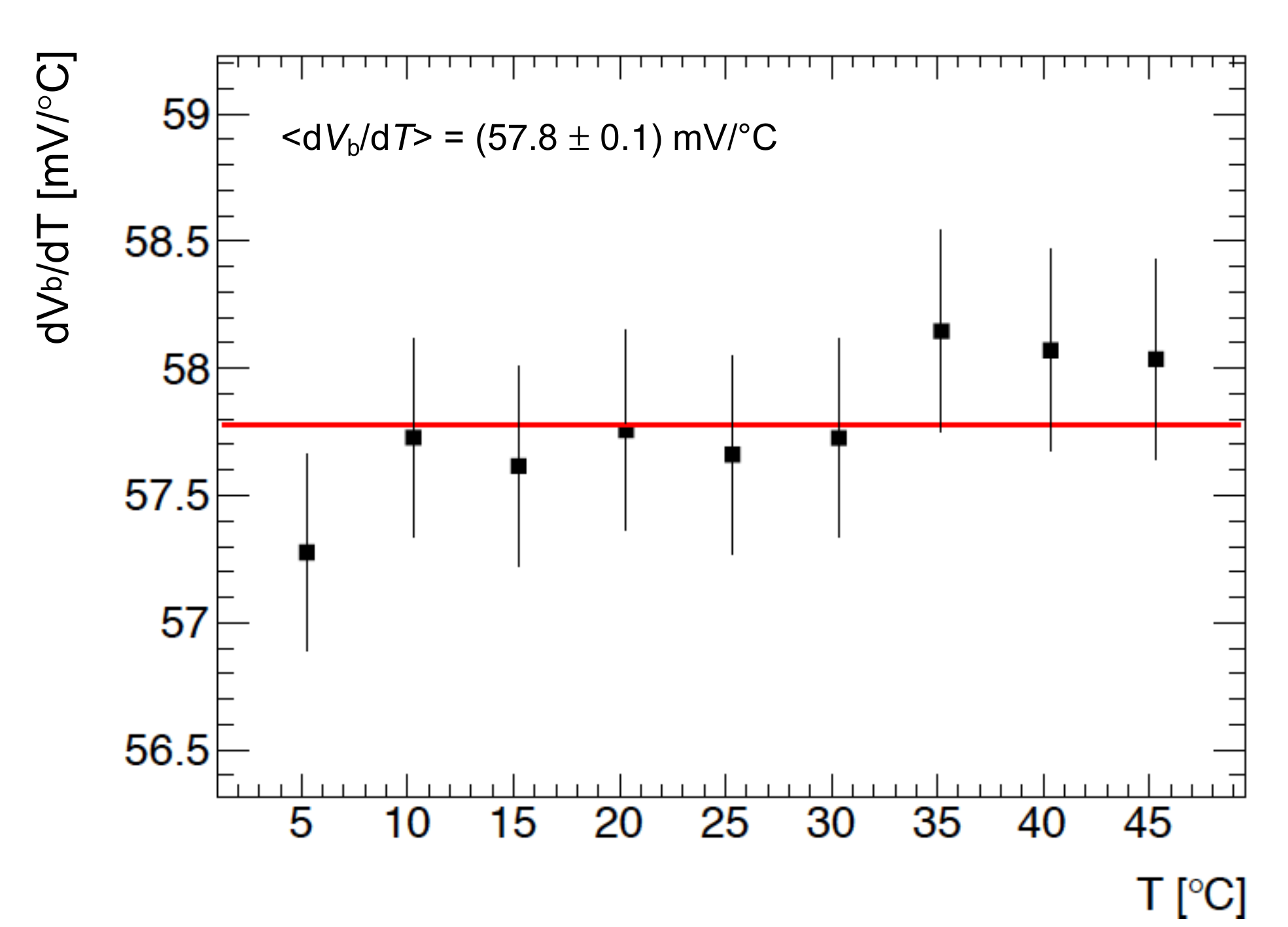}
\includegraphics[width=70 mm]{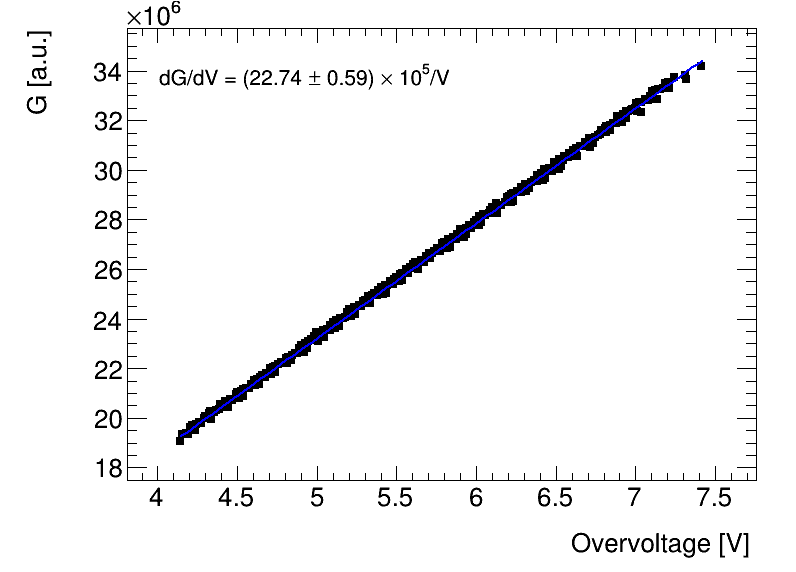}

\caption{Left: Measurements of the $dV_{\rm b}/dT$ slopes (points with error bars)  versus  temperature  for a Hamamatsu MPPC S13360-1325b with the
result of a line fit overlaid (solid red line). Right: measurements of gain versus overvoltage combining the measurements at all temperatures.}

\label{fig:dvdt}
\end{figure}

For illustrative purpose, we present also the $G$-versus-$V_{\rm b}$ measurements for different $T$ and $G$-versus-$T$ measurements for different $V_{\rm b}$ with the results of the one-dimensional fits overlaid for which the gain was determined from the waveform-integrated photoelectron spectrum  with the first gain fit model.  For example, Figs.~~\ref{fig:dgdv} (left) and \ref{fig:dgdt} (left) depict these measurements for the Hamamatsu  S13360-1325b MPPC, respectively.
Figures~\ref{fig:dgdv} (right) and \ref{fig:dgdt} (right)  show the extracted slopes  $dG/dV_{\rm b}$ versus $T$ and $dG/dT$ versus  $V_{\rm b}$, respectively.
Both $dG/dV_{\rm b}$ and  $dG/dT$  respectively decrease with $T$ and $V_{\rm b}$. However, the average values
\begin{eqnarray}
dG(V_{\rm b},~T)/dV_{\rm b}& =&~(4.636\pm 0.002_{\rm stat})\times 10^6 /V \\
 dG(V_{\rm b},~T)/dT &=&-(2.678\pm 0.004_{\rm stat})\times 10^5 /^\circ C. 
\end{eqnarray}
\noindent 
provide a good approximation since  deviations from non-uniformity are less than $\pm 0.6\%$  in the  5$^{\circ}$C to 45$^{\circ}$C temperature range and less than $\pm 0.4\%$  in the 55.5 to 58.5~V range, respectively. The averages are consistent with the result of the two-dimensional fit.  Figure~\ref{fig:dvdt} (left) shows the  $dV_{\rm b}/dT$ values versus $T$ that result from dividing the $dG/dT$ average value by each 
$dG/dV_{\rm b}$ value. The error bars are dominated by the error on  $dG/dT$. A simple line fit yields
\begin{equation}
\langle dV_{\rm b}/dT \rangle =(57.8 \pm 0.1_{\rm stat} ) \rm ~mV/^\circ C,
\end{equation}
The error is underestimated since the correlation among the individual  $dV_{\rm b}/dT$ points is not accounted for. Representative plots for the other MPPC types are shown in appendix~\ref{sec:appHamamatsuBV}.
Figure~\ref{fig:dvdt} (right) shows the measured gain versus overvoltage for all temperatures combined for the S13360-1325b MPPC. The gain depends linearly on overvoltage independent of temperature.


\subsection{Measurements of KETEK  and CPTA SiPMs}
For KETEK and CPTA SiPMs we proceed in a similar way. In addition to the two-dimensional fits of the gain versus $V_{\rm b}$ and versus $T$, we also performed one-dimensional fits of the $G$-versus-$V_{\rm b}$ and $G$-versus-$T$ measurements and extracted the slopes  $dG/dV_{\rm b}$  and $dG/dT$.
While all tested CPTA SiPMs and the KETEK experimental devices W12 work in the 5$^{\circ}$C to 40$^{\circ}$C temperature range, 
the KETEK PM3350 SiPMs do not work properly at temperatures above 30$^{\circ}$C. Furthermore, we noticed that
the gain  of  CPTA 975 SiPM  changed during the bias voltage scan when the temperature was set to  44$^{\circ}$C. This gain change, however, did not affect its performance afterwards. Representative plots for the KETEK W12, PM3350 and CPTA SiPMs are depicted in appendices~\ref{sec:appKetekBV} and ~\ref{sec:appCPTABV}, respectively.

\subsection{Analytical Description of the Bias-Voltage-versus-Temperature Dependence} 
\label{sec:VT}
The functional dependence of  $V_{\rm b}$ on temperature can be approximated by an analytic expression. The gain change $\Delta G$ is given by:
\begin{equation}
\Delta G(V_{\rm b},~T)=\frac{dG(V_{\rm b},~T)}{dT}\cdot \Delta T + \frac{ dG(V_{\rm b},~T)}{ dV_{\rm b}}\cdot \Delta V_{\rm b}.
\end{equation}
For constant gain, $\Delta G(V_{\rm b},T)=0$  yielding
\begin{equation}
\Delta V_{\rm b}/ \Delta T=-\frac{dG(V_{\rm b},~T)/dT}{dG(V_{\rm b},~T)/dV_{\rm b}}.
\label{eq:5}
\end{equation}

\begin{figure}
\centering
\includegraphics[width=2.7in]{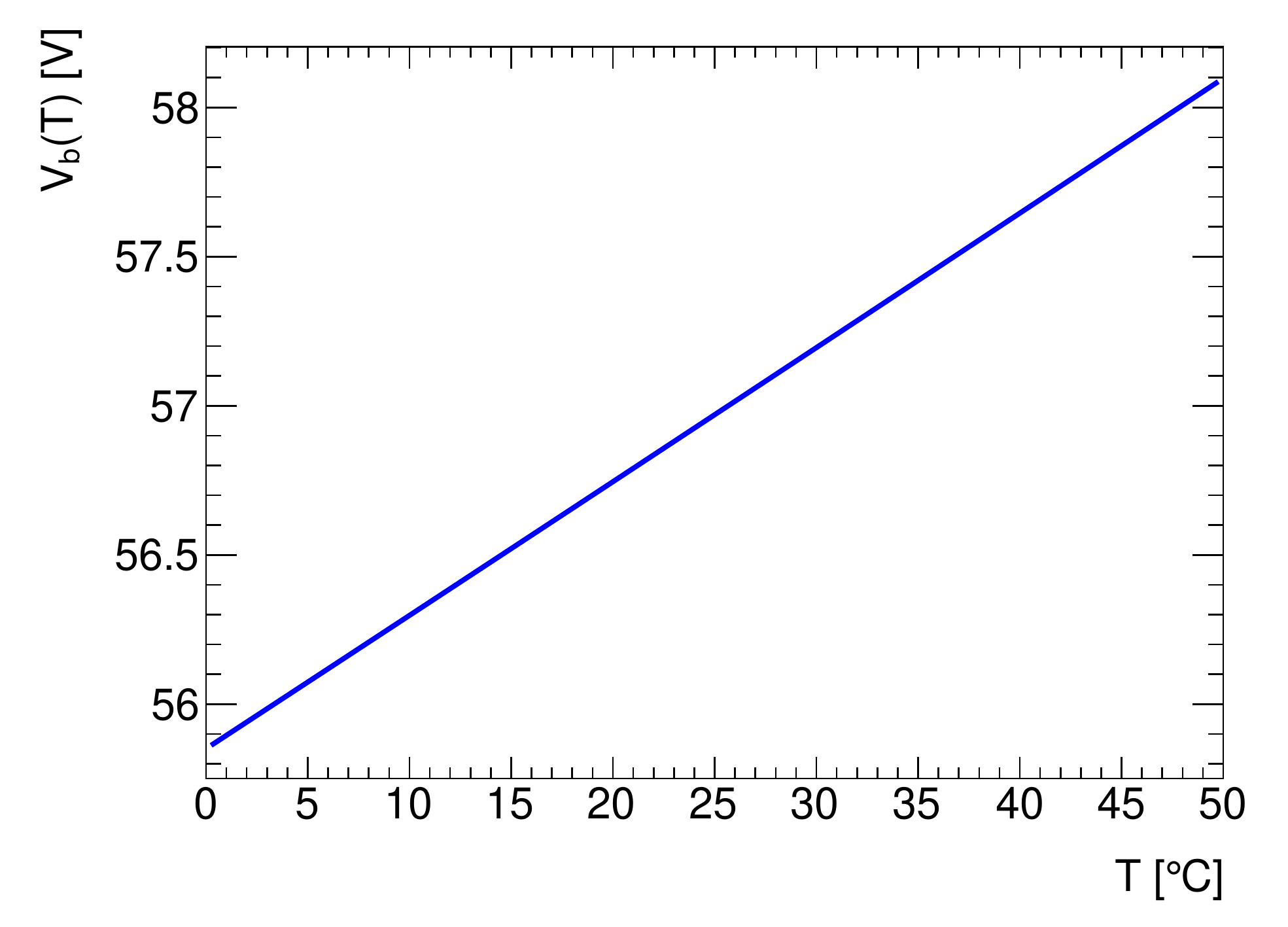}
\caption{\label{fig:A1_ch1_temp4} Calculated dependence of bias voltage versus temperature for Hamamatsu MPPC S13360-1325b using eq.~\ref{eq:9}.}
\end{figure}

Since we observe a linear dependence for  both $dG/dV_{\rm b}$ versus $T$ and $dG/dT$ versus $V_{\rm b}$, we model them by first-order polynomials
\begin{equation}
\frac{ dG(V_{\rm b},~T)}{dT}= a + b\cdot V_{\rm b}(T)
\label{eq:6}
\end{equation}  
\begin{equation}
\frac{dG(V_{\rm b},~T)}{dV_{\rm b}}= c + d\cdot T
\label{eq:7}
\end{equation}
\noindent where $a (c)$ are offsets and $b(d)$ are slope parameters determined from fits. Plugging eqs.~\ref{eq:6} and ~\ref{eq:7} into eq~\ref{eq:5} yields
\begin{equation}
\frac{dV_{\rm b}}{dT}=- \frac{a + b\cdot V_{\rm b}(T)}{c + d\cdot T}
\label{eq:8}
\end{equation}
This is a first-order differential equation. The solution for $b\neq 0$ and $d\neq 0$ is simply
\begin{equation}
V_{\rm b}(T) = -\frac{a}{b} + \frac{K}{(c + d\cdot T)^{\frac{b}{d}}}
\label{eq:9}
\end{equation}
\noindent where $K$ is an integration constant. For $b=0$  and $d$=0, the solution becomes 
\begin{equation}
V_{\rm b}(T) = -\frac{a}{c} \cdot T +K.
\label{eq:10}
\end{equation}
For small values of $b$ and $d$, the linear dependence is a good approximation. This is basically what we observe for Hamamatsu and CPTA sensors. 
For example, Fig.~\ref{fig:A1_ch1_temp4} shows $V_{\rm b}(T)$ for a Hamamatsu MPPC S13360-1325b. The dependence looks rather linear since the parameters $b$ and $d$ are small. The gain variations over the entire $V_{\rm b}$ range and entire temperature range are less than $\pm 1\%$.

For $b \neq 0$ and $d=0$, the solution becomes
\begin{equation}
V_{\rm b}(T) = -\frac{a}{b} +K \cdot \exp{\Big({-\frac{b \cdot T}{c }}\Big)},
\end{equation}
while for $b\ne 0$ and $d=0$ we get 
\begin{equation}
V_{\rm b}(T) = K- \frac{a \log {(c+d \cdot T)}}{d},
\end{equation}
The solution for quadratic dependences of $dG/dV_{\rm b}$ and $dG/dT$ is shown in appendix~\ref{sec:app-VT}.  

\section{Gain  Stabilization Measurements}
\label{sec:gainstab}

We performed the gain stabilization studies by simultaneously readjusting the bias voltage of four SiPMs  with the bias voltage regulator using a single compensation parameter $dV_{\rm b}/dT$. We selected this value such that it provided the best match for all four SiPMs based on the results of the bias voltage scans. 
Since we did not have sufficient time before a stabilization run to perform bias voltage scans at all selected temperature points, we determined  non-optimal   $dV_{\rm b}/dT$ values in some cases, particularly for KETEK SiPMs. Typically, we did half the bias voltage scans before and half after the stabilization runs.  We started and ended a stabilization run at $25^{\circ}$C as a reference. 
Then, we jumped to the maximum temperature (typically 48$^\circ$C and ramped down to $1^{\circ}$C in steps of $5.0^{\circ}$C or $2.5^{\circ}$C.  In the $20- 30^{\circ}$C temperature range, we reduced the step size to  $2.0^{\circ}$C. Since stable temperature is reached after 15 minutes, we typically stayed 30 minutes at each temperature point where  we recorded at least $\sim 40$ samples with 50000 waveforms each. 
We fit the gain of the 40 samples with a Gaussian function to extract the mean value and the $rms$ error.

The deviation from uniformity is defined as 
\begin{equation}
\frac{\Delta G}{G_0}=  \pm(dG/dT)\times \frac{\Delta T}{G_0}. 
\end{equation}
where $\Delta T= 5.0^{\circ}$C and  $G_0$ is the gain at nominal bias voltage at $25^{\circ}$C.

\subsection{Gain Stabilization of Hamamatsu MPPCs}
\begin{figure}[H]
\centering 
\includegraphics[width=2.9in]{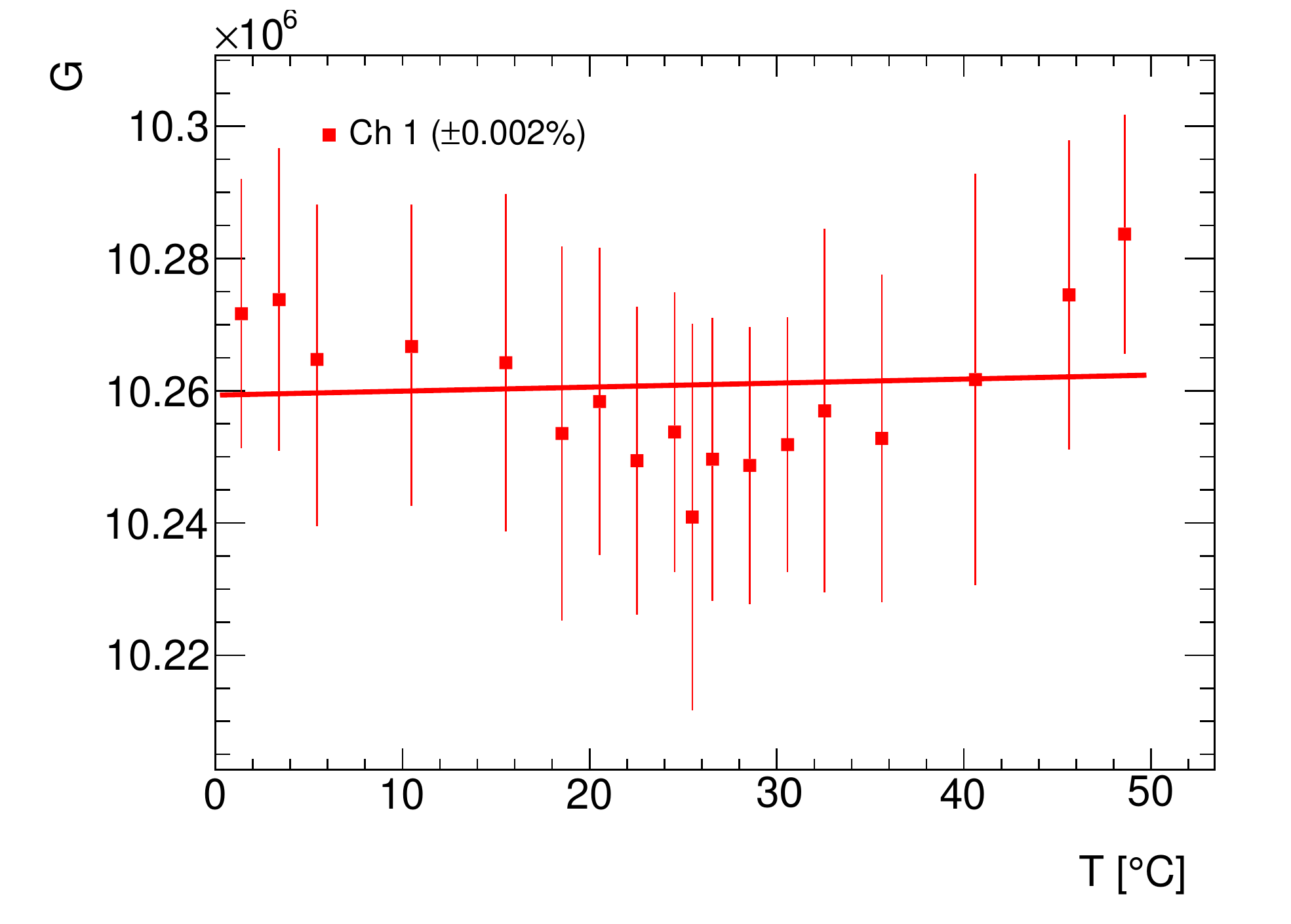}
\includegraphics[width=2.9in]{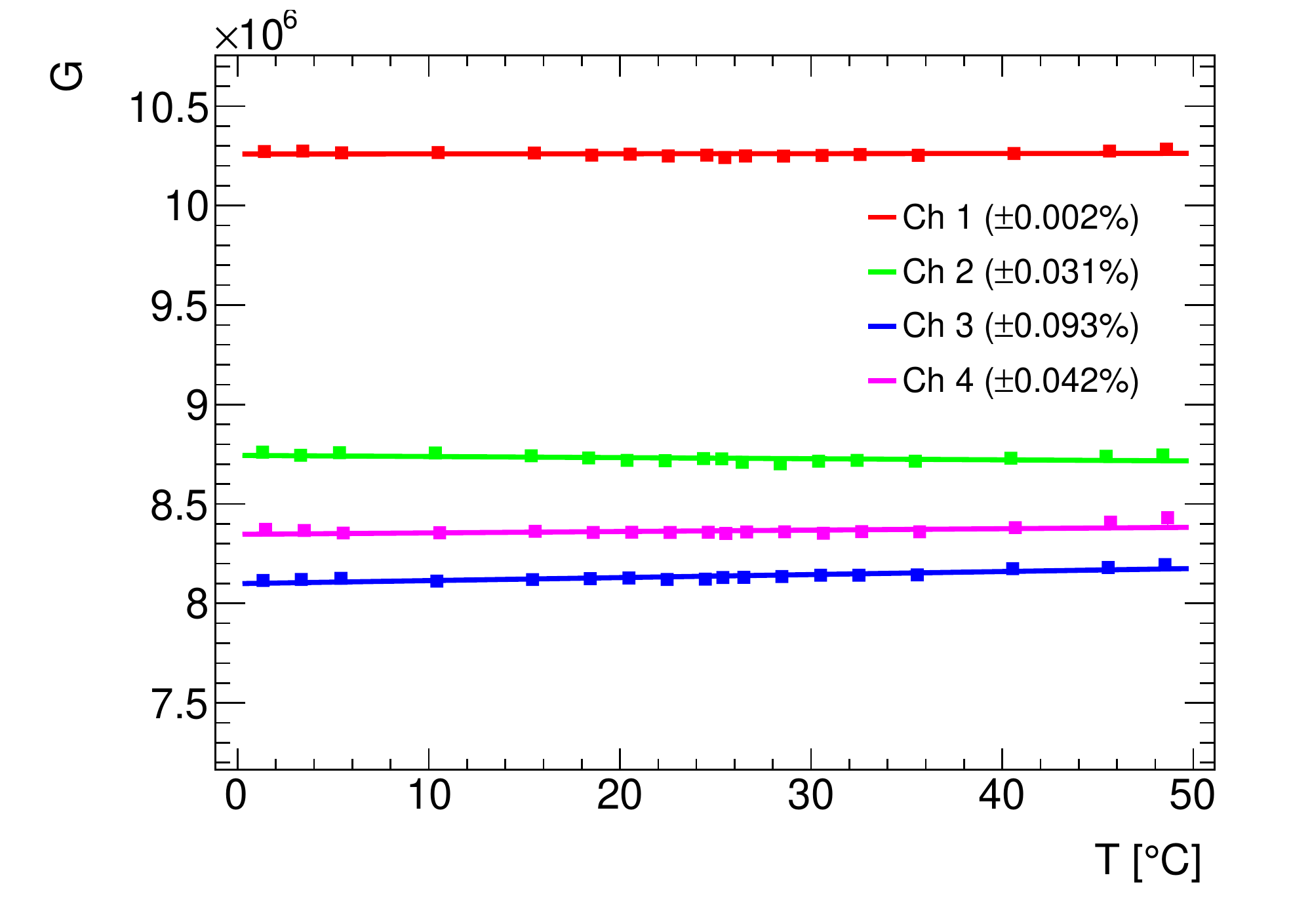}\\
\includegraphics[width=2.9in]{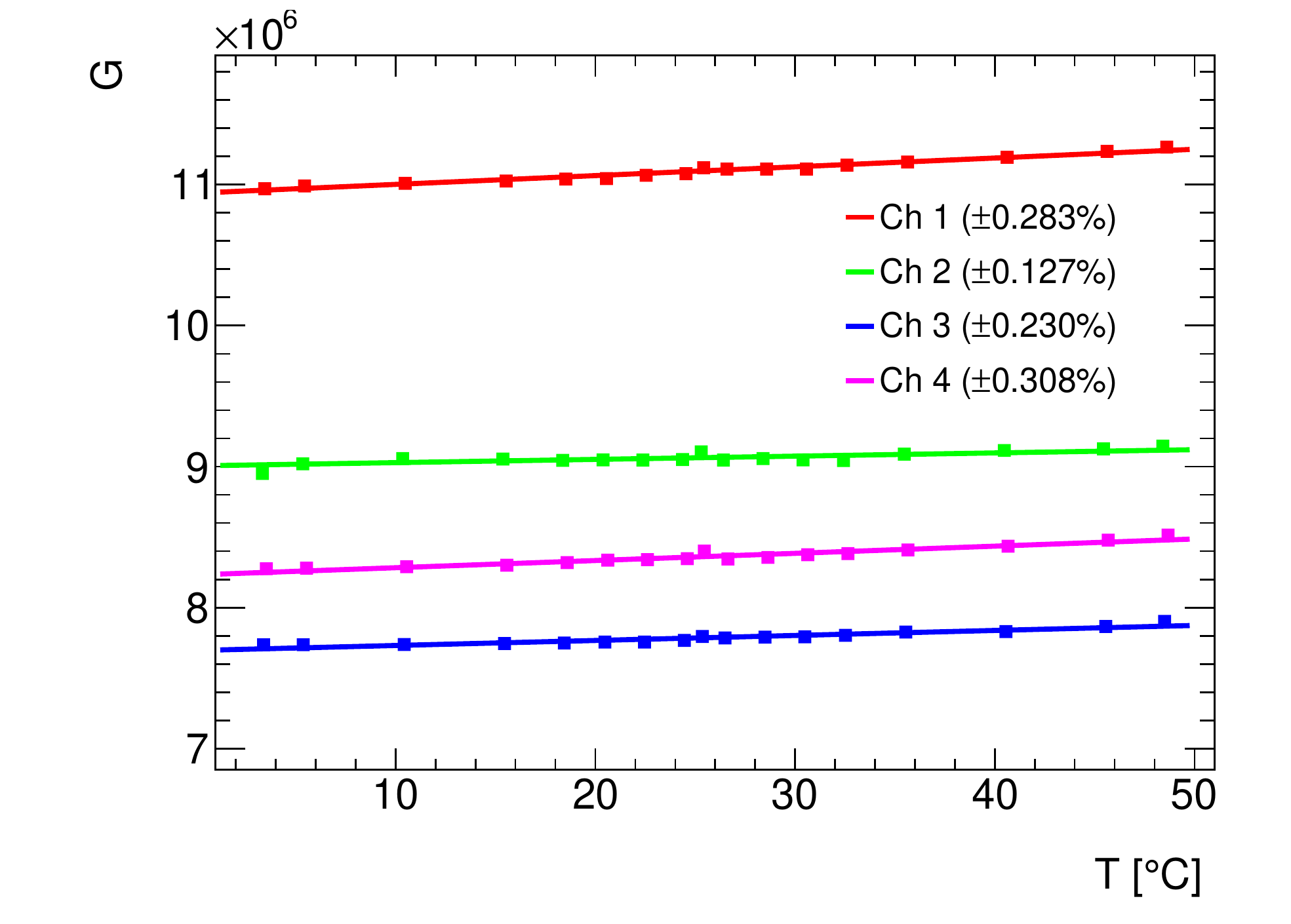}
\includegraphics[width=2.9in]{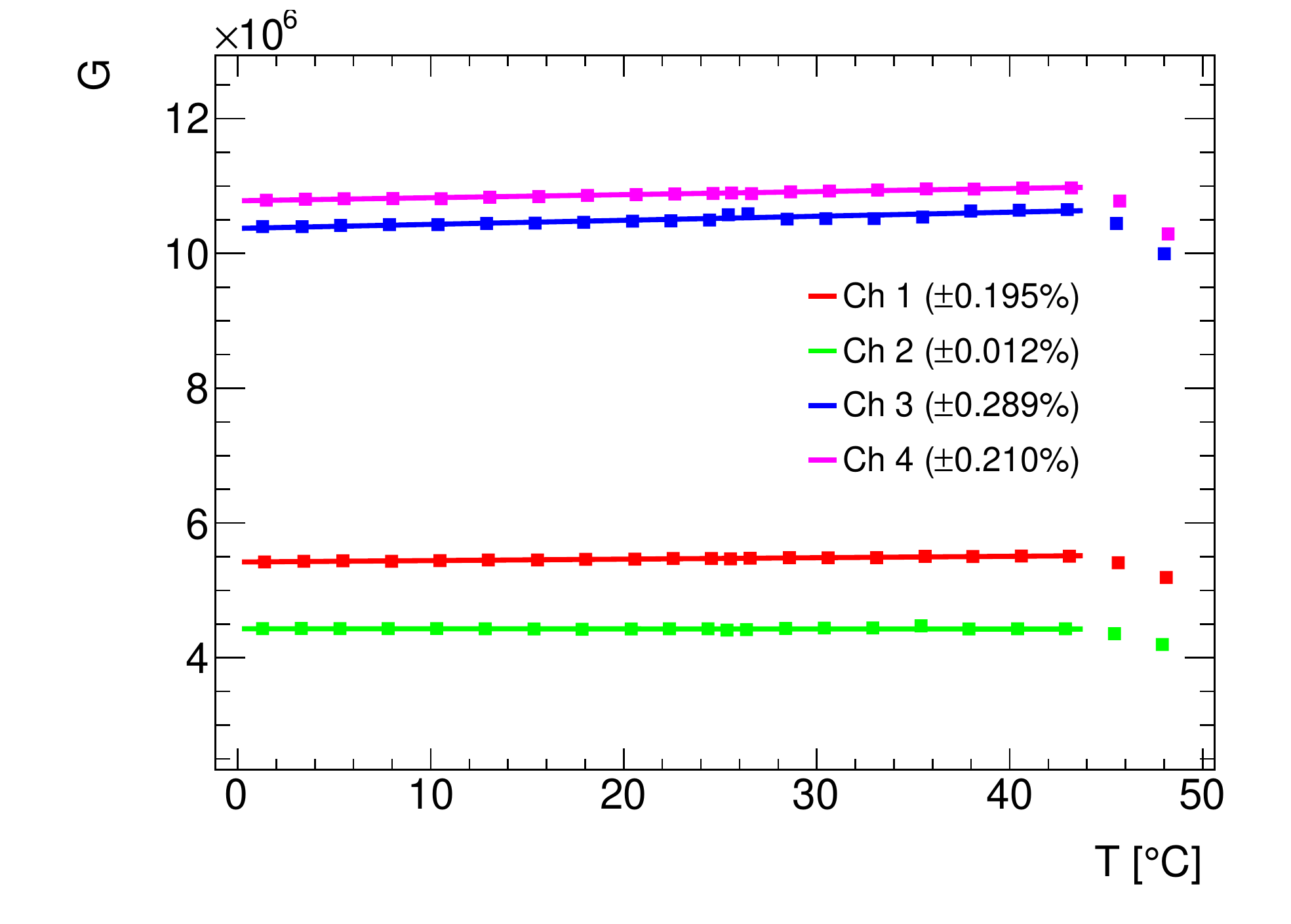}
\caption{ Measurements of stabilized gain versus temperature for Hamamatsu MPPCs without trenches in which the gain is extracted from the photoelectron spectra 
with the second gain fit model. Top left: A1-20 MPPC; top right:  all $\rm A-$type MPPCs; bottom left: all $\rm B-$type MPPCs;  bottom right: all S12571  MPPCs.}
\label{fig:Ham_notr}
\end{figure}

We tested 18 Hamamatsu MPPCs in five batches in the temperature range 1 - 48$^{\circ}$C. 
As an example, Fig.~\ref{fig:Ham_notr} (top left) shows the gain-versus-temperature dependence for MPPC A1-20 after compensation. The plot shows an expanded scale. The error bars are  of the order of $\pm 0.2\%$ and the measurements agree well with a uniform distribution. 
Figures~\ref{fig:Ham_notr} (top right, bottom left, bottom right) show the stabilized gain versus temperature for all MPPCs without trenches, A-type, B-type and  S12571, respectively. Here, the gain was determined with the second gain fit model. Corresponding results obtained with the first gain fit model are depicted in appendix~\ref{sec:app-gainstab}. 
 Figure~\ref{fig:Ham_tr} shows the results for sensors with trenches. Here, the gain was determined with the first gain fit model. We fit the temperature dependence of the stabilized gain of each MPPC with a linear function in the  $20 -30^\circ$C temperature range. Table~\ref{tab:non-uniformity} summarizes the measured $\Delta G/G_0$ values at $20^\circ$C and $30^\circ$C.
 All  tested Hamamatsu MPPCs show stable gain satisfying our requirement of $\Delta G/G_0<  \pm 0.5\% $ in  the 20$^{\circ}$C to 30$^{\circ}$C temperature range.  Most of the MPPCs actually satisfy this criterion in the fully tested   temperature range ($1-48^\circ$C). For the MPPCs without trenches the two fit methodologies yield consistent results (see appendix~\ref{sec:app-gainstab}). For MPPCs with trenches, the second gain fit model yields poor fits at certain $V_{\rm b}$ and $T$ values because of the pronounced tails on the right-hand side of the photoelectron peaks.

\begin{figure}[H]
\centering 
\includegraphics[width=2.9in]{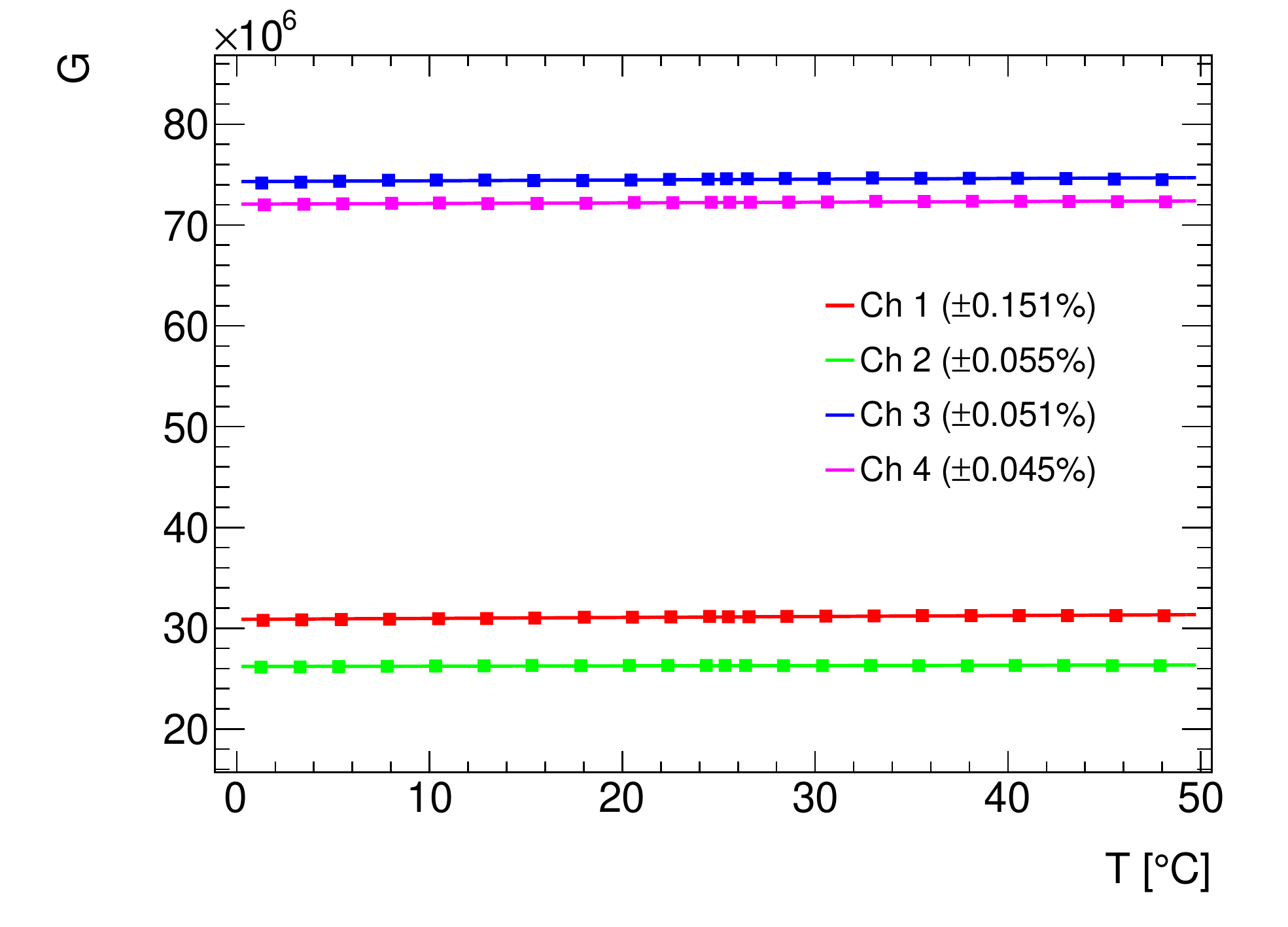}
\includegraphics[width=2.9in]{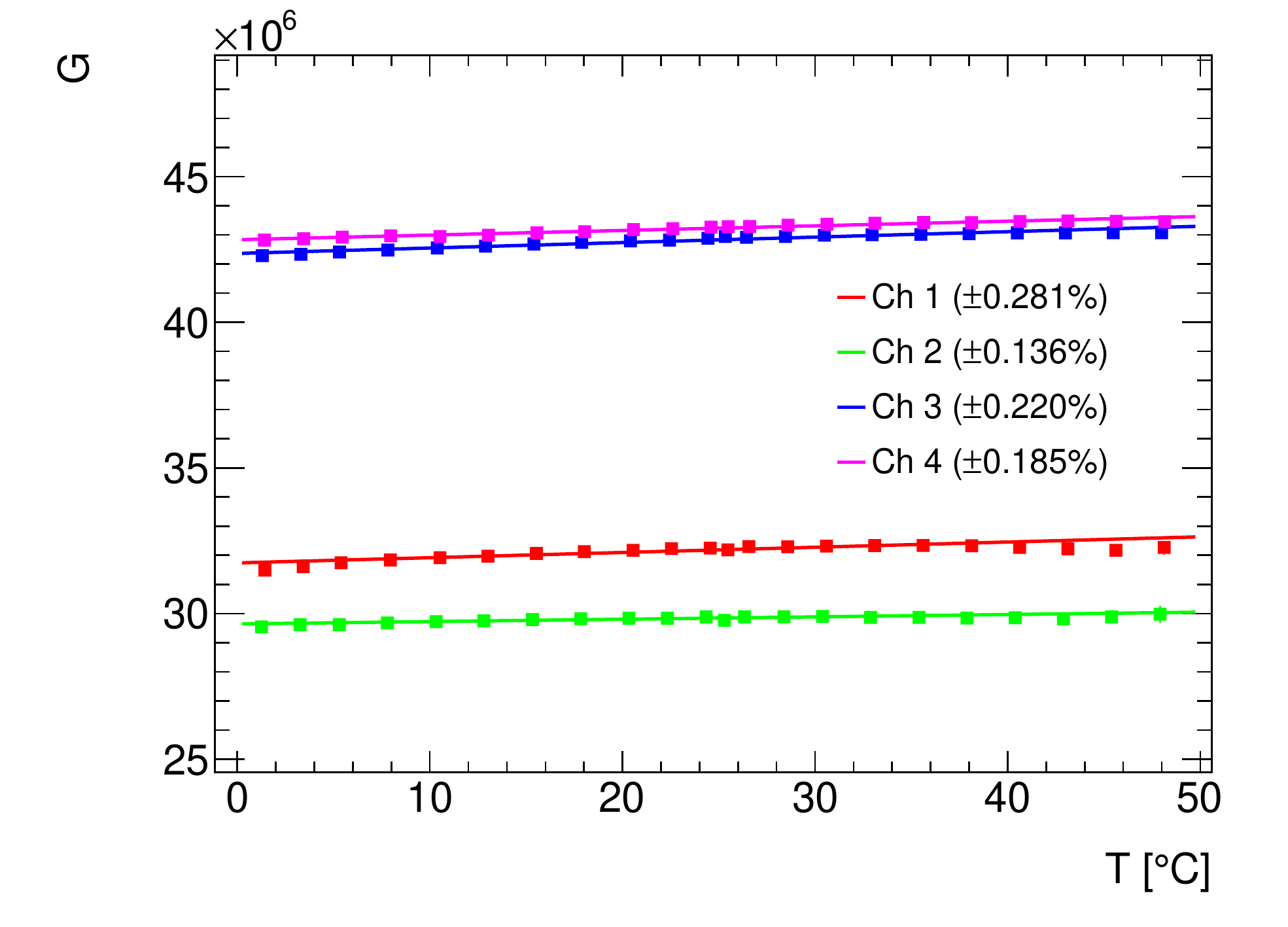}
\caption{Measurements of stabilized gain versus temperature  for Hamamatsu MPPCs with trenches in which the gain is extracted from the photoelectron spectra 
with the first gain fit model.  Left:  S13360-1325 and LCT MPPCs. Right: all S13360 MPPCs. }
\label{fig:Ham_tr}
\end{figure}

\begin{table}[htbp]
\begin{center}
\caption{Measured gain deviations $\Delta G/G_0$ from uniformity in the $\rm 20 - 30^\circ C$ temperature range. For $\rm A-$type, $\rm B-$type, S12571, KETEK and CPTA MPPCs, the $\Delta G/G_0$  values are obtained from fits with the second gain fit model. 
For LCT and S13360 MPPCs, the $\Delta G/G_0$  values are obtained from fits with the first gain fit model.}
\begin{tabular}{|l|c|c|c|c|}  
\hline \hline
SiPM   &Ch1 $\Delta G /G_0$  & Ch2 $\Delta G/G_0$ &  Ch3 $\Delta G/G_0$ & Ch4 $\Delta G/G_0$  \\
 & & & &  \\ \hline \hline
Hamamatsu   & A1-20    & A2-20   & A1-15  & A2-15  \\
A sensors &   ${\bf \pm 0.002\%}$  &  ${\bf \pm 0.031\%}$ &   ${\bf \pm 0.093\%}$&  ${\bf \pm 0.042\%}$ \\ \hline
Hamamatsu  &B1-20  & B2-20 & B1-15  & B2-15  \\
 B sensors&  ${\bf \pm 0.283\%}$ &${\bf \pm 0.127\%}$&  ${\bf \pm 0.230\%}$ &  ${\bf \pm 0.308\%}$ \\ \hline
Hamamatsu &   010a  & 010b  & 015a & 015b \\
 S12571- &   ${\bf \pm 0.195\%}$& ${\bf \pm 0.012\%}$ & ${\bf \pm 0.289\%}$  &  ${\bf \pm 0.210\%}$\\  \hline
Hamamatsu   &3025a & 3025b  & 1325a  & 1325b  \\ 
S13360-&  ${\bf \pm 0.281\%}$ &  ${\bf \pm 0.136\%}$ &  ${\bf \pm 0.220\%}$ & ${\bf \pm 0.185\%}$ \\ \hline
Hamamatsu  &1325a & 1325b  & LCT4\#6       &   LCT4\#9  \\   
 S13360/LCT& ${\bf \pm 0.151\%}$  & ${\bf  \pm 0.055 \%}$  & ${\bf  \pm 0.051 \%}$  & ${\bf  \pm 0.045 \%}$ \\    \hline \hline
KETEK &W12A  & W12B  & $\rm PM3350\#1$ &  $ \rm PM3350\#2$ \\
 W12/PM3350&  ${\bf \pm 0.585\%}$ &  ${\bf \pm 0.791\%}$ &  ${\bf \pm 1.616\%}$ &   ${\bf \pm 1.429\%}$ \\\hline
KETEK & PM3350\#5  & PM3350\#6 &  PM3350\#7  &  PM3350\#8 \\
 PM3350& ${\bf \pm 1.408\%}$ &  ${\bf \pm 1.392\%}$ &   ${\bf \pm 1.650\%}$ &   ${\bf \pm 1.644\%}$  \\ \hline \hline
CPTA & $\#857$  & $\#922$ & $\#975$  &$ \#1065$  \\ 
&  ${\bf \pm 0.017\%}$ &  ${\bf \pm 0.307\%}$ &  ${\bf \pm 0.161\%}$ & ${\bf \pm 0.032\%}$ \\ \hline \hline
\end{tabular}
\vskip -0.2cm
\label{tab:non-uniformity}
\end{center}
\end{table}

\subsection{Gain Stabilization of KETEK SiPMs}
\begin{figure}
\centering 
\includegraphics[width=2.9in]{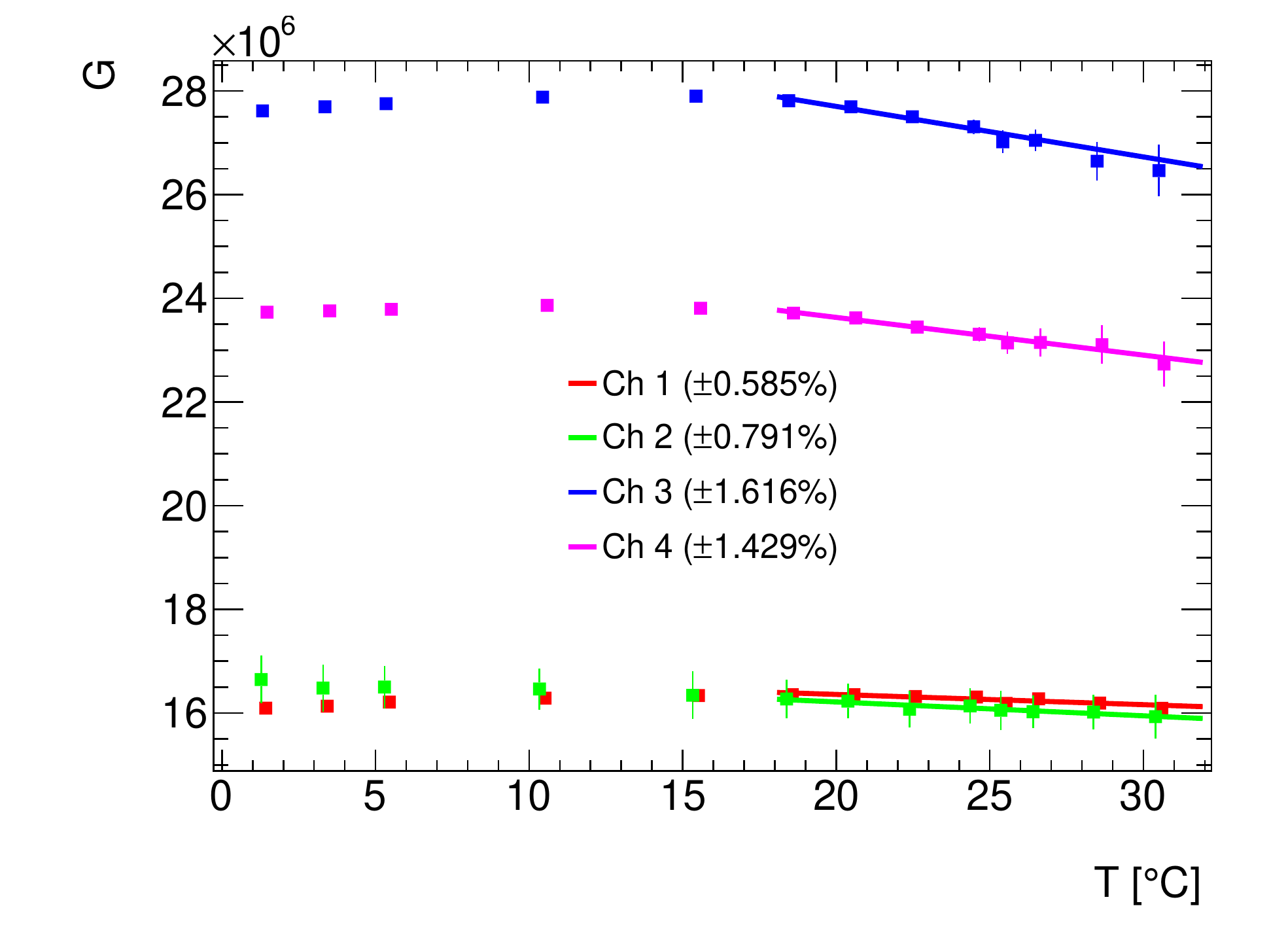}
\includegraphics[width=2.9in]{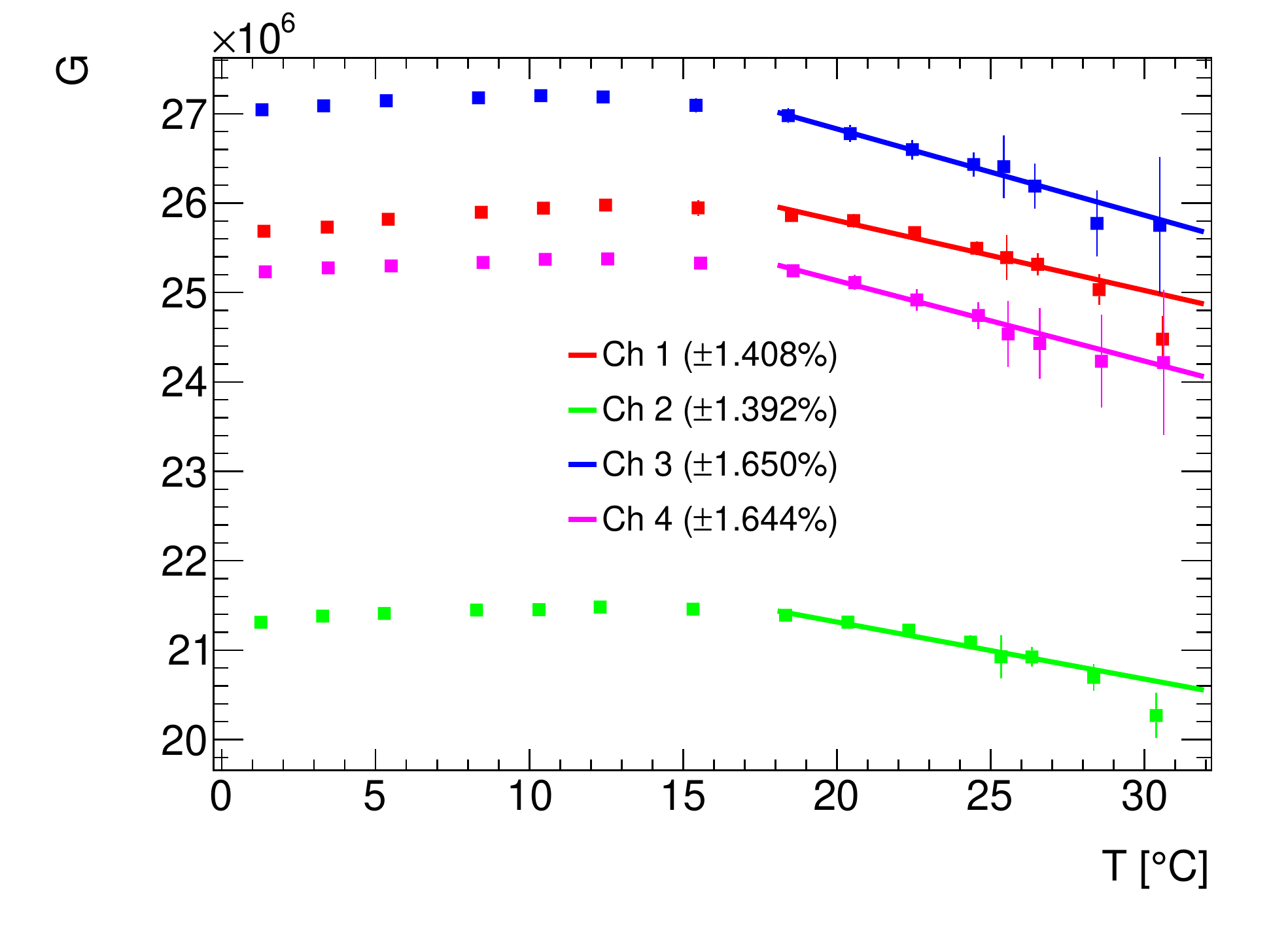}
\caption{\label{fig:4ch_PM_W12_m} Measurements of stabilized gain versus temperature for KTEK SiPMs. Left: W12A, W12B, PM3350\#1 and PM3350\#2 SiPMs. Right: four PM3350 SiPMs (\#5 to \#8). }
\end{figure}

We tested the eight SiPMs from KETEK in two batches using a compensation parameter of $18.3~\rm mV/^\circ C$. 
Above 30$^{\circ}$C, the  KETEK PM3350 SiPMs do not produce photoelectron spectra with separated photoelectron peaks. 
The W12 SiPMs show the same behavior above 40$^{\circ}$C. 
Figures~\ref{fig:4ch_PM_W12_m} (left and right) show the gain-versus-temperature measurements for the KETEK SiPMs after stabilization. The temperature dependence of the gain is rather complex.  At low temperatures ($\rm 1 - 18^\circ C$) the gain rises slowly while in the $\rm 18 - 22^\circ$C range it remains constant before decreasing again. For PM3350 SiPMs the slope in the $\rm 20 - 30^\circ$C temperature region is steeper than that of the W12 experimental devices. For such a complex temperature dependence it is difficult to determine the compensation parameter from the entire temperature range accurately.  
The compensation parameter we selected used only some of temperature points from which we determined a too low value of $dV_{\rm b}/dT$. The correct procedure should have been to extract $dV_{\rm b}/dT$ from temperature points in the $20 -30^\circ$C temperature range. Table ~\ref{tab:non-uniformity} summarizes our $\Delta G/G_0$ results for all tested KETEK SiPMs. In fact, no tested KETEK SiPM satisfies our requirements of  $\Delta G/ G_0< \pm 0.5\%$.  For the PM3350 SiPMs, the observed  $\Delta G/ G_0$ values are between $\pm 1-2\%$, while for the W12 experimental devices they are close to the goal  of $\pm 0.5\%$.

\subsection{Gain stabilization of CPTA SiPMs}  
\begin{figure}
\centering 
\includegraphics[width=2.9in]{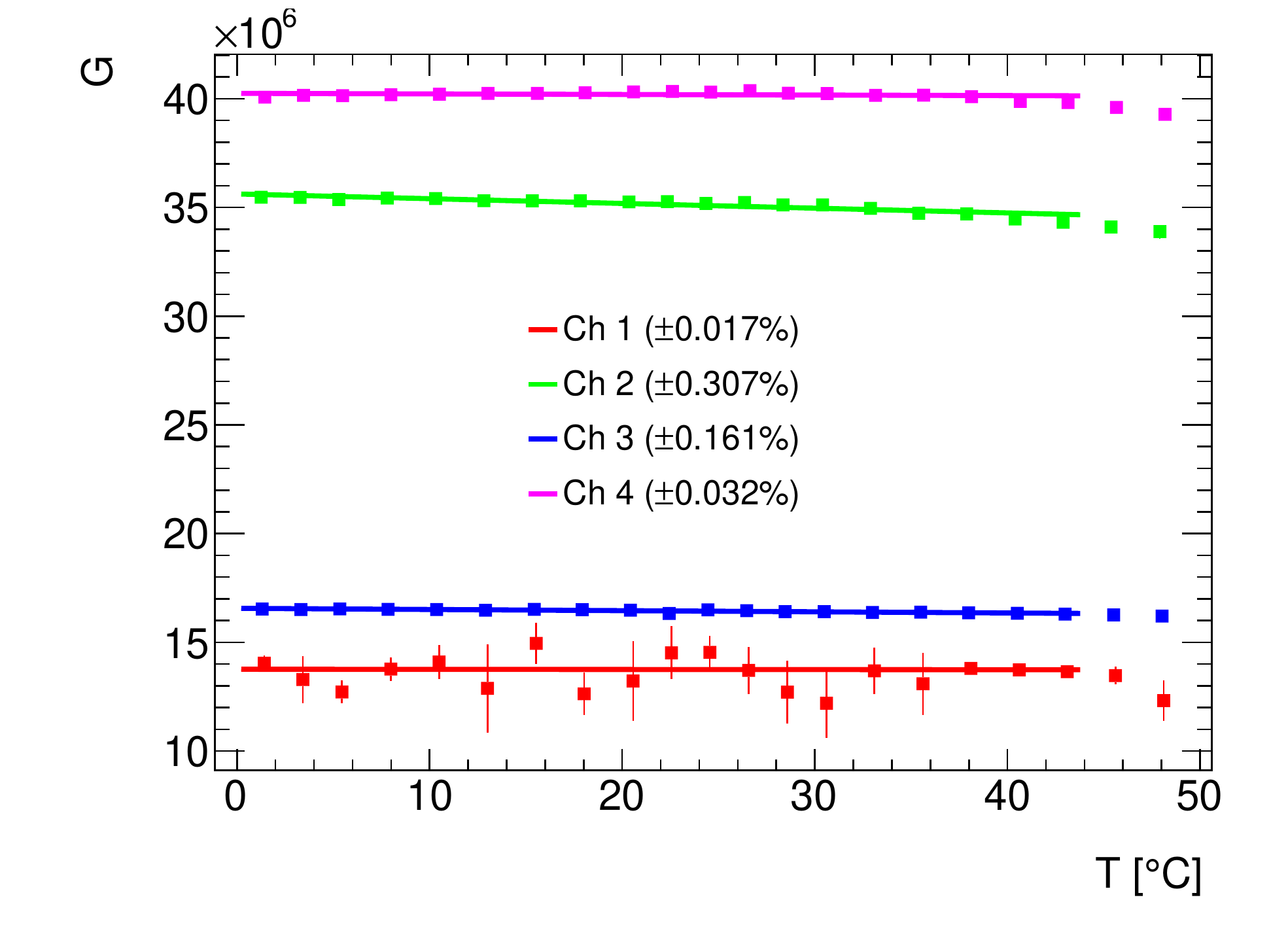}
\caption{\label{fig:CPTA1} Measurements of stabilized gain versus temperature for CPTA SiPMs \#857, \#922, \#975 and \#1065.}
\end{figure}

We used a compensation parameter of  $dV_{\rm b}/dT=\rm 21.2~ mV/^\circ C$ to stabilize the four CPTA SiPMs  simultaneously in the  $\rm 1 - 48^\circ C$ temperature range.   
Figure~\ref{fig:CPTA1} shows the gain stabilization measurements. The gain is nearly uniform up to $\rm 30^\circ C$.  The  SiPMs \#922 and \#1065 work well; SiPM \#857 was rather noisy and SiPM \#975 changed gain after operation at $T=\rm 45^\circ C$ but worked well afterwards with a new gain. All  CPTA SiPMs satisfy our requirements.
Table ~\ref{tab:non-uniformity} summarizes our results for $\Delta G/G_0$ in the  $20 -30^\circ$C temperature region.

\section{Conclusions}
\label{sec:conclusion}
We successfully completed gain stabilization tests for 30 SiPMs demonstrating that batches of similar-type SiPMs can be stabilized with one compensation parameter $dV_{\rm b}/dT$. Even SiPMs with slightly different nominal bias voltages can be stabilized by adjusting the applied bias voltage appropriately with a voltage divider.  All 18 Hamamatsu MPPCs, six with trenches and 12 without trenches, satisfy the goal of $\Delta G/G_0 < \pm 0.5\%$ in the $\rm 20- 30^\circ C$ temperature range. In fact,  most  MPPCs satisfy this requirement in the extended temperature range of $\rm 1- 48^\circ C$. 
Gain stabilization of KETEK SiPMs is more complex since  the signals are rather long and are  affected by afterpulsing. The temperature range is limited to  $\rm 1- 30^\circ C$. We did not succeed in stabilizing any of the eight KETEK SiPMs tested. The $V(T)$ behavior is more complex  requiring  individual $dV_{\rm b}/dT$ values to stabilize the gain of four SiPMs in the  $\rm 20- 30^\circ C$ temperature range. We may have been more successful if we had determined the compensation parameter just in the $\rm 20- 30^\circ C$ temperature range. Gain stabilization of CPTA SiPMs works well. All  four SiPMs satisfy our criterion despite the fact that the LED light had to be absorbed by the scintillator and/or wavelength-shifting fiber before reaching the SiPM. Thus, this demonstrates that our procedure is applicable to a full tile/SiPM setup. 
 In the analog HCAL for ILC, the bias voltage  adjustment will be implemented on the electronics boards. Gain stabilization looks promising if the temperature is well measured and SiPMs with similar properties are stabilized with one $dV_{\rm b}/dT$ compensation parameter.

\acknowledgments
This  work  was  supported by the H2020 project AIDA-2020, GA no. 654169.  It  has  been
supported by the Norwegian Research Council and by the Ministry of Education, Youth and Sports
of the Czech Republic under the project LG14033. We acknowledge the contributions of Erik van der Kraaij in the early stage of this study. 
We would like to thank L. Linssen, Ch. Joram, W. Klempt, and D. Dannheim  for using the E-lab and for supplying electronic equipment. We further would like to thank the team of the climate chamber at CERN for their  assistance and support.

\appendix
\section{Properties of Tested SiPMs}
\label{sec:appSiPM}

Table~\ref{tab:PropSiPMs} summarizes characteristic properties of the SiPMs we tested in our gain stabilization study.
\begin{table}[H]
\begin{center}
\caption{Properties of tested SiPMs including sensitive area, pixel pitch, number of pixels, nominal bias voltage and gain. For Hamamatsu and KETEK SiPMs, operating voltage and SiPM gain are specified for 25$^{\circ}$C, while for
CPTA SiPMs  the specifications are for 22$^{\circ}$C.}
\label{tab:PropSiPMs}
\begin{tabular}{|l|c|c|c|c|c|c|}
\hline
\textbf{Manufacturer} & \textbf{Sensitive area} &  \textbf{Pixel pitch} &  \textbf{$\#$pixels} &
\textbf{Nominal} & \textbf{Typical} & 
\textbf{Serial $\#$}
\\
\textbf{and Type $\#$} &\textbf{$[\rm mm^2]$} &  \textbf{$[{\mu \rm m} ]$} &  & $V_{\rm b}$ [V] &
\textbf{ $G~ [\times 10^5] $} &
 \\ [0.7ex]
 \hline
Hamamatsu & & & & & &
\\
A1-20  & $1 \times 1$ & $20$ & 2500 & $66.7$ & $2.3$ & A1-20
\\
A1-15  & $1 \times 1$ & $15$ & 4440 & $67.2$ & $2.0$ & A1-15
\\
A2-20  & $1 \times 1$ & $20$ & 2500 & $66.7$ & $2.3$ & A2-20
\\
A2-15  & $1 \times 1$ & $15$ & 4440 & $67.2$ & $2.0$ & A2-15
\\
B1-20  & $1 \times 1$ & $20$ & 2500 & $73.3$ & $2.3$ & B1-20
\\
B1-15  & $1 \times 1$ & $15$ & 4440 & $74.2$ & $2.0$ & B1-15
\\
B2-20  & $1 \times 1$ & $20$ & 2500 & $73.4$ & $2.3$ & B2-20
\\
B2-15  & $1 \times 1$ & $15$ & 4440 & $74.0$ & $2.0$ & B2-15
\\
S13360-1325a  & $1.3 \times 1.3$ & $25$ & 2668 & $57.2$ & $7.0$ & 10143
\\
S13360-1325b  & $1.3 \times 1.3$ & $25$ & 2668 & $57.1$ & $7.0$ & 10144
\\
S13360-3025a  & $3 \times 3$ & $25$ & 14400 & $57.7$ & $7.0$ & 10103
\\
S13360-3025b   & $3 \times 3$ & $25$ & 14400 & $57.0$ & $7.0$ & 10104
\\
LCT4  & $1 \times 1$ & $50$ & 400 & $50.8$ & $16.0$ & \#6
\\
LCT4  & $1 \times 1$ & $50$ & 400 & $51.0$ & $16.0$ & \#9
\\
S12571-010a  & $1 \times 1$ & $10$ & 10000 & $69.8$ & $1.4$ & 271
\\
S12571-010b  & $1 \times 1$ & $10$ & 10000 & $69.9$ & $1.4$ & 272
\\
S12571-015a  & $1 \times 1$ & $15$ & 4489 & $68.1$ & $2.3$ & 136
\\
S12571-015b  & $1 \times 1$ & $15$ & 4489 & $68.0$ & $2.3$ & 137
\\ [0.7ex]
 \hline
CPTA & $1 \times 1$ & $40$ & 796 & 33.4 & 7.1 & 857
\\
& $1 \times 1$ & $40$ & 796 & 33.1 & 6.3 & 922
\\
& $1 \times 1$ & $40$ & 796 & 33.3 & 6.3 & 975
\\
& $1 \times 1$ & $40$ & 796 & 33.1 & 7.0 & 1065
\\ [0.7ex]
\hline
KETEK  & & & & & & \\
W12A & $3 \times 3$ & 20 & 12100 & 28 & 5.4 & A
\\
W12B & $3 \times 3$ & 20 & 12100 & 28 & 5.4 & B
\\
PM3350 & $3 \times 3$ & 50 & 3600 & 29.5 & 20 & $\#1$
\\
PM3350 & $3 \times 3$ & 50 & 3600 & 29.5 & 20 & $\#2$
\\
PM3350 & $3 \times 3$ & 50 & 3600 & 29.5 & 20 & $\#5$
\\
PM3350 & $3 \times 3$ & 50 & 3600 & 29.5 & 20 & $\#6$
\\
PM3350 & $3 \times 3$ & 50 & 3600 & 29.5 & 20 & $\#7$
\\
PM3350 & $3 \times 3$ & 50 & 3600 & 29.5 & 20 & $\#8$
\\
\hline

\end{tabular}
\end{center}
\end{table}
\newpage

\section{Comparison of the two Gain Fit Models}
\label{sec:appfits}

\begin{figure}[htbp!]
\centering 
\includegraphics[width=150 mm]{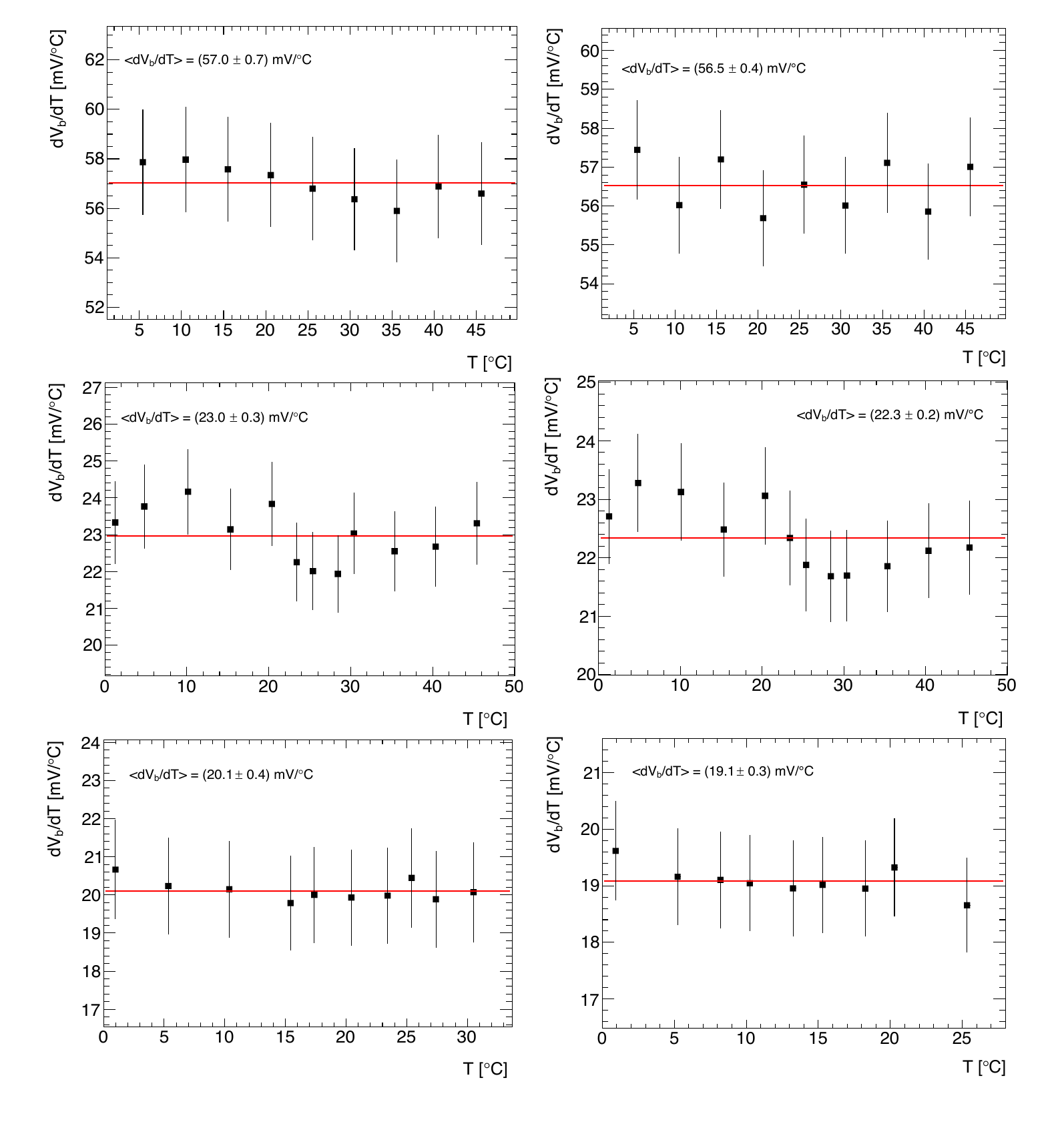}
\caption{Left column: $dV_{\rm b}/dT$ measurements obtained with the first gain fit model; right column:  $dV_{\rm b}/dT$ measurements obtained 
with the second  gain fit model  for top: Hamamatsu B2-20 MPPC; middle: for CPTA-\#1065 SiPM;  bottom: for KETEK PM3350\#6 SiPM.}
\label{fig:comparefits}
\end{figure}

We performed fits of the photoelectron spectra in the bias voltage scans with both gain fit models. We show results for three SiPMs,  the Hamamatsu MPPC B2-20, CPTA SiPM 1065 and KETEK SiPM $\rm PM3350\#6$.
Figure~\ref{fig:comparefits}  shows the resulting  $dV_{\rm b}/dT$ measurements for the three SiPMs for both gain fit methodologies. 
 For Hamamatsu MPPCs without trenches, the $ dV_{\rm b}/dT$ values for the two gain fit methodologies are in good agreement. For the KETEK and CPTA SiPMs 
 we get consistent results. 

\section{Bias Voltage Scans for different types of  SiPMs}
\label{sec:appBV}
We show results of bias voltage scans for one detector of each SiPM type using 1-D fits.

\subsection{Hamamtsu MPPCs}
\label{sec:appHamamatsuBV}

\begin{figure}[!htb]
\centering 
\includegraphics[width=2.9in]{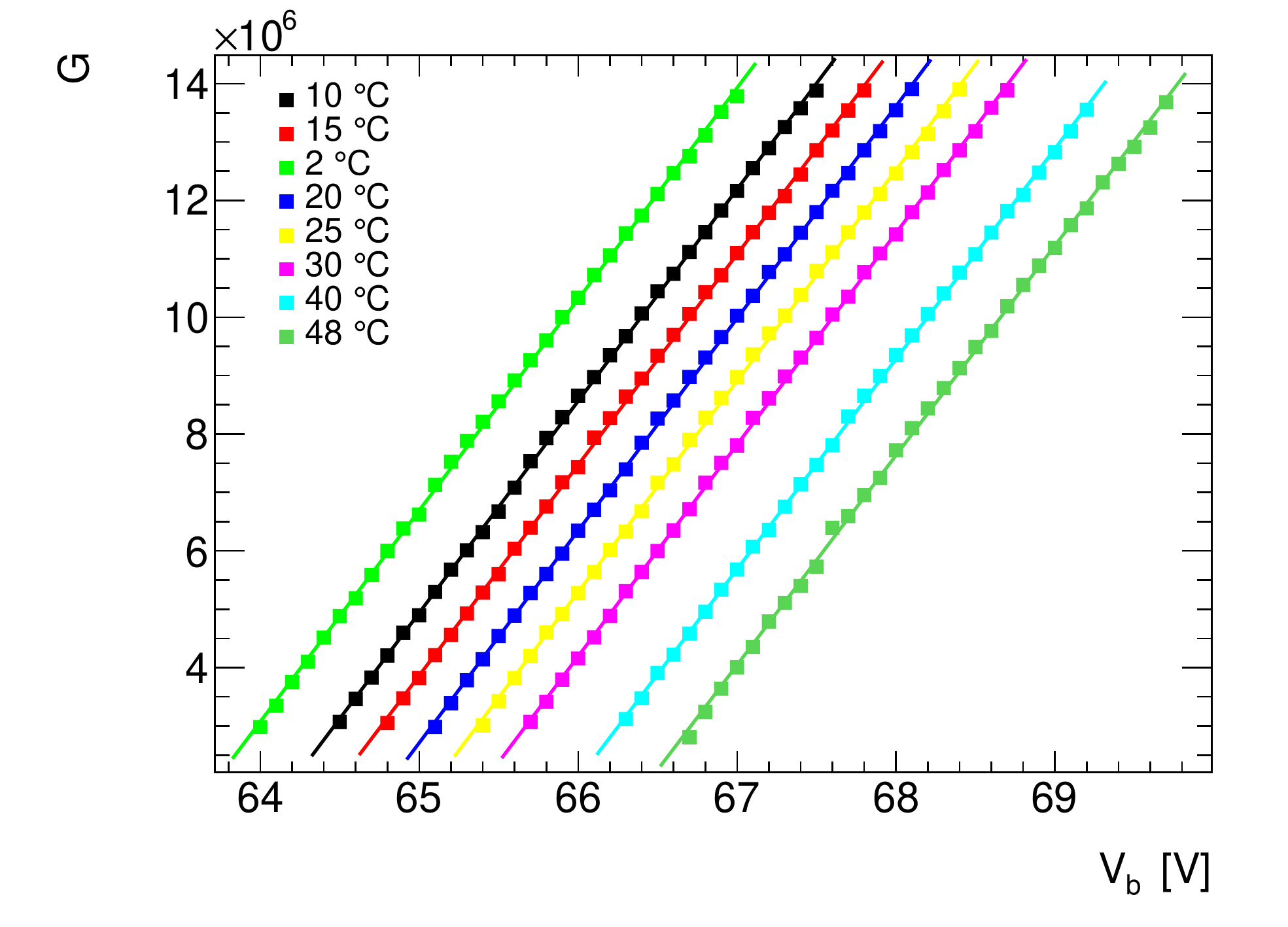}
\includegraphics[width=2.9in]{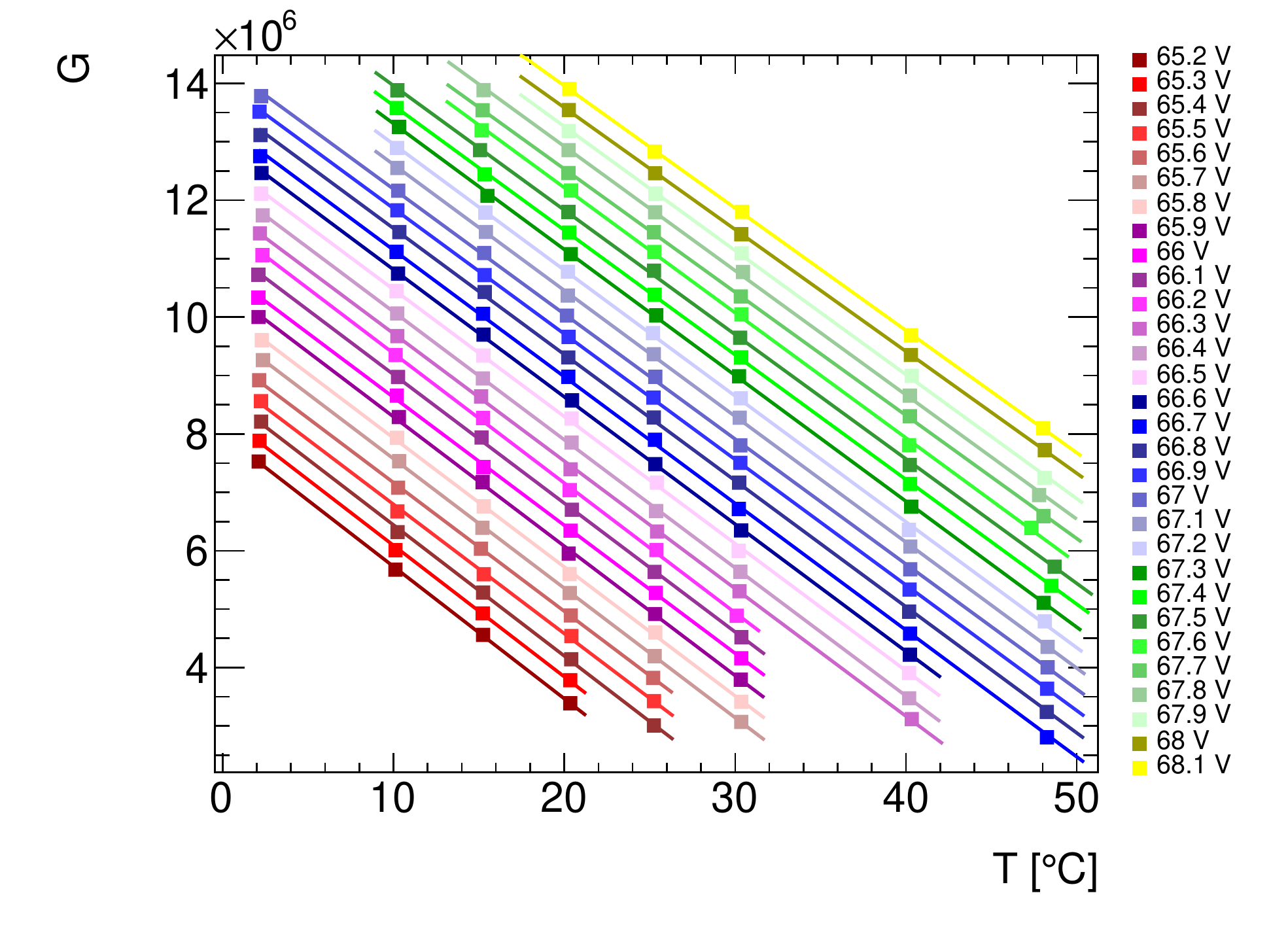}\\
\includegraphics[width=2.9in]{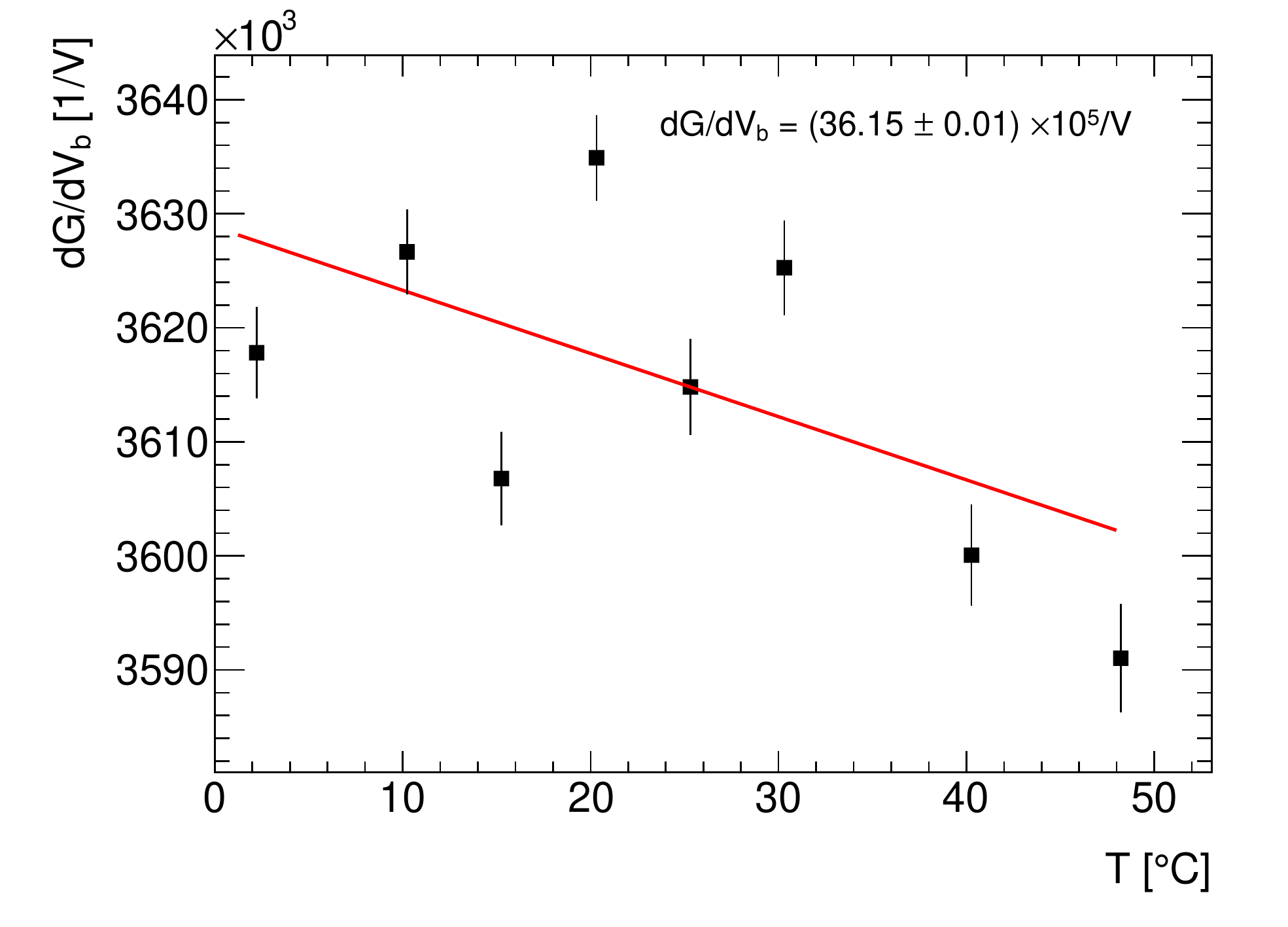}
\includegraphics[width=2.9in]{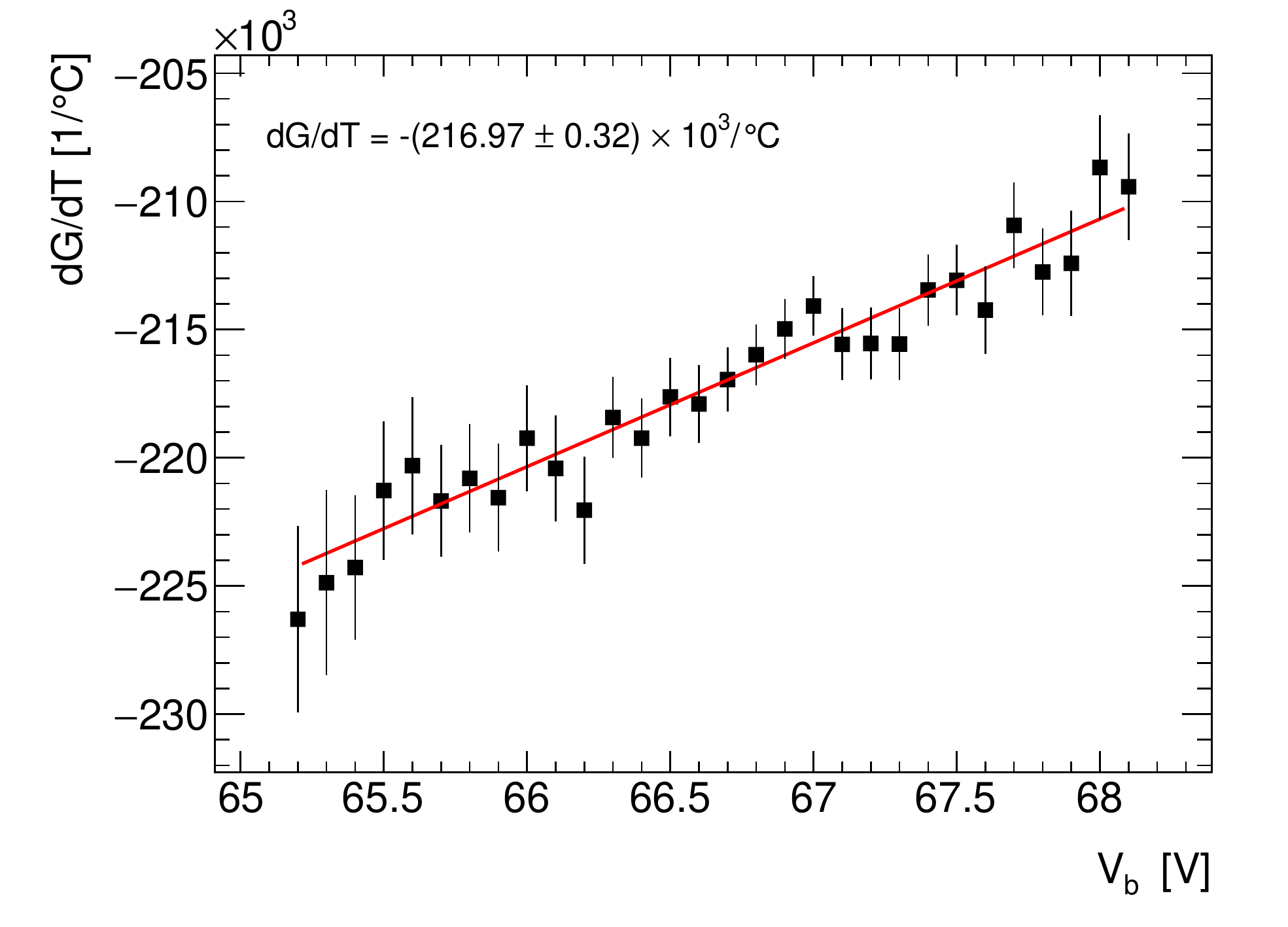}\\
\includegraphics[width=2.9in]{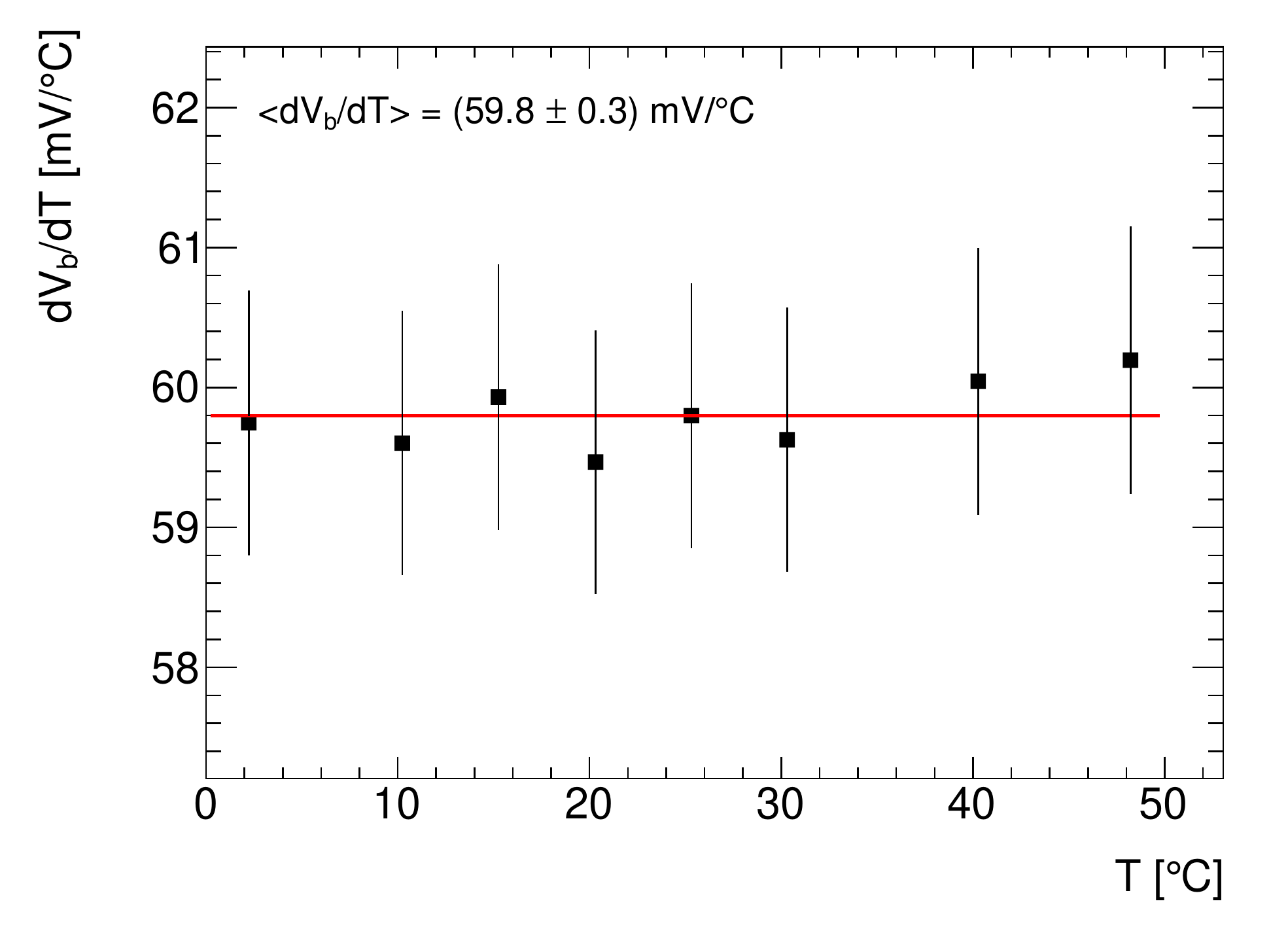}
\includegraphics[width=2.9in]{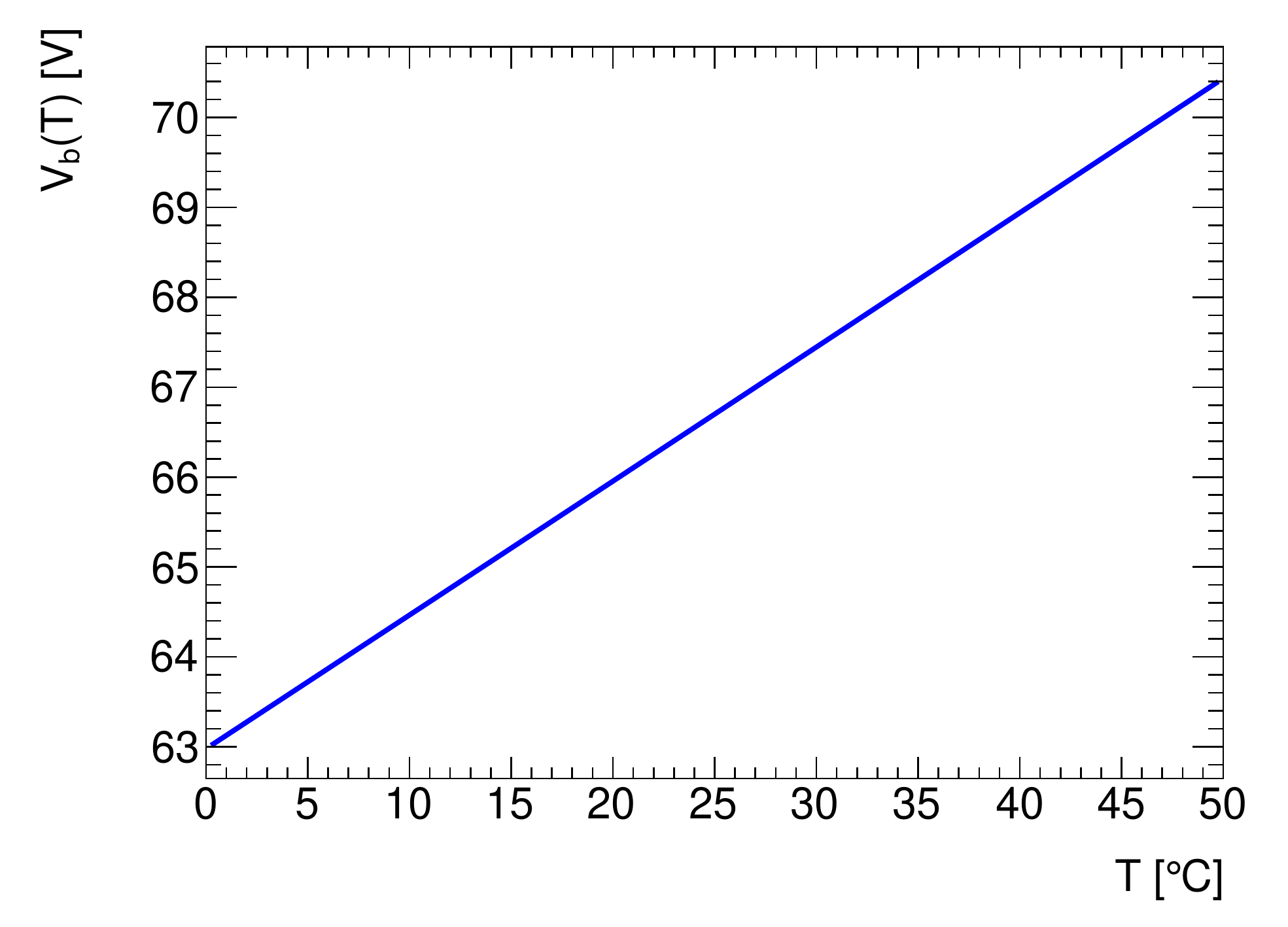}
\caption{\label{fig:A2_ch2} Measurements of $G$ versus $V_{\rm b}$ for fixed $T$ (top left), $G$ versus  $T$ for fixed $V_{\rm b}$ (top right), $dG/dV_{\rm b}$ versus  $T$ (middle left), $dG/dT$ versus $V_{\rm b}$  (middle right), $dV_{\rm b}/dT$ versus $T$ (bottom left) and distribution $V_{\rm b}(T)$ versus $T$ (bottom right) for Hamamatsu detector A2-20. Points with error bars show data and solid lines show  fit results. }
\end{figure}

\begin{figure}[!htb]
\centering 
\includegraphics[width=2.9in]{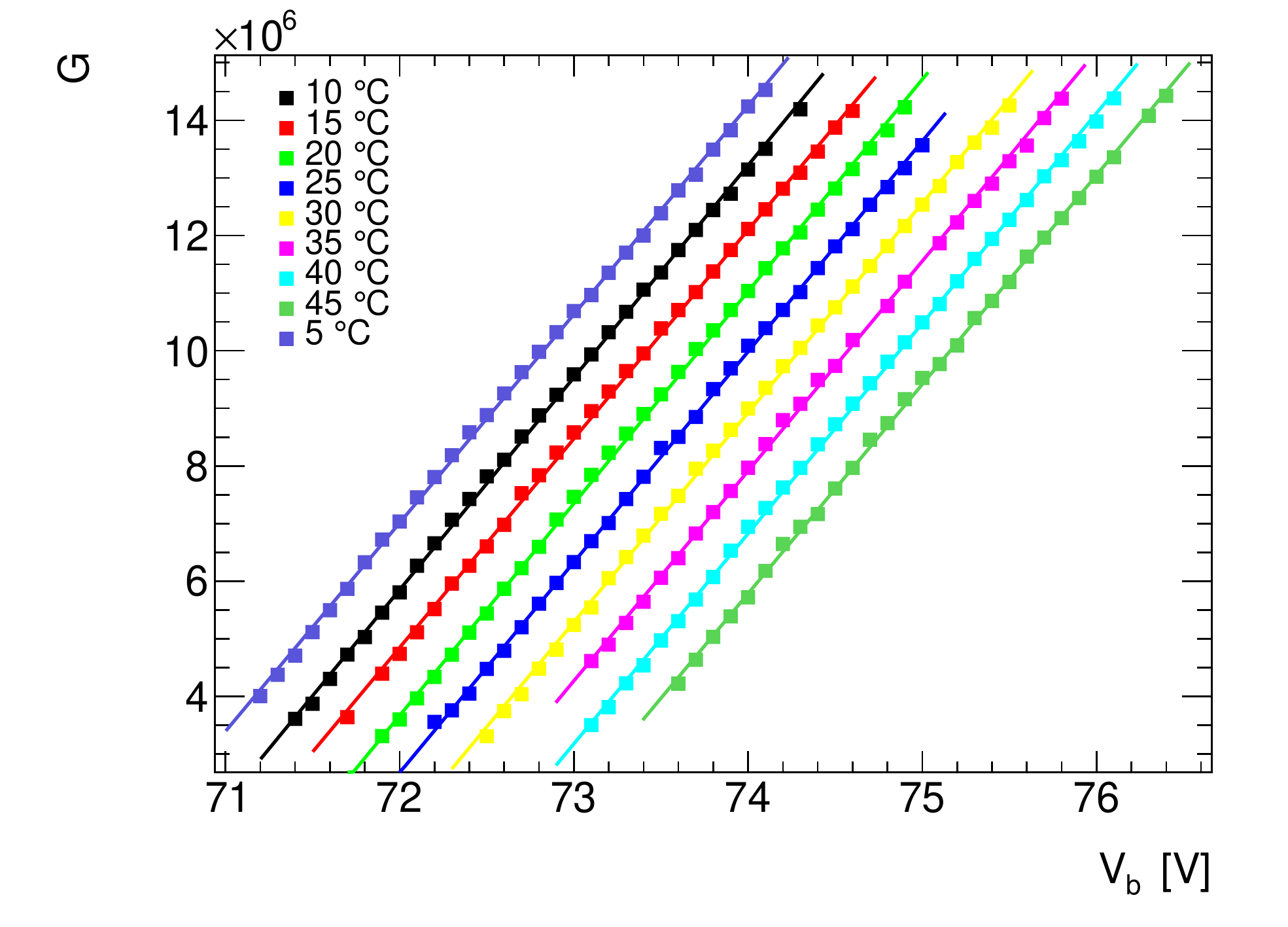}
\includegraphics[width=2.9in]{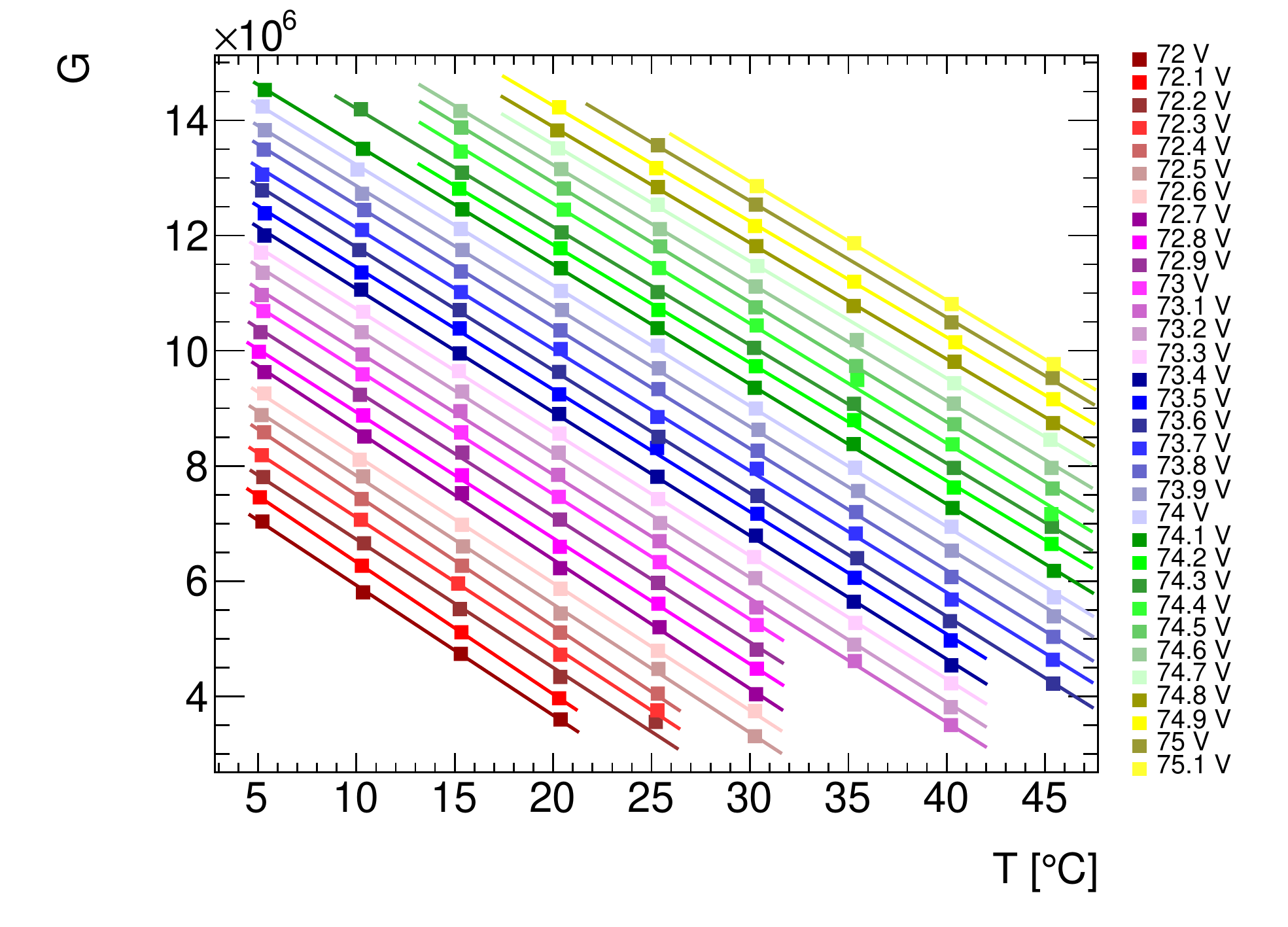}\\
\includegraphics[width=2.9in]{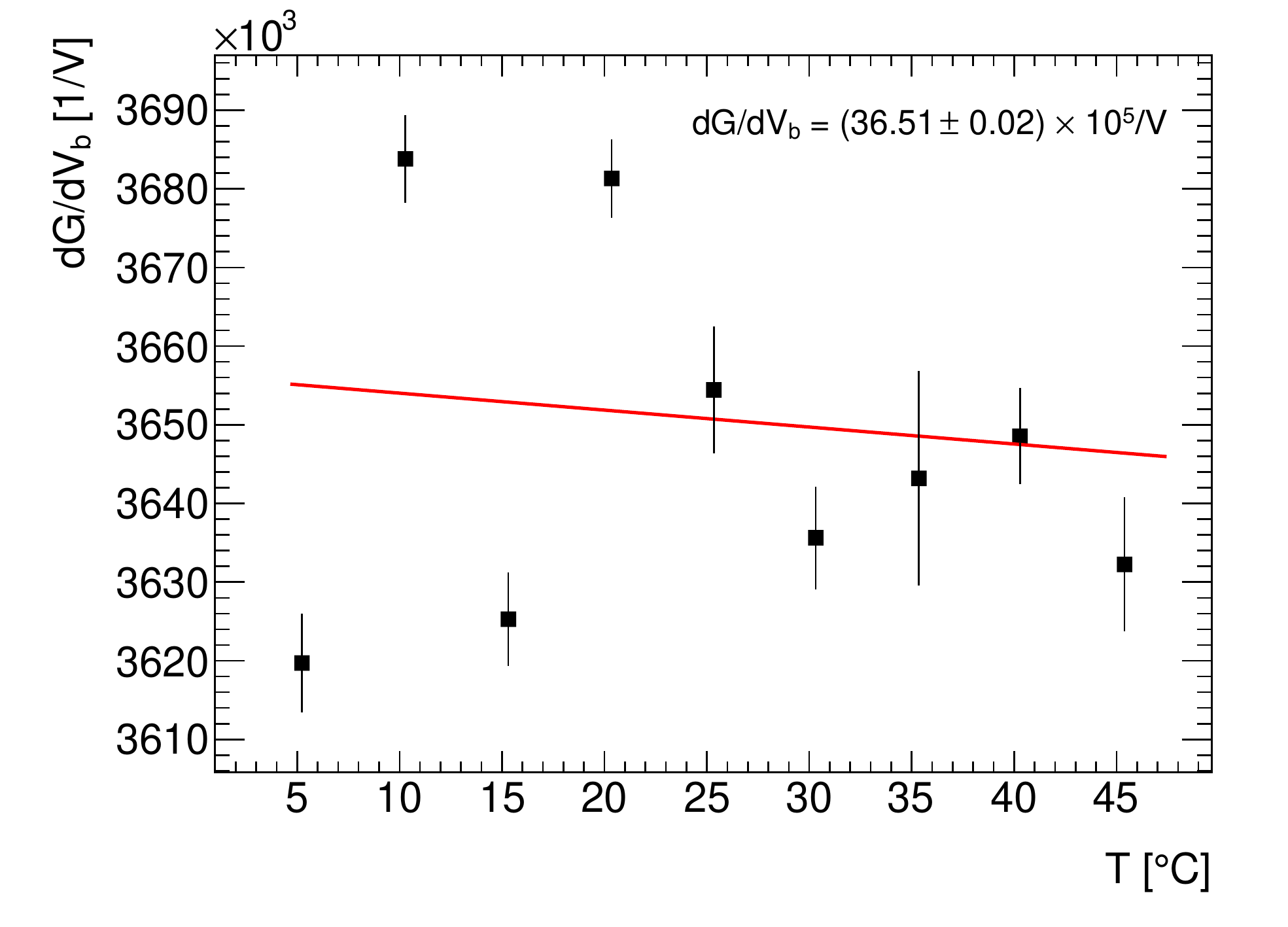}
\includegraphics[width=2.9in]{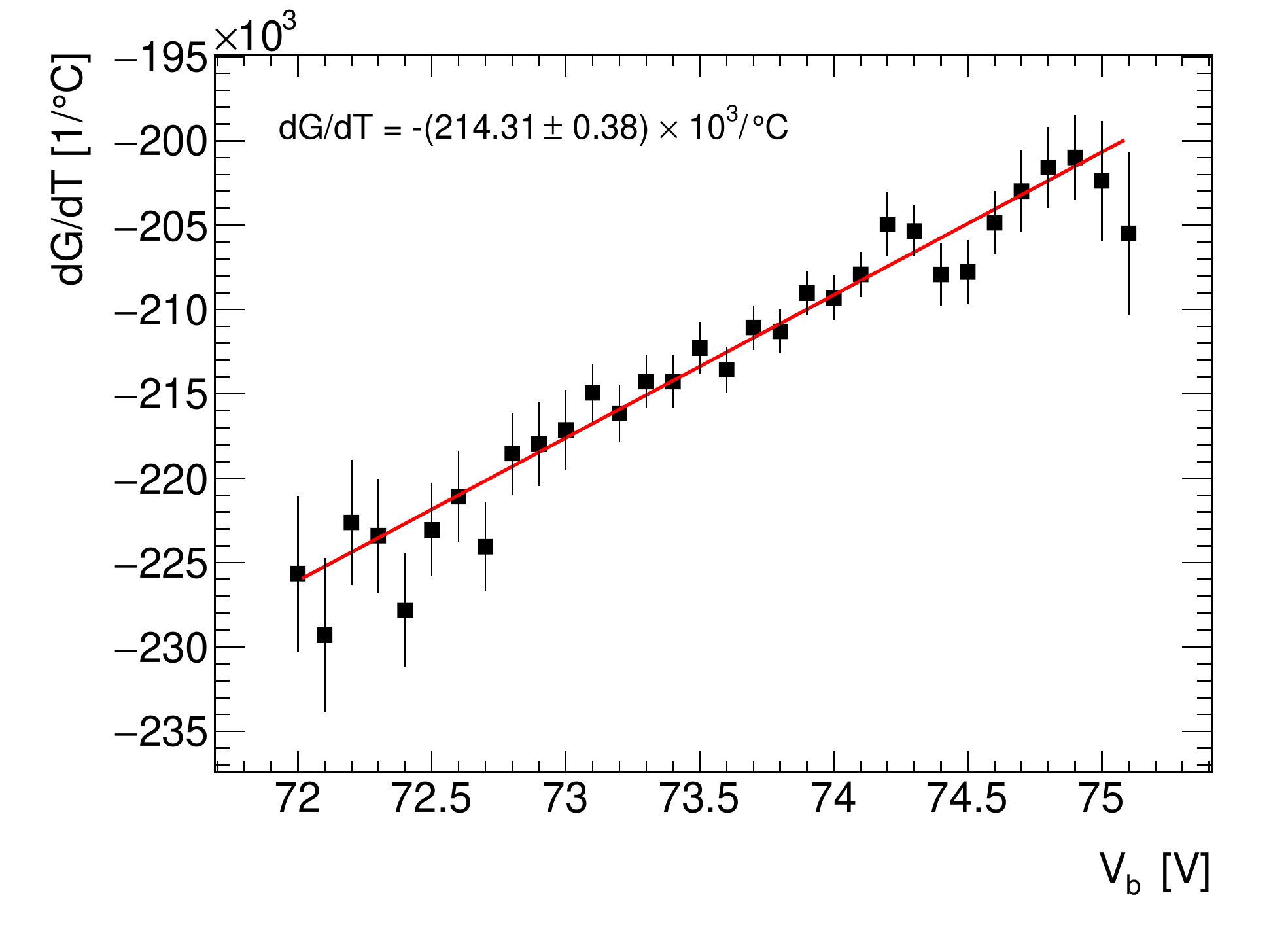}\\
\includegraphics[width=2.9in]{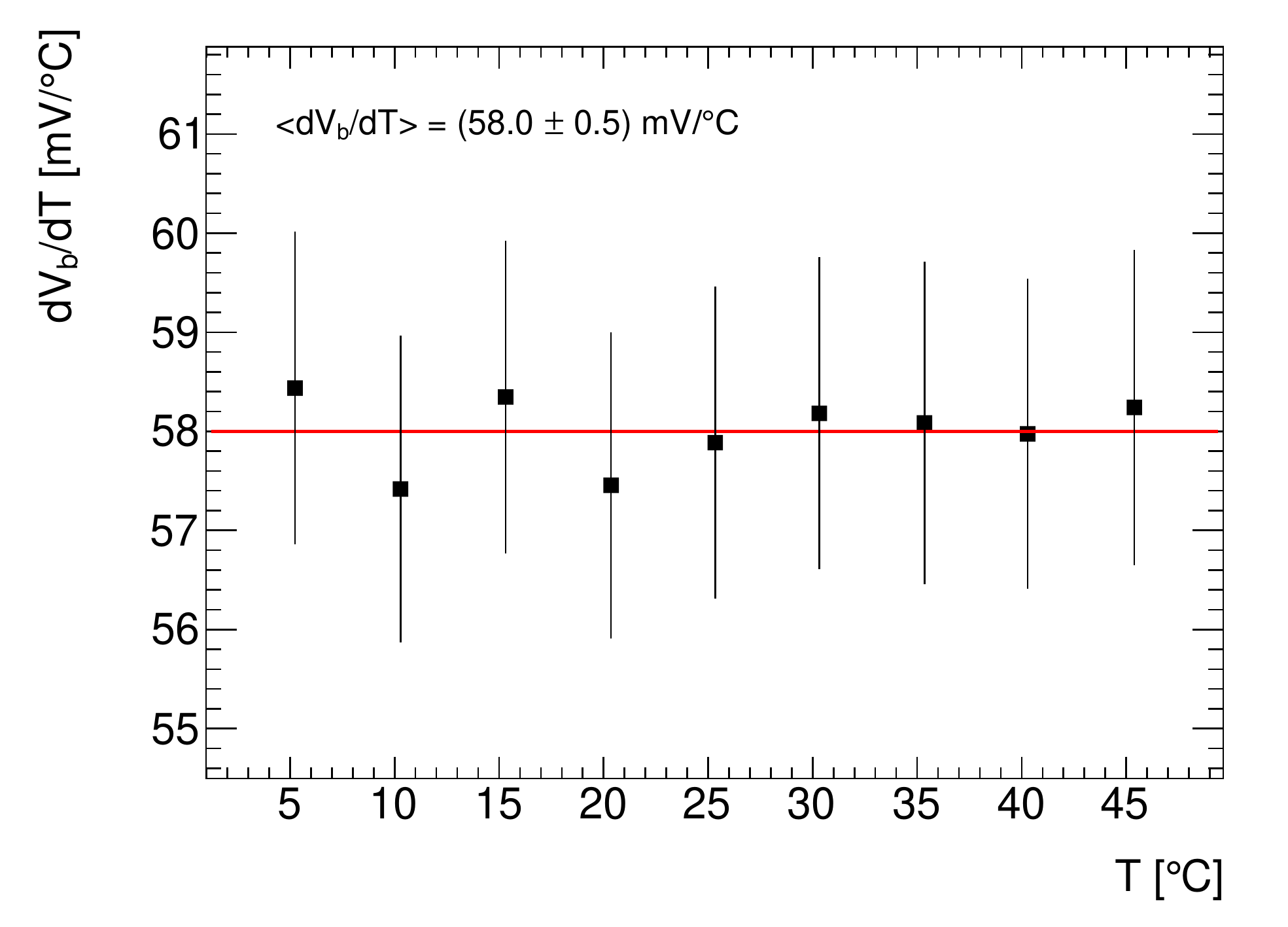}
\includegraphics[width=2.9in]{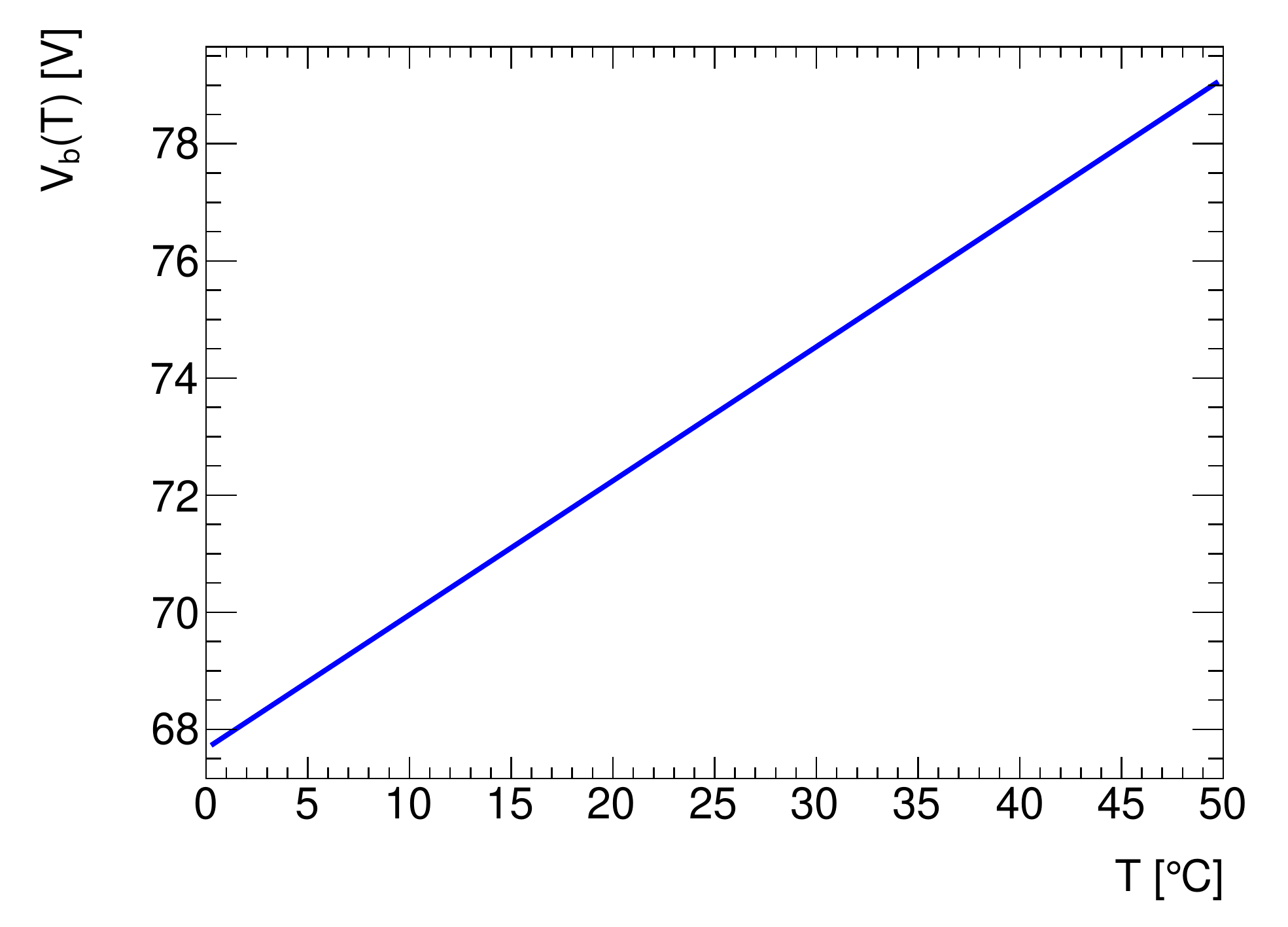}
\caption{\label{fig:B2_ch2} 
Measurements of $G$ versus $V_{\rm b}$ for fixed $T$ (top left), $G$ versus  $T$ for fixed $V_{\rm b}$ (top right), $dG/dV_{\rm b}$ versus  $T$ (middle left), $dG/dT$ versus $V_{\rm b}$  (middle right), $dV_{\rm b}/dT$ versus $T$ (bottom left) and distribution $V_{\rm b}(T)$ versus $T$ (bottom right) for  Hamamatsu detector B2-20. Points with error bars show data and solid lines show  fit results. }
\end{figure}

\begin{figure}[!htb]
\centering 
\includegraphics[width=2.9in]{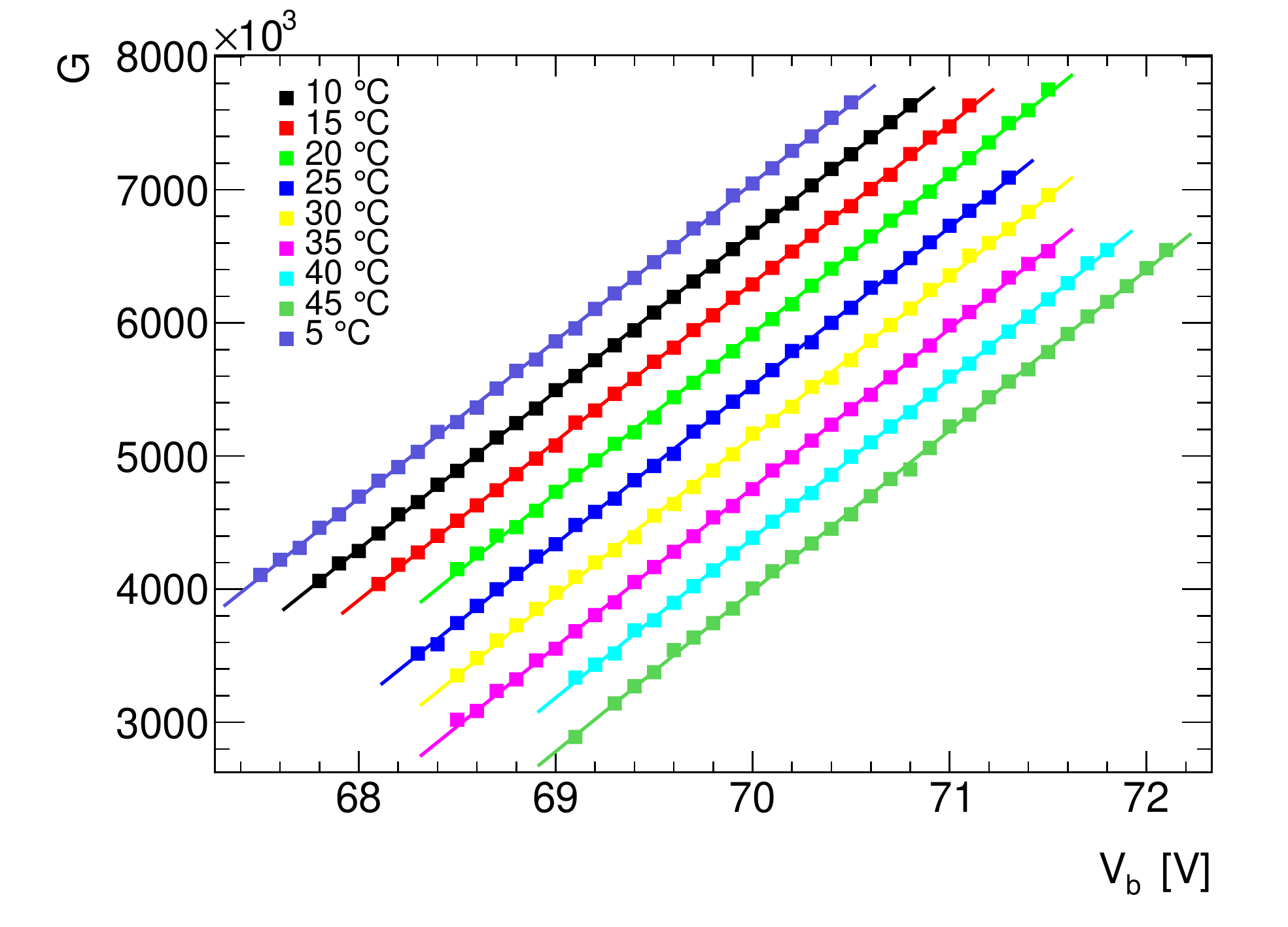}
\includegraphics[width=2.9in]{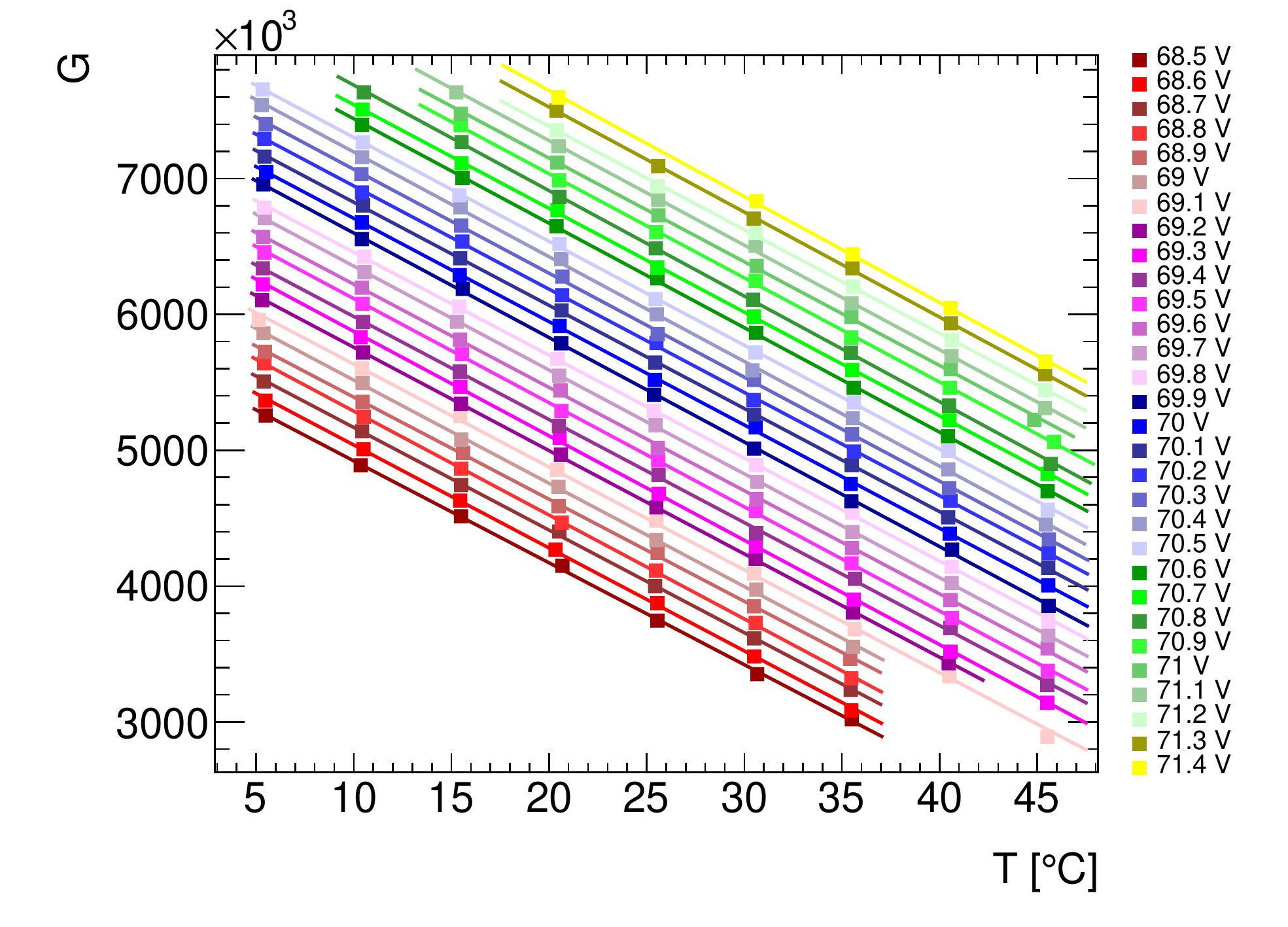}\\
\includegraphics[width=2.9in]{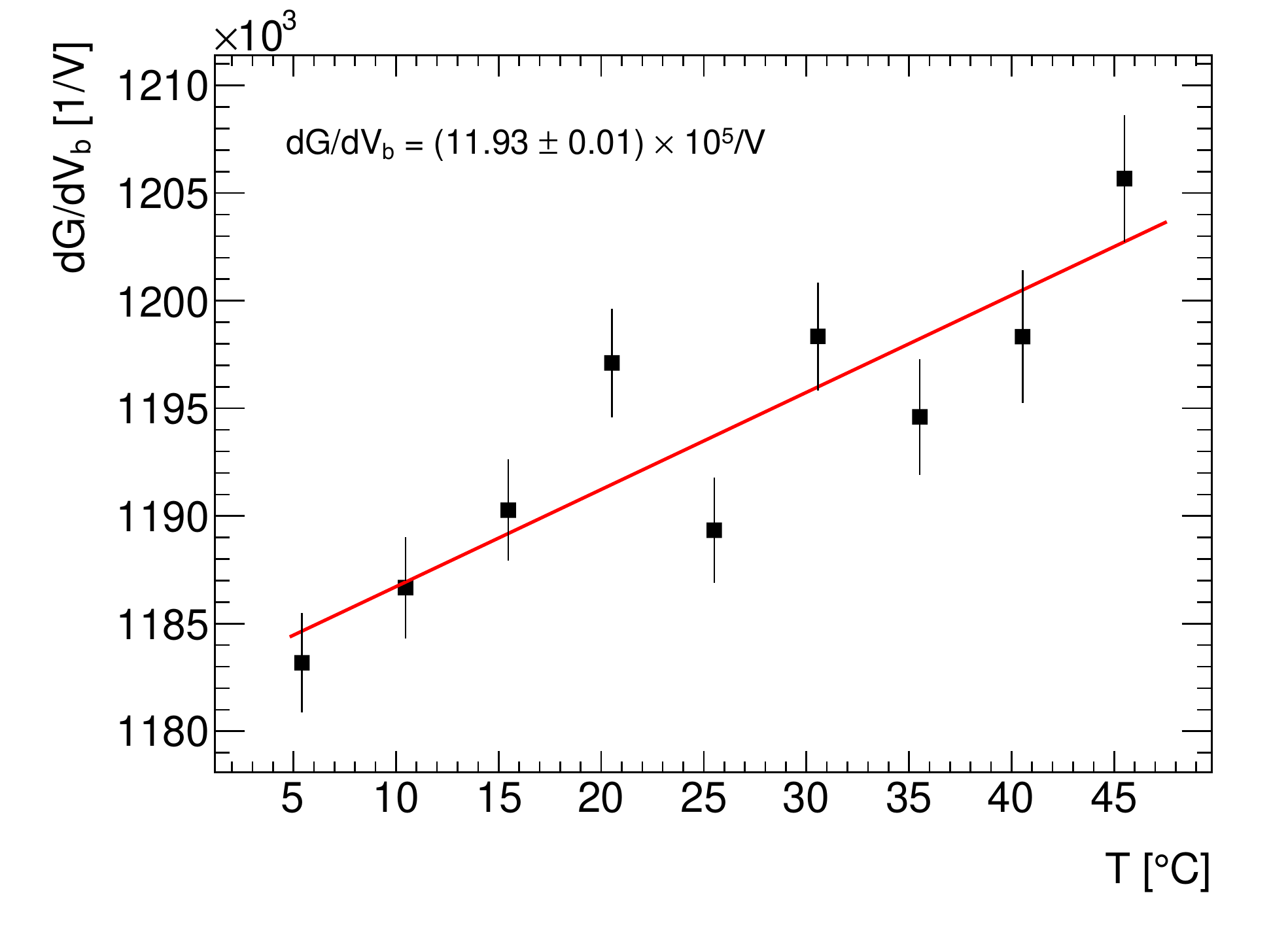}
\includegraphics[width=2.9in]{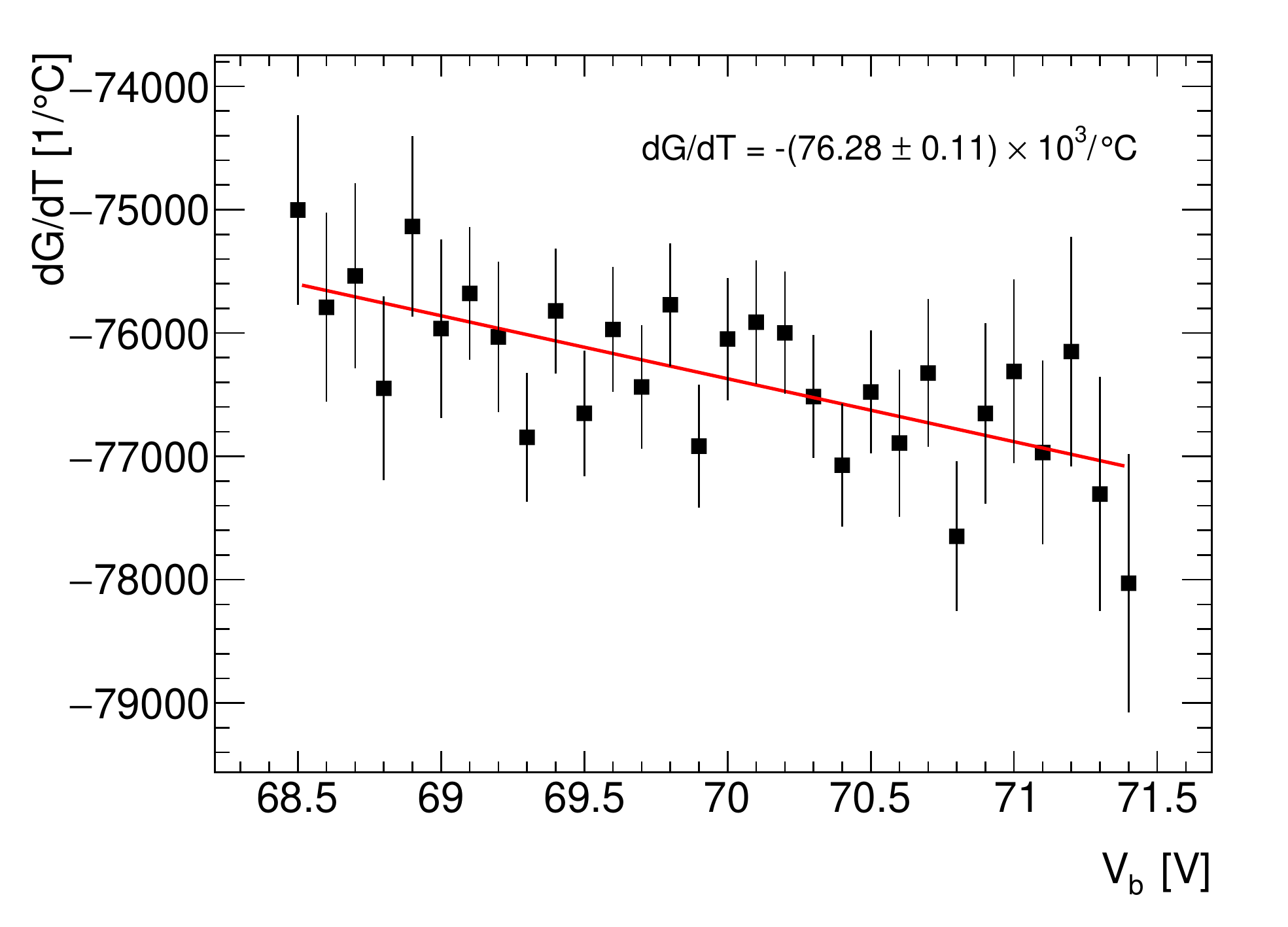}\\
\includegraphics[width=2.9in]{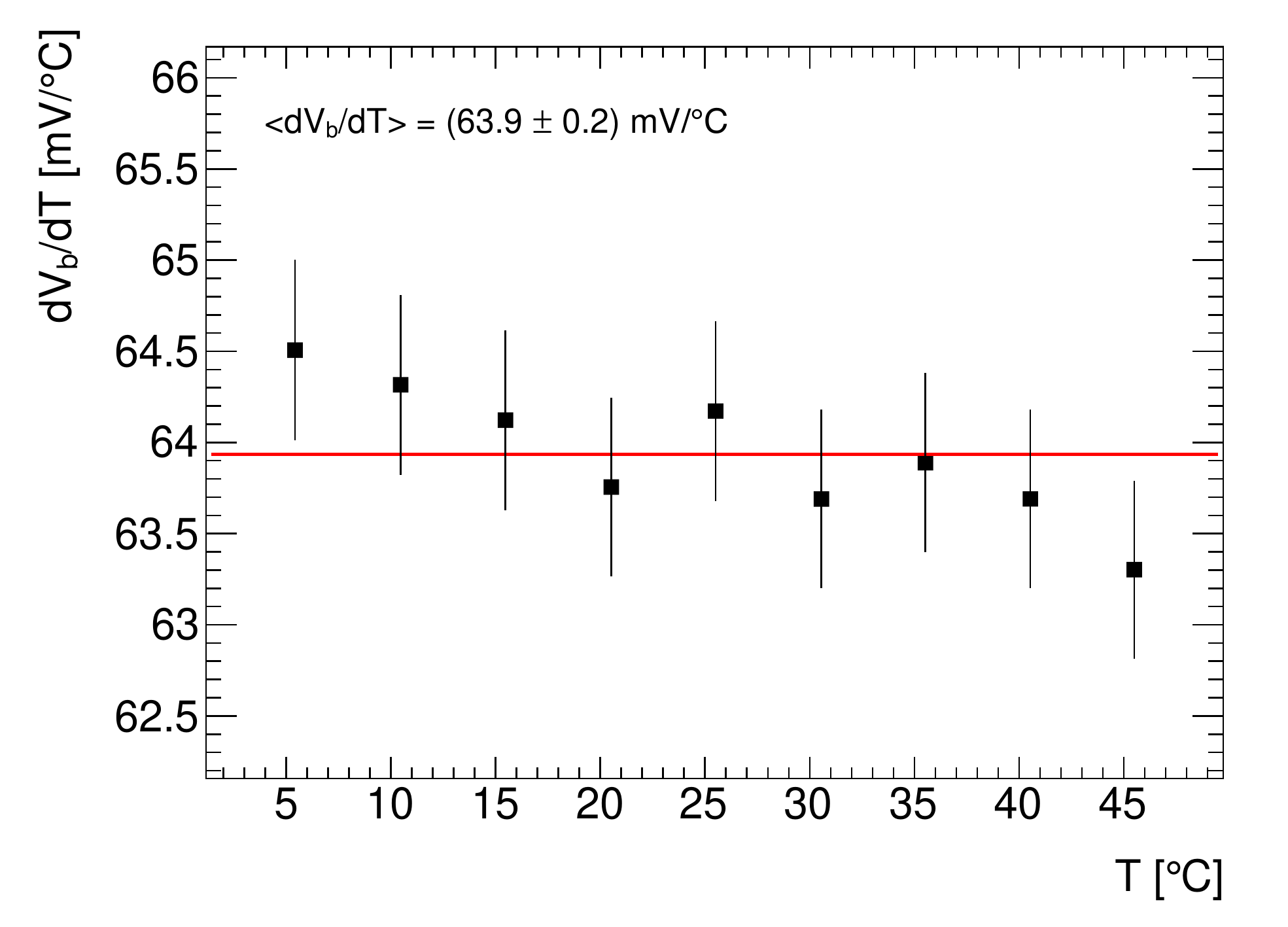}
\includegraphics[width=2.9in]{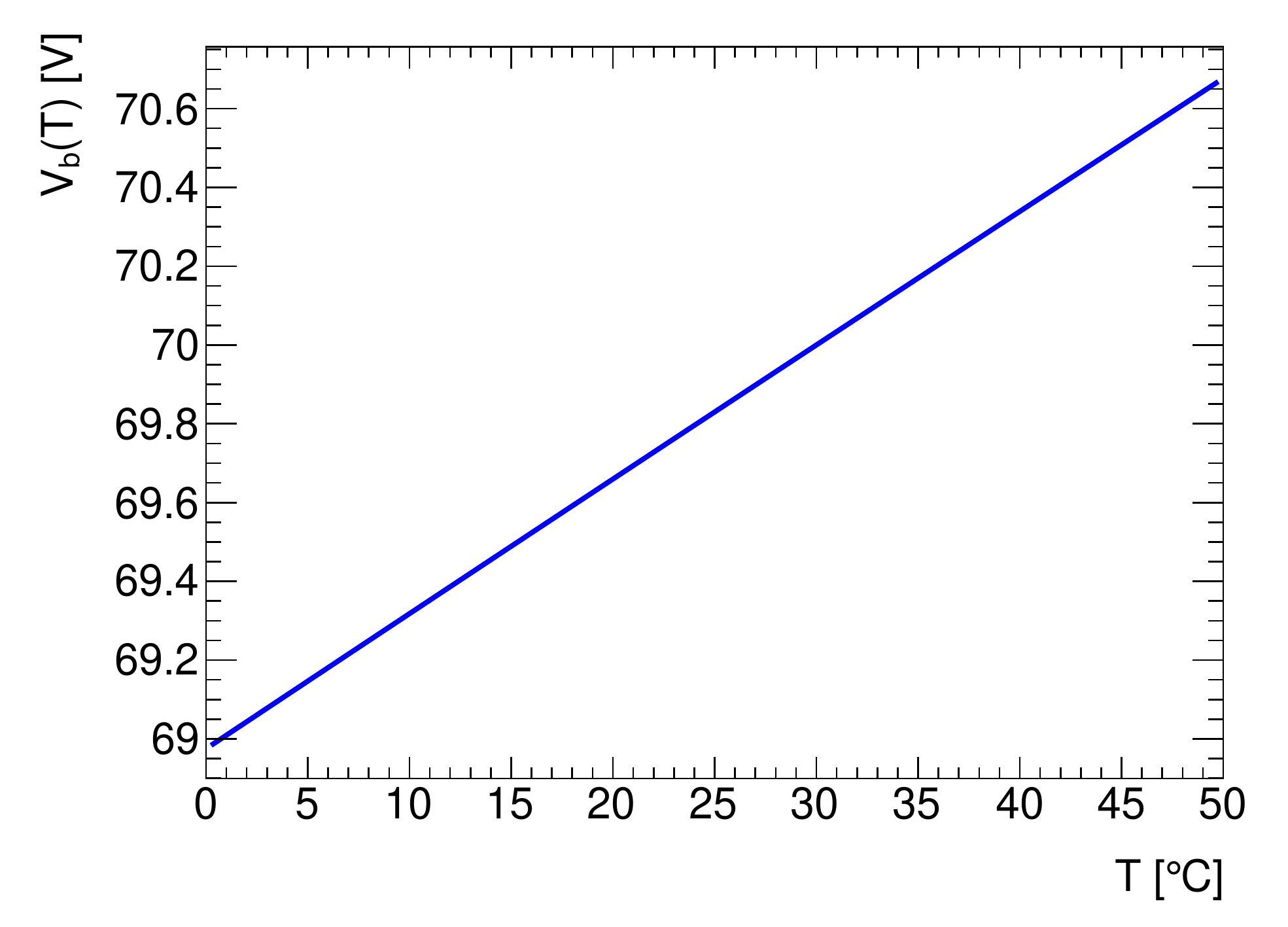}
\caption{\label{fig:S12571_010c_ch1}  Measurements of $G$ versus $V_{\rm b}$ for fixed $T$ (top left), $G$ versus  $T$ for fixed $V_{\rm b}$ (top right), $dG/dV_{\rm b}$ versus  $T$ (middle left), $dG/dT$ versus $V_{\rm b}$  (middle right), $dV_{\rm b}/dT$ versus $T$ (bottom left) and distribution $V_{\rm b}(T)$ versus $T$ (bottom right) for  Hamamatsu detector S12571-010a. Points with error bars show data and solid lines show fit results. }
\end{figure}

Figures~\ref{fig:A2_ch2} -~\ref{fig:S12571_010c_ch1} show the results for Hamamatsu MPPCs A2-20, B2-20 and S12571-010a, respectively. 
\newpage
\newpage

The results for $dG/dV_{\rm b}$ versus $T$ and $dG/dT$ versus $V_{\rm b}$ show small linear deviations from a constant value, while extracted values for $dV_{\rm b}/dT$ versus $T$ are consistent with being constant. For all Hamamatsu MPPCs,  the $G$ versus $V_{\rm b}$ and $G$ versus $T$ lines are rather parallel and are spread apart. 

\newpage
\subsection{KETEK SiPMs}
\label{sec:appKetekBV}
\begin{figure}[!htb]
\centering 
\includegraphics[width=2.9in]{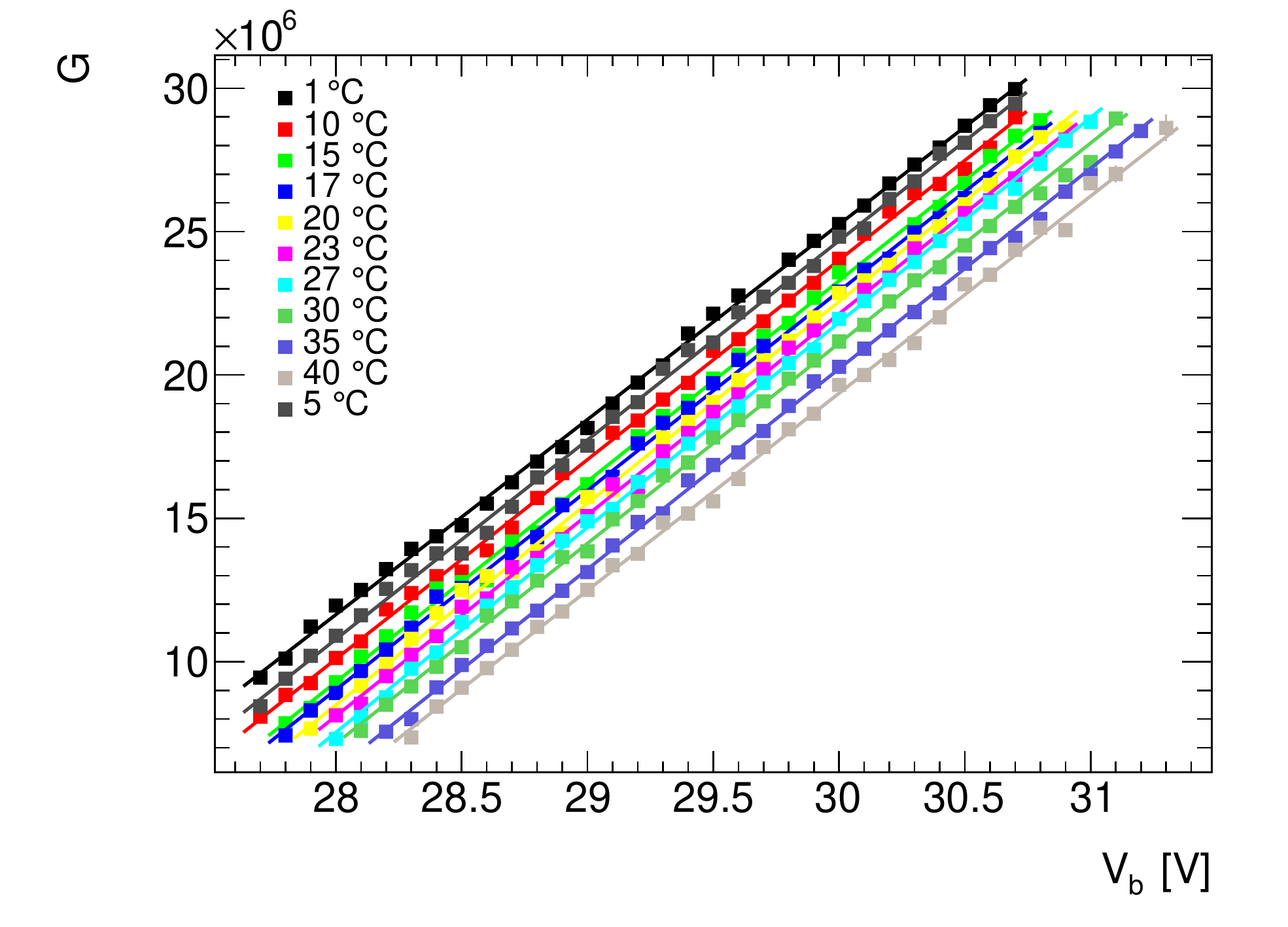}
\includegraphics[width=2.9in]{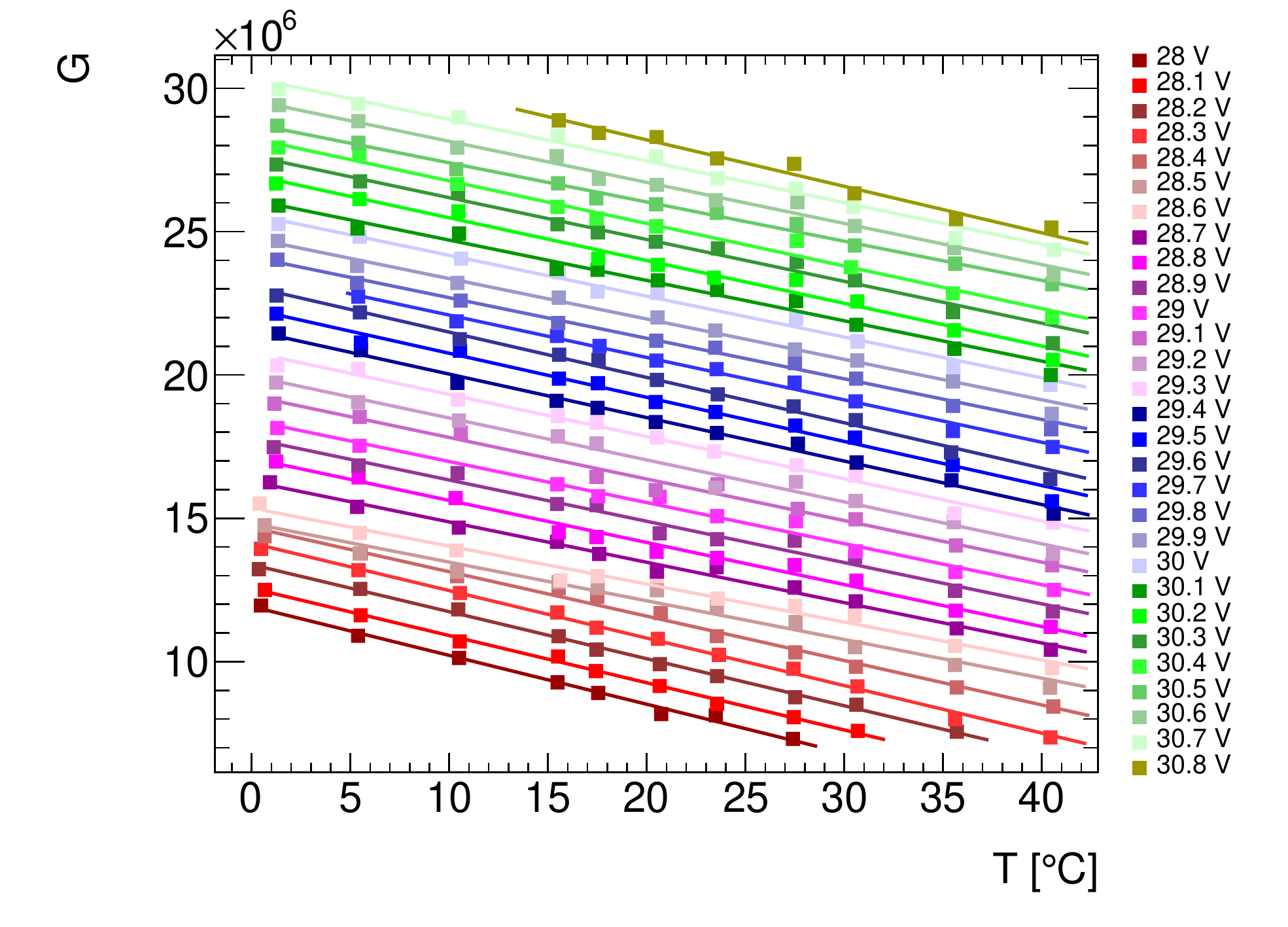}\\
\includegraphics[width=2.9in]{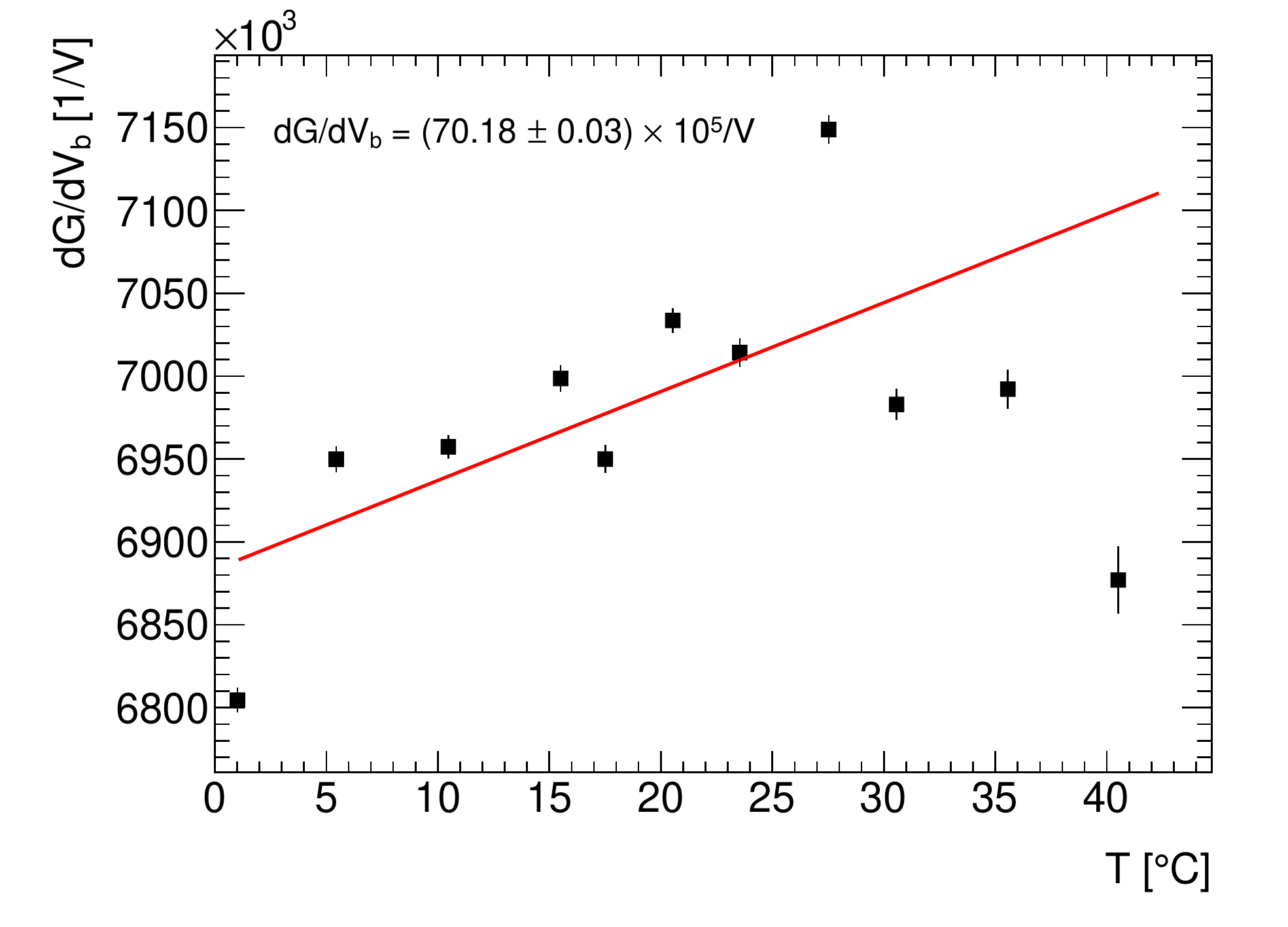}
\includegraphics[width=2.9in]{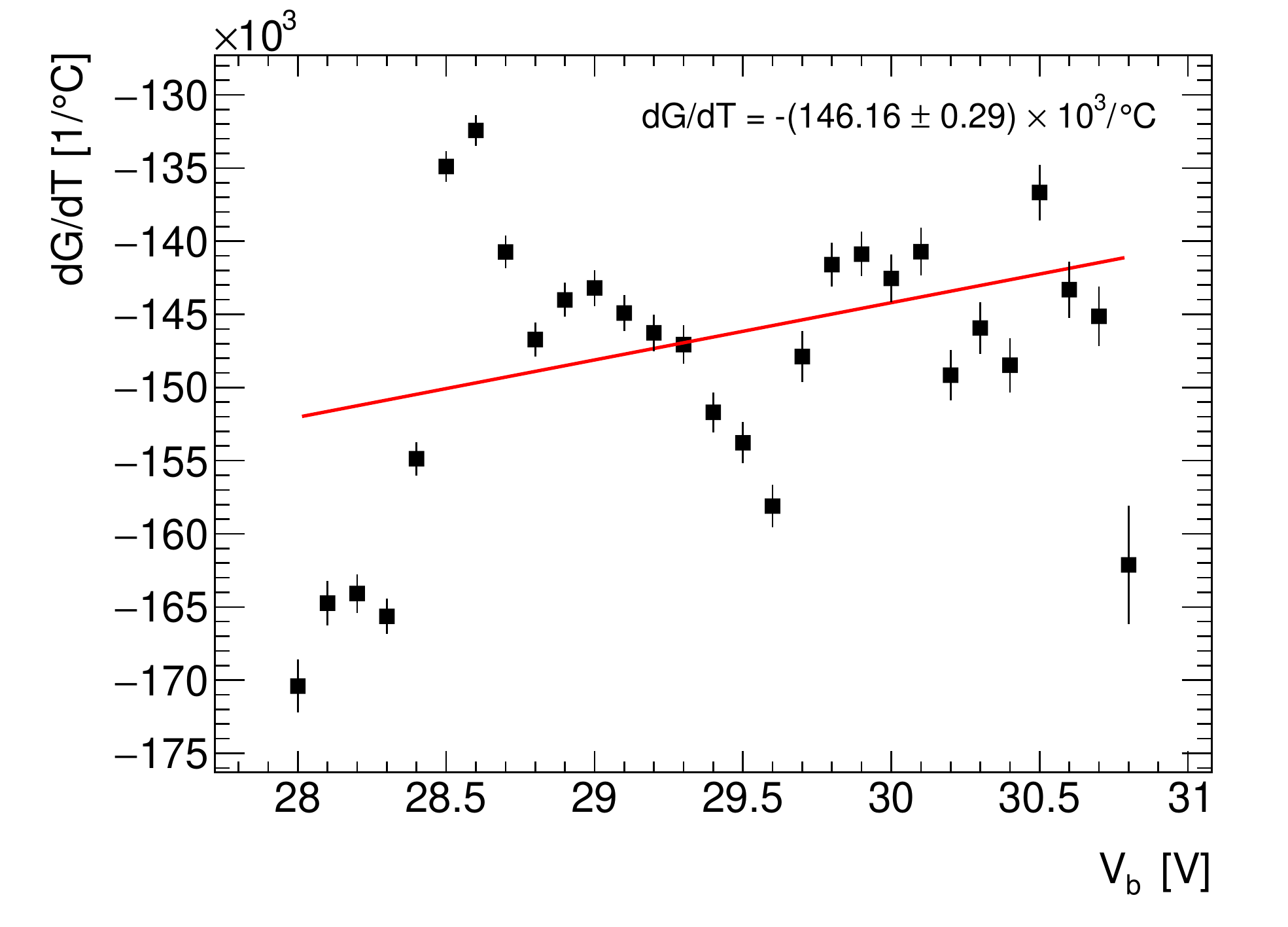}\\
\includegraphics[width=2.9in]{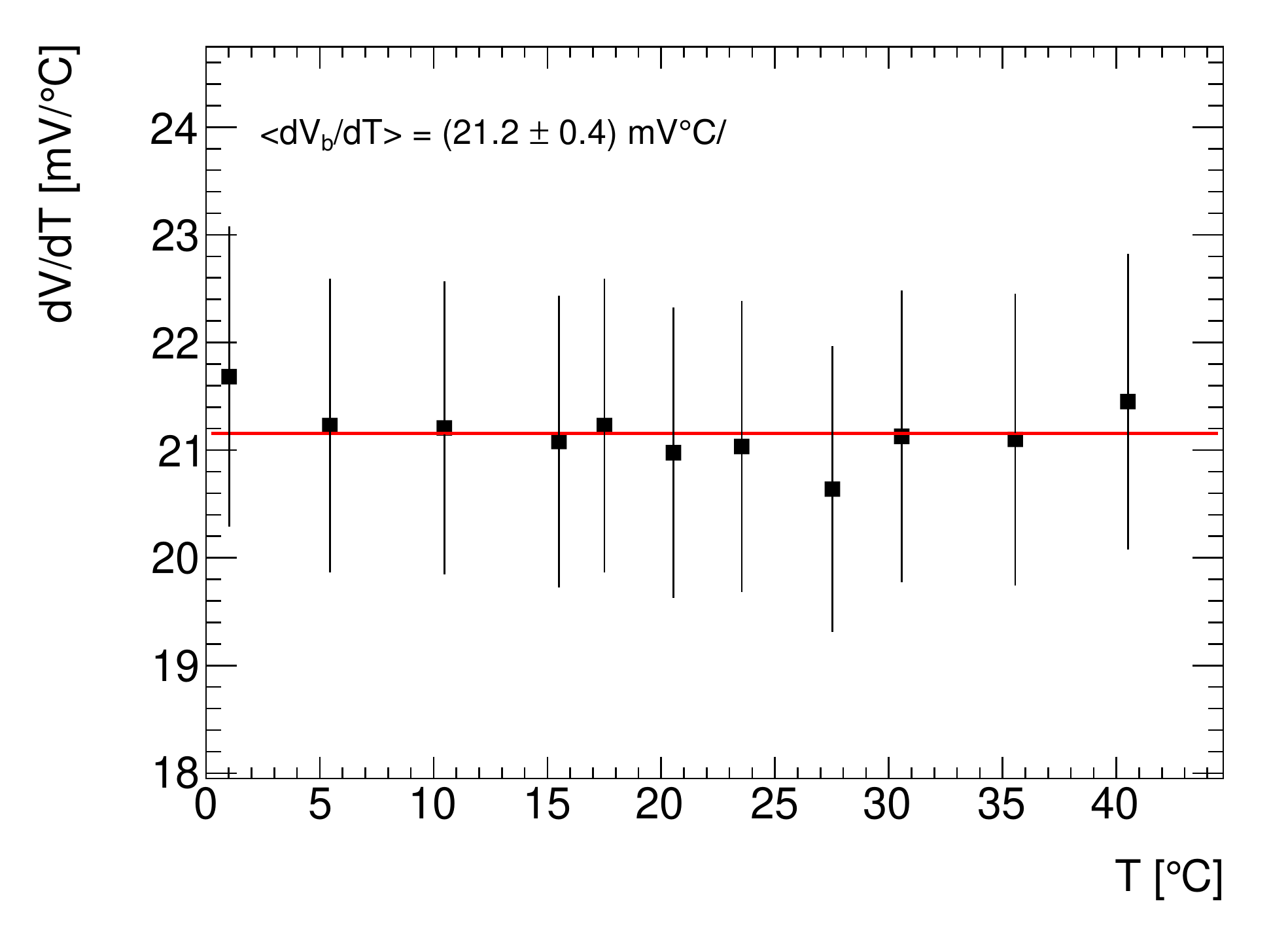}
\includegraphics[width=2.9in]{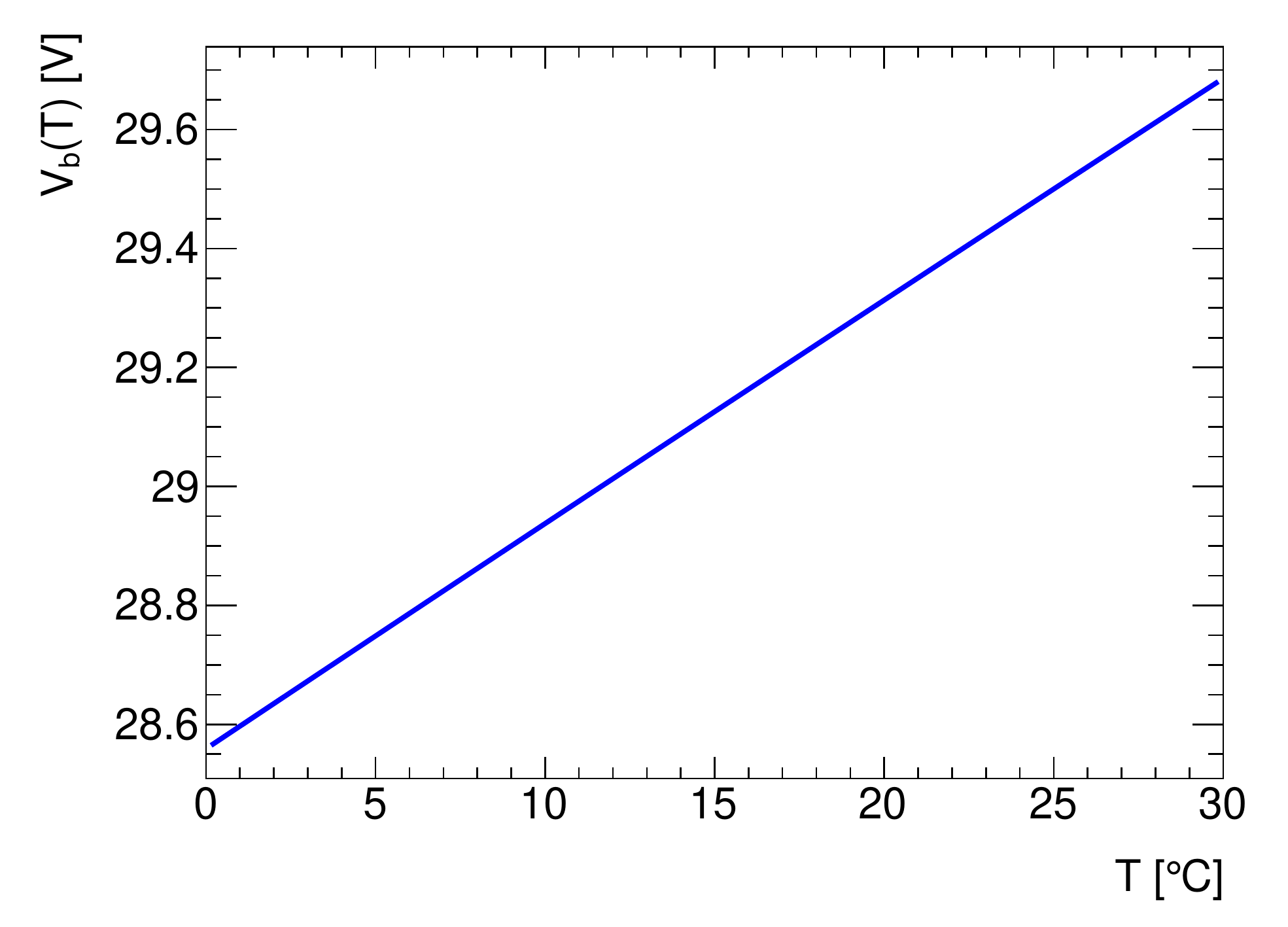}
\caption{\label{fig:W12_ch2} 
Measurements of $G$ versus $V_{\rm b}$ for fixed $T$ (top left), $G$ versus  $T$ for fixed $V_{\rm b}$ (top right), $dG/dV_{\rm b}$ versus  $T$ (middle left), $dG/dT$ versus $V_{\rm b}$  (middle right), $dV_{\rm b}/dT$ versus $T$ (bottom left) and distribution $V_{\rm b}(T)$ versus $T$ (bottom right) for the
KETEK W12B SiPM. Points with error bars show data and solid lines show  fit results.}
\end{figure}

\begin{figure}[!htb]
\centering 
\includegraphics[width=2.9in]{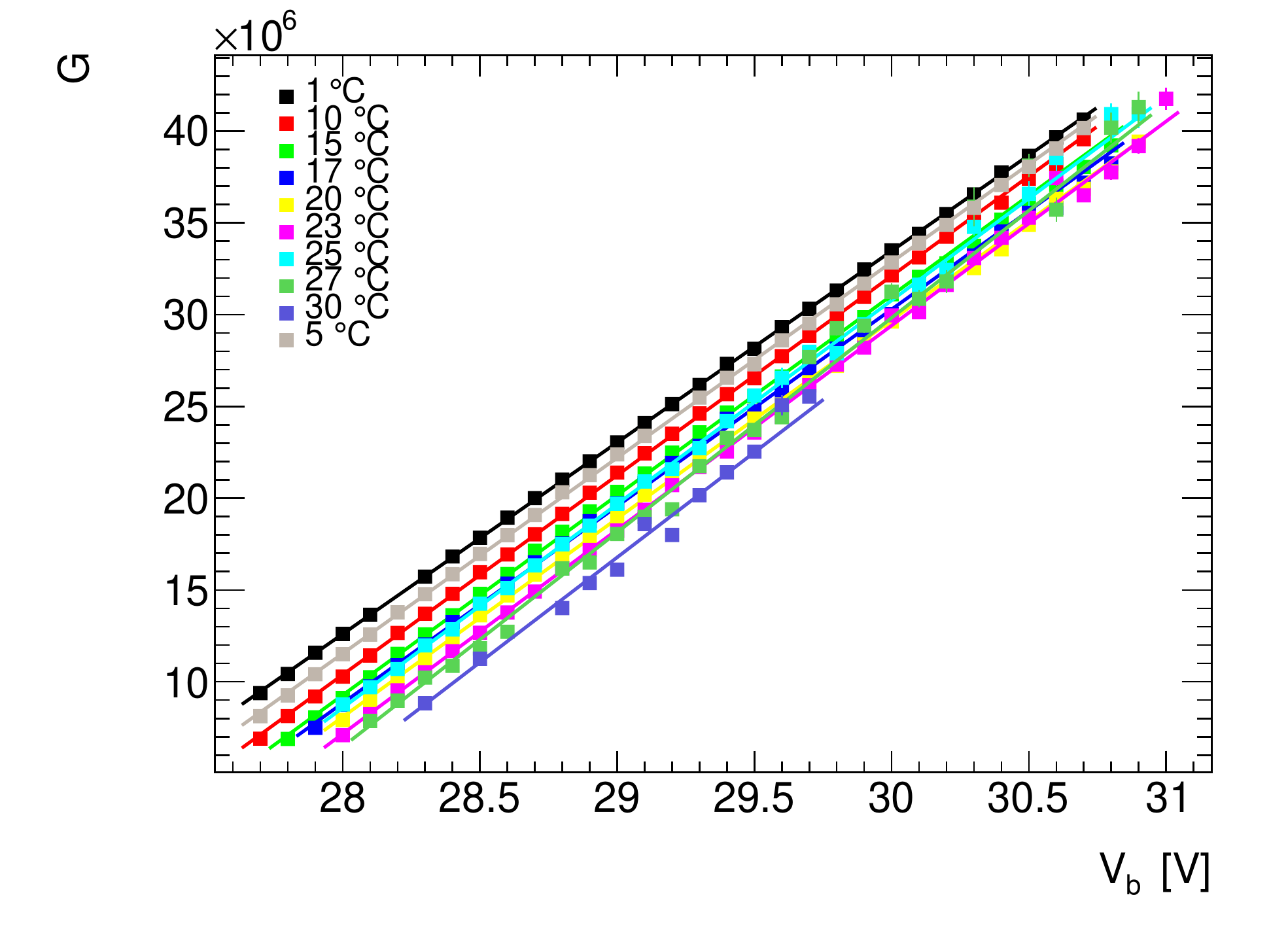}
\includegraphics[width=2.9in]{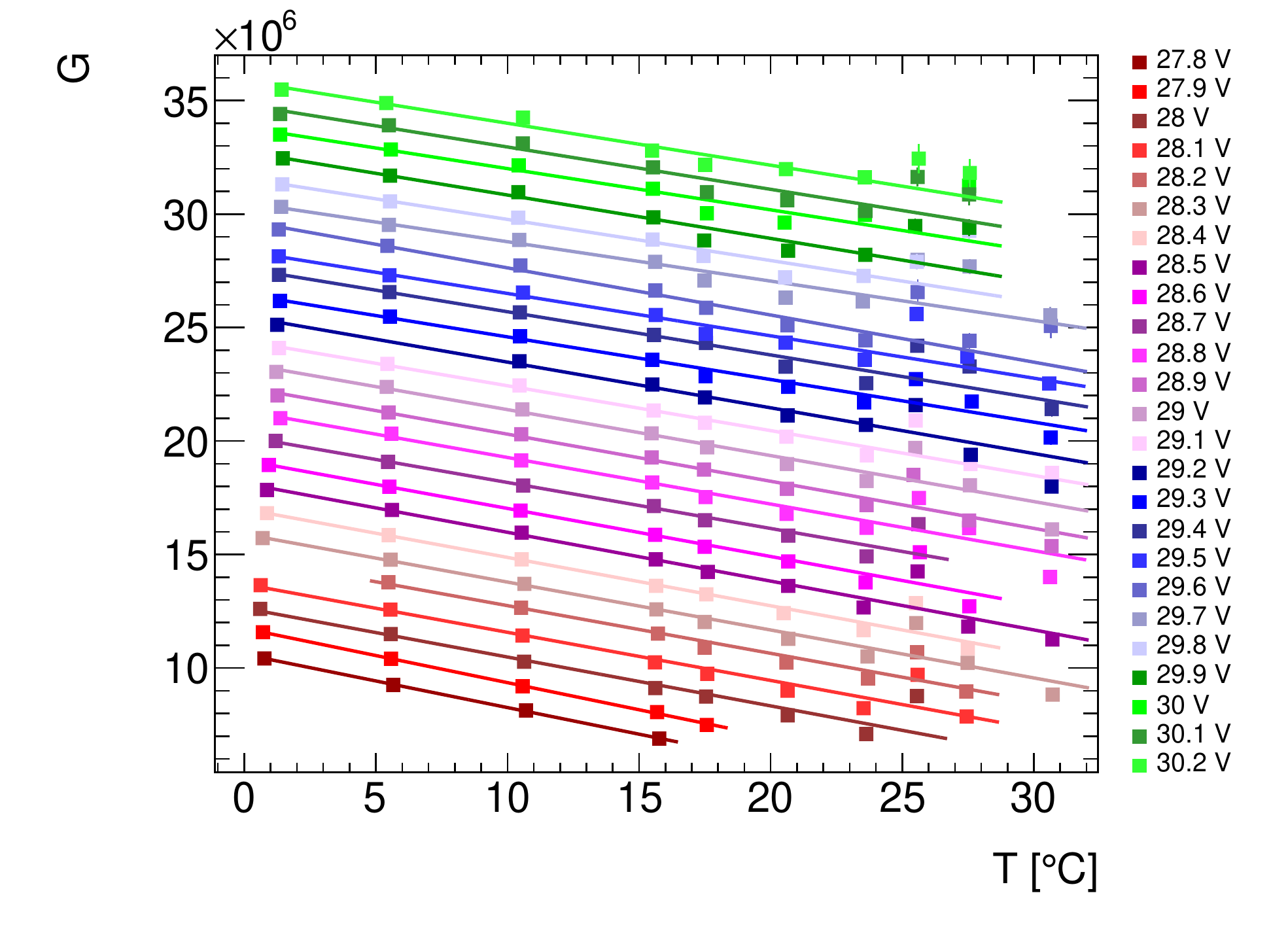}\\
\includegraphics[width=2.9in]{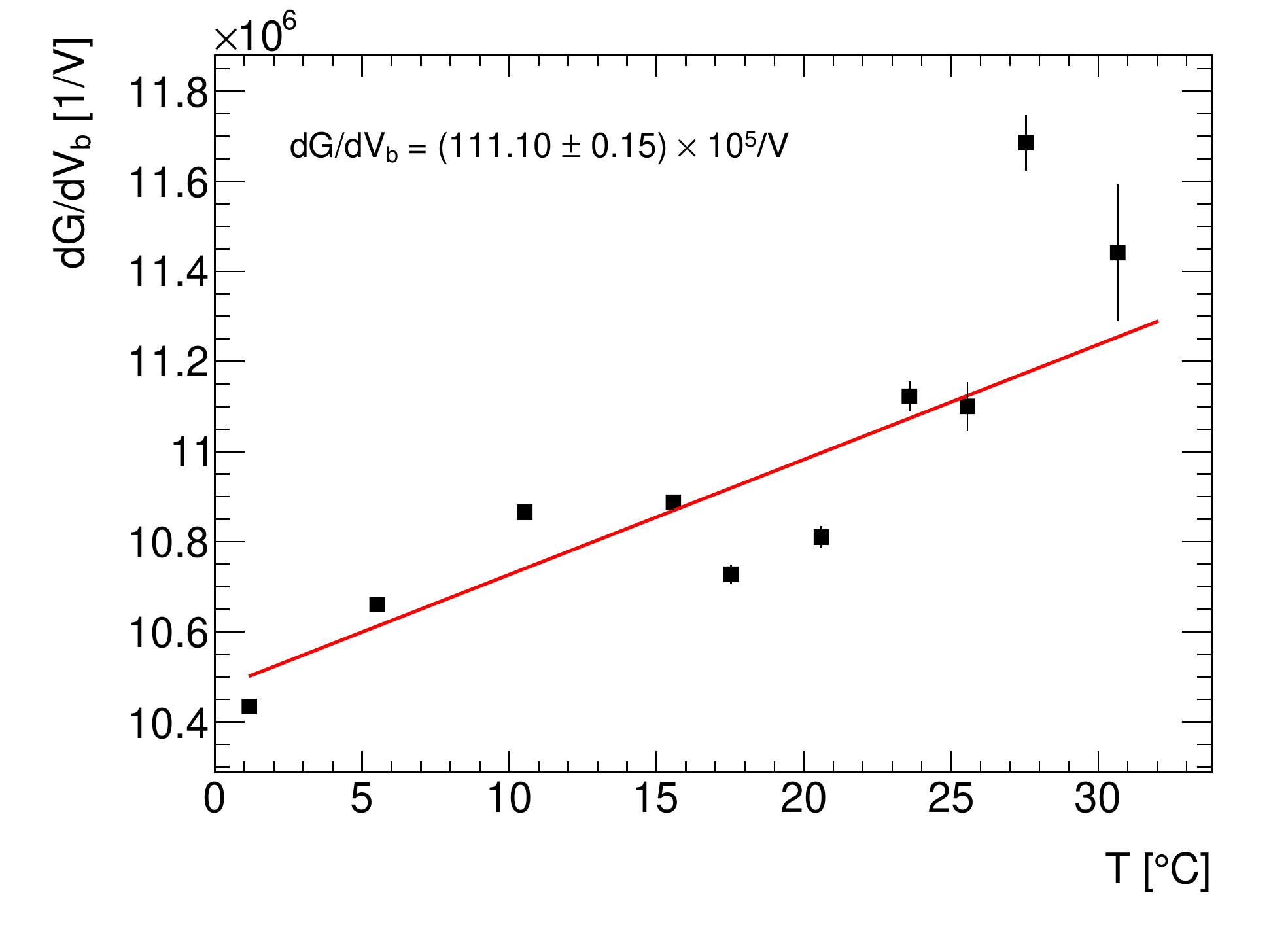}
\includegraphics[width=2.9in]{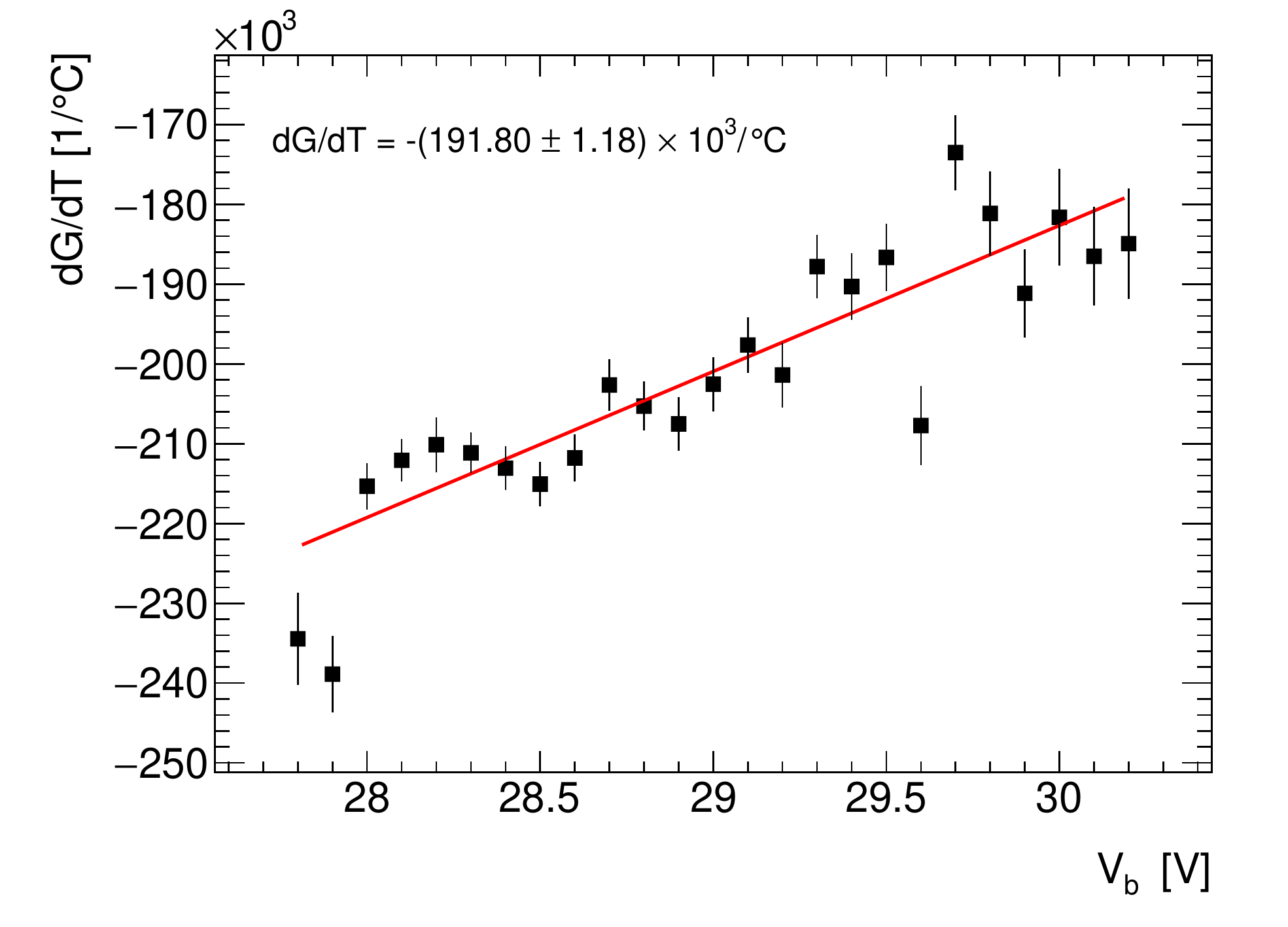}\\
\includegraphics[width=2.9in]{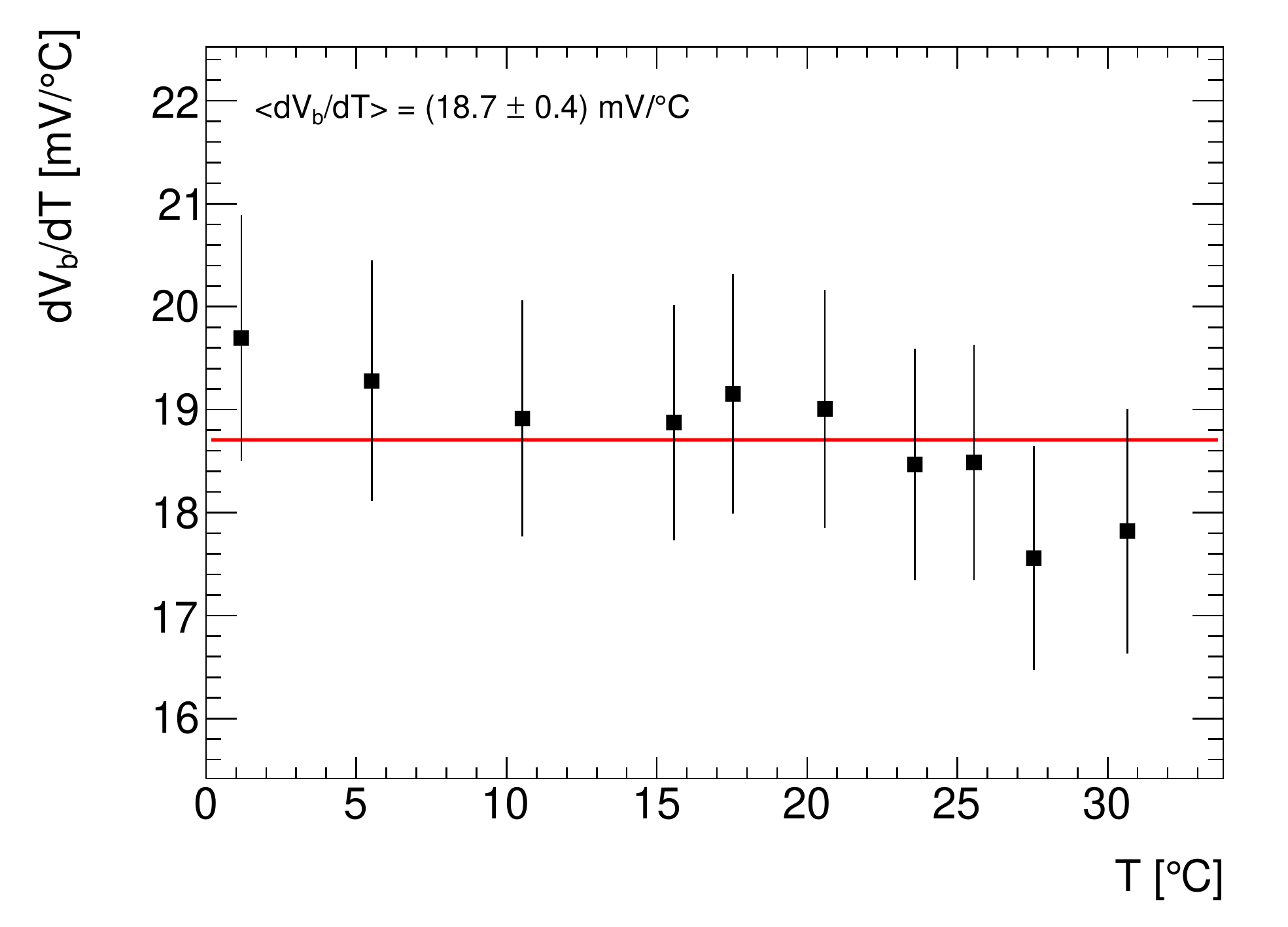}
\includegraphics[width=2.9in]{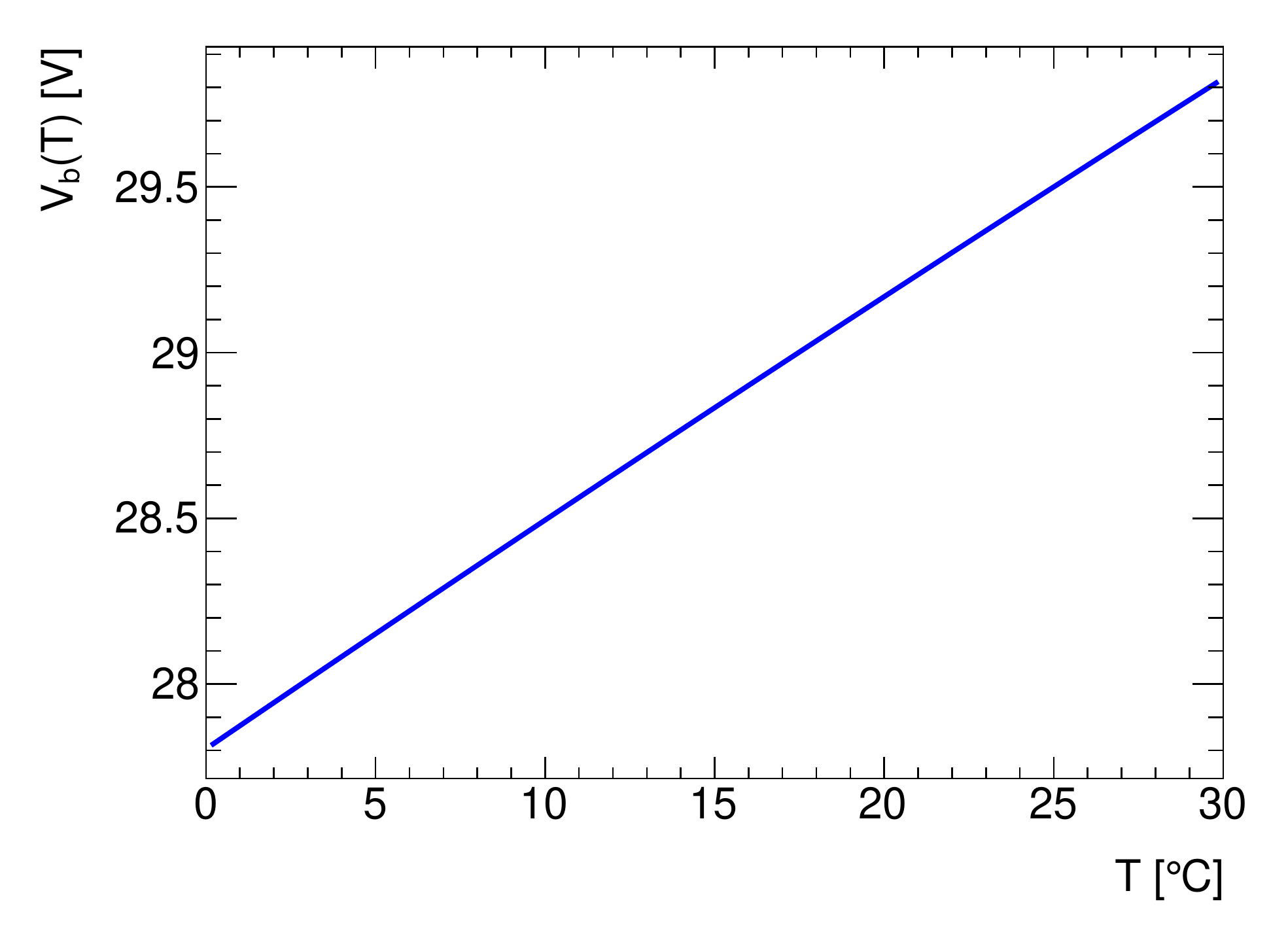}
\caption{\label{fig:PM3350_2_ch4} 
Measurements of $G$ versus $V_{\rm b}$ for fixed $T$ (top left), $G$ versus  $T$ for fixed $V_{\rm b}$ (top right), $dG/dV_{\rm b}$ versus  $T$ (middle left), $dG/dT$ versus $V_{\rm b}$  (middle right), $dV_{\rm b}/dT$ versus $T$ (bottom left) and distribution $V_{\rm b}(T)$ versus $T$ (bottom right) for the KETEK PM3350\#2 SiPM. Points with error bars show data and solid lines show fit results.}
\end{figure}

Figures~\ref{fig:W12_ch2} and ~\ref{fig:PM3350_2_ch4} show the results for KETEK SiPMs W12B and PM3350\#2, respectively. The results for $dG/dV_{\rm b}$ versus $T$ and $dG/dT$ versus $V_{\rm b}$ show larger linear deviations from uniformity ($\sim \pm 5\%$ and $\sim \pm 10\%$, respectively).

\newpage

For the KETEK experimental devices, the $G$ versus $V_{\rm b}$ and $G$ versus $T$ curves are nearly parallel. However, they are more closely spaced than
the corresponding lines for Hamamatsu MPPCs. For the PM3350 SiPMs, the slopes at higher temperature are steeper than those at lower temperature. 

\newpage
\subsection{CPTA SiPMs}
\label{sec:appCPTABV}
\begin{figure}[!htb]
\centering 
\includegraphics[width=2.9in]{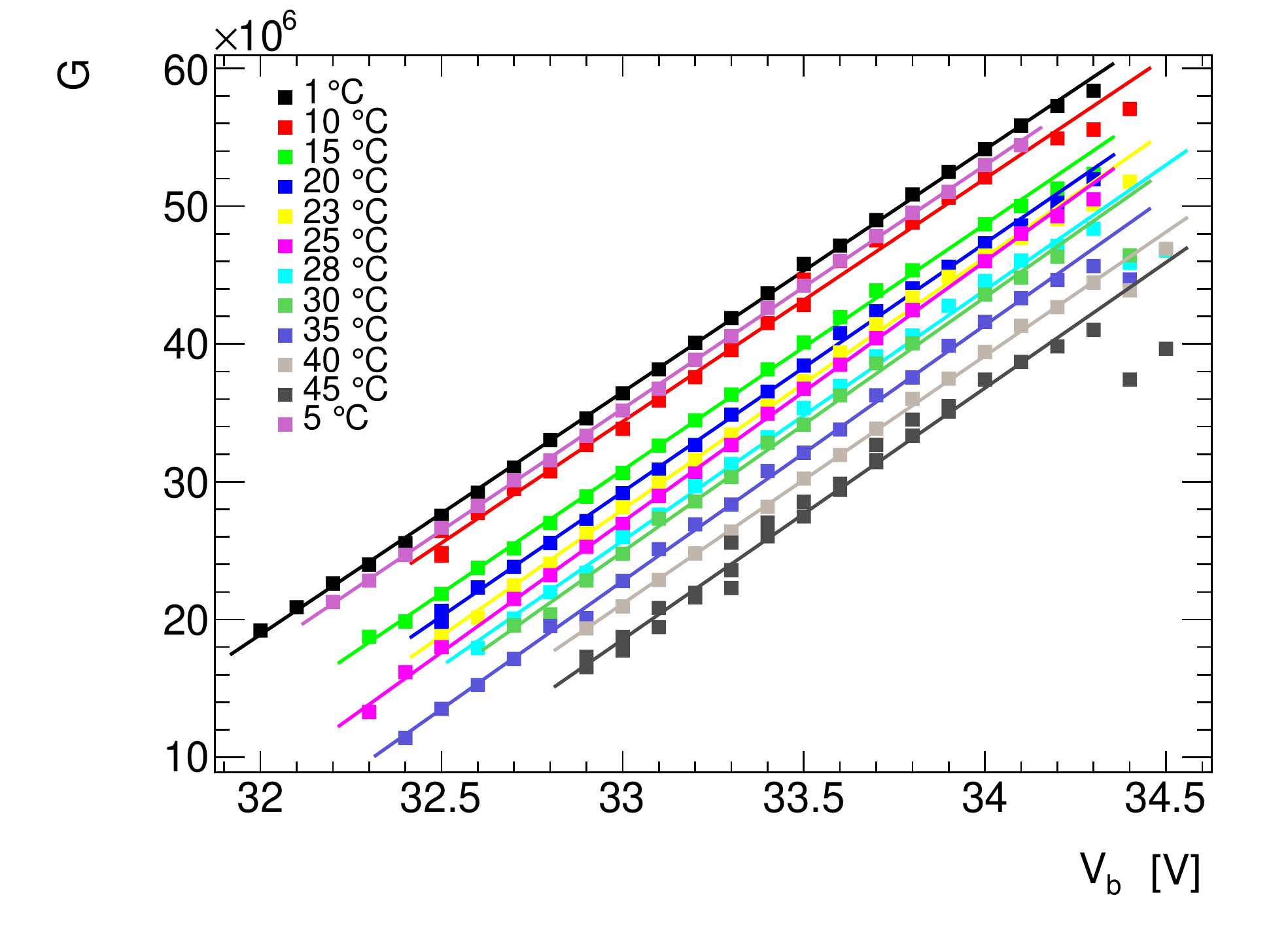}
\includegraphics[width=2.9in]{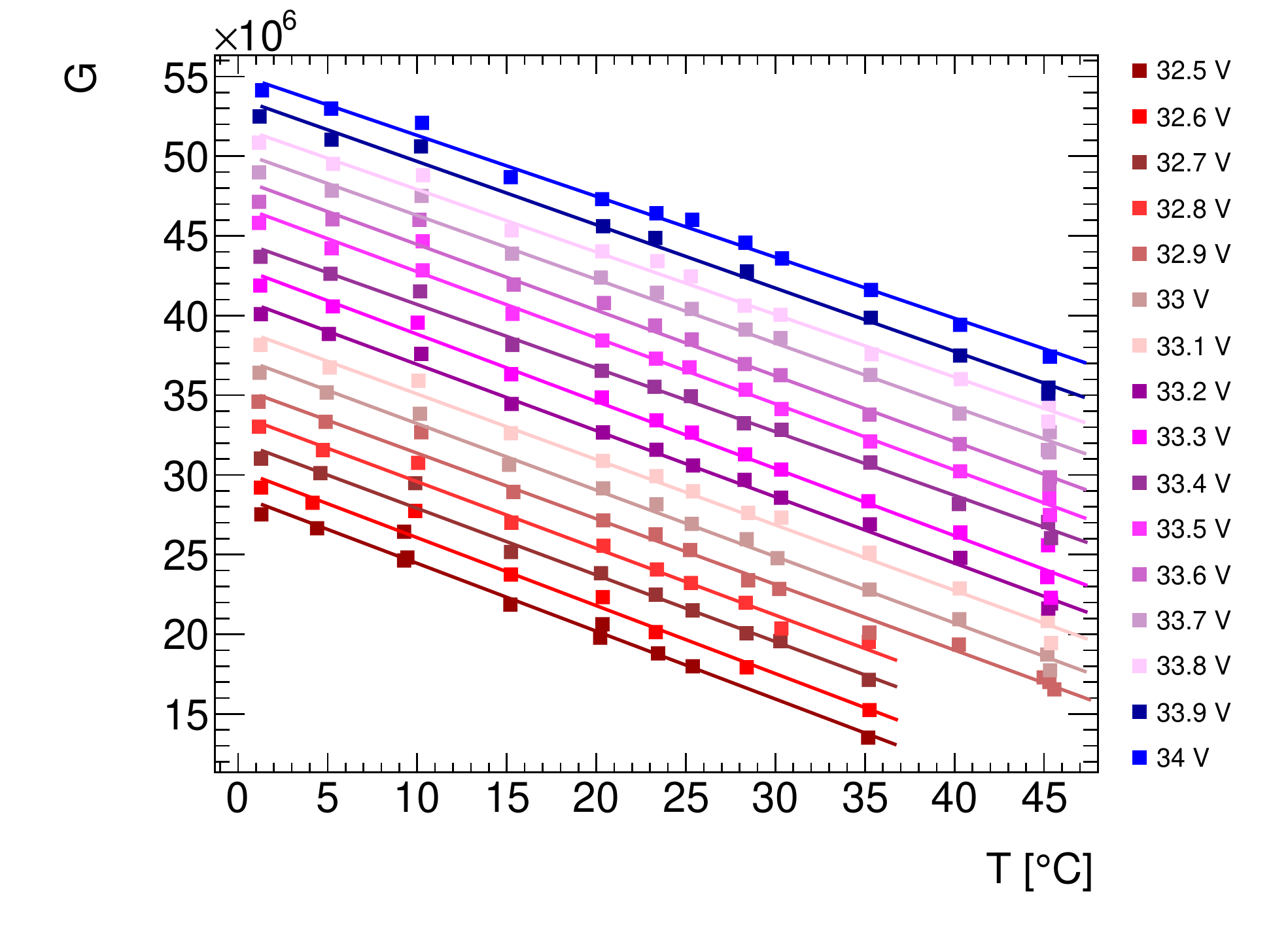}\\
\includegraphics[width=2.9in]{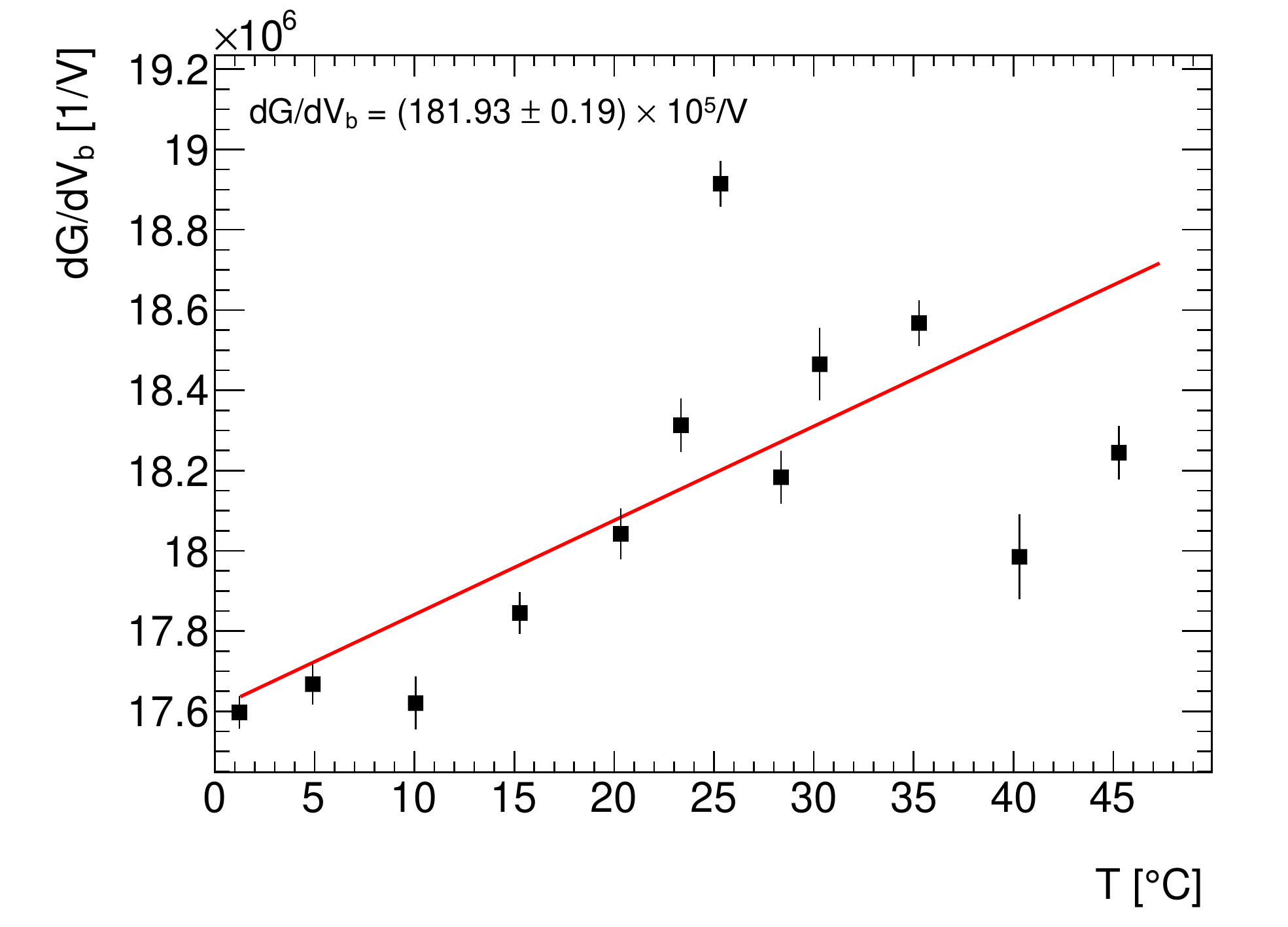}
\includegraphics[width=2.9in]{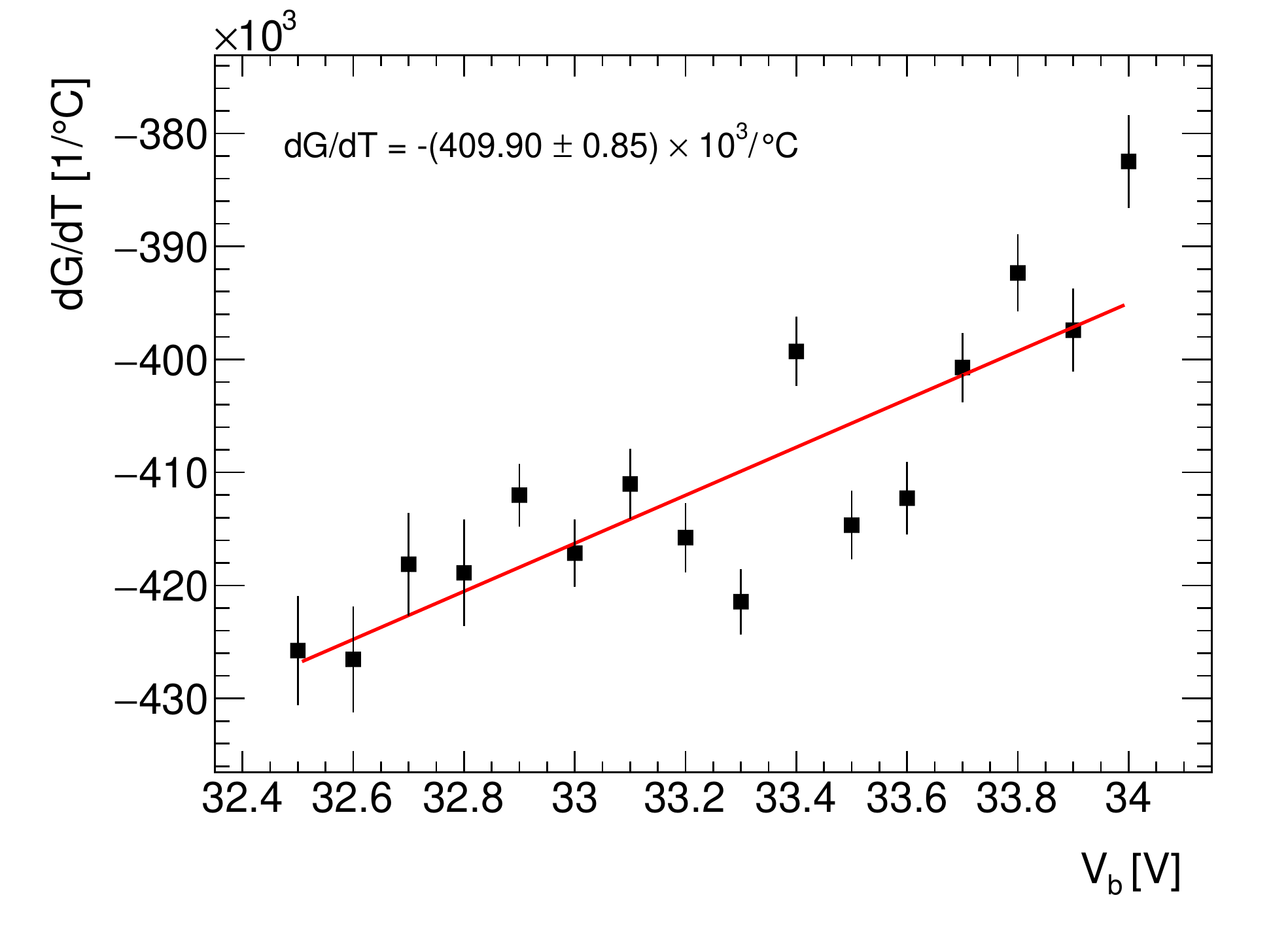}\\
\includegraphics[width=2.9in]{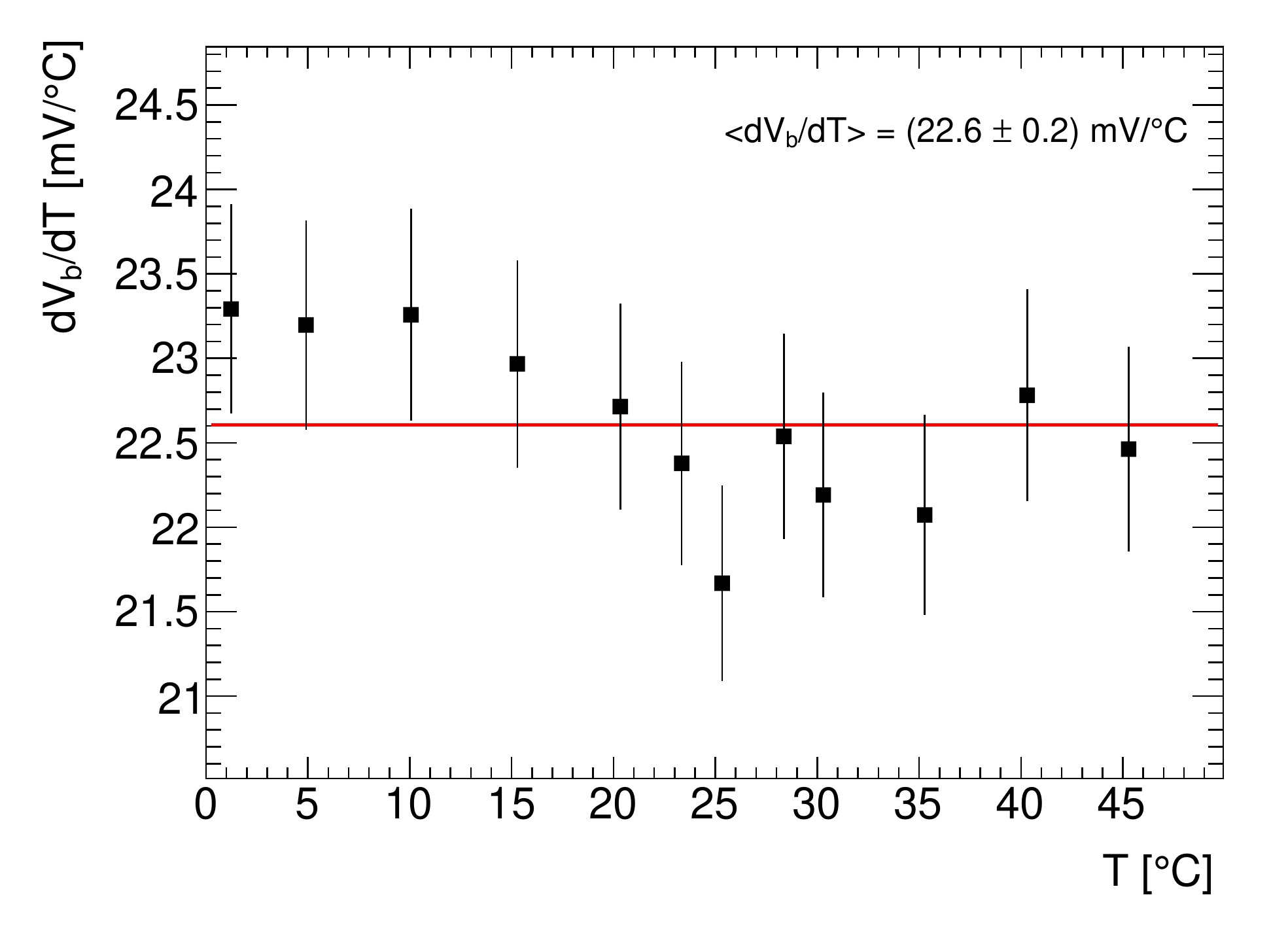}
\includegraphics[width=2.9in]{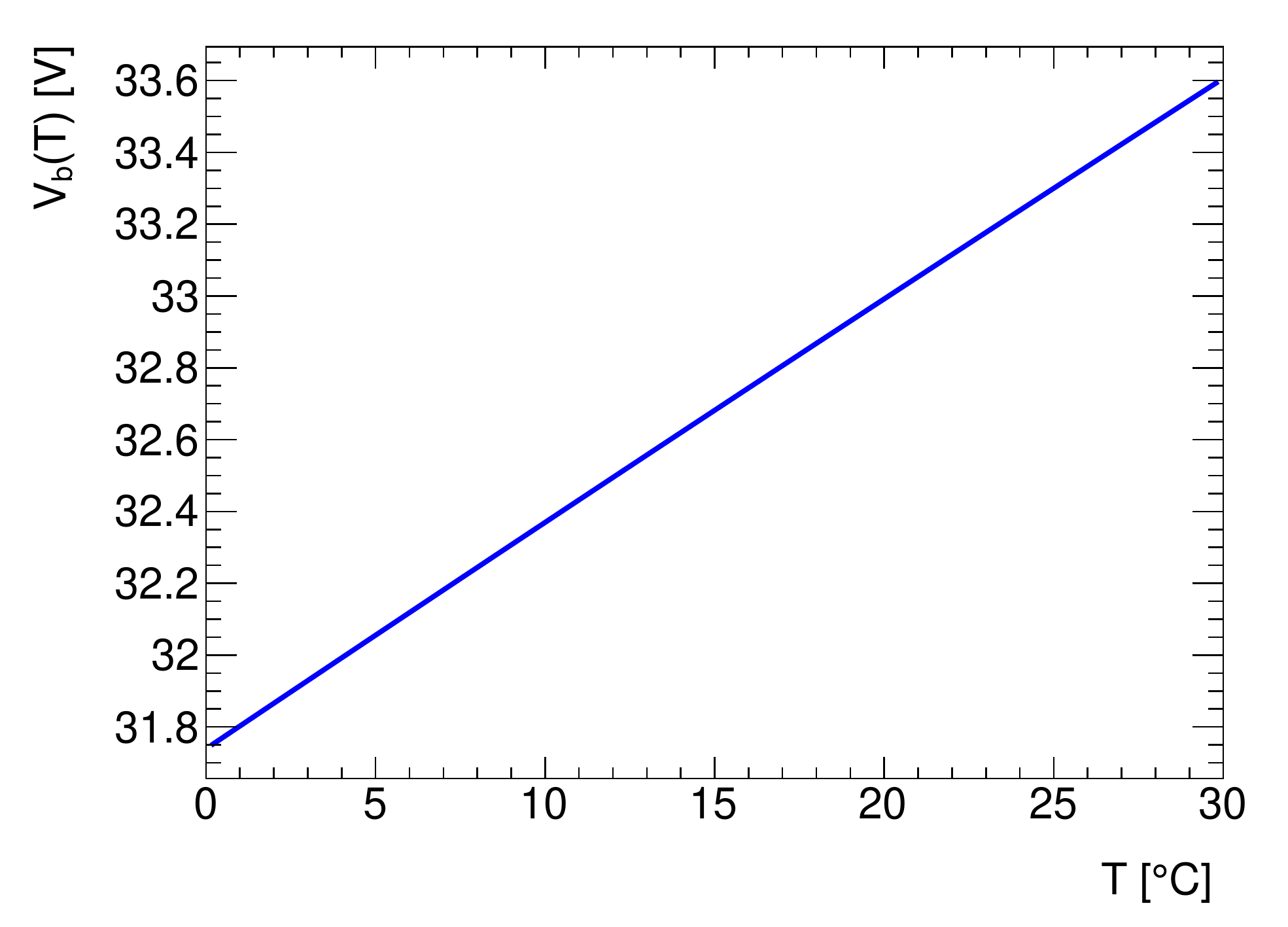}
\caption{\label{fig:1065} 
 Measurements of $G$ versus $V_{\rm b}$ for fixed $T$ (top left), $G$ versus  $T$ for fixed $V_{\rm b}$ (top right), $dG/dV_{\rm b}$ versus  $T$ (middle left), $dG/dT$ versus $V_{\rm b}$  (middle right), $dV_{\rm b}/dT$ versus $T$ (bottom left) and distribution $V_{\rm b}(T)$ versus $T$ (bottom right) for the CPTA\#1065 SiPM. Points with error bars show data and solid lines show  fit results. }
\end{figure}

Figure~\ref{fig:1065} shows the results for the CPTA  \#1065 SiPM. 
The extracted $dV_{\rm b}/dT$ versus $T$ dependence is consistent with being constant.

For CPTA, the $G$ versus $V_{\rm b}$ and $G$ versus $T$ lines are rather parallel and are more spread out than those curves for KETEK SiPMs.

\newpage
\section{Temperature Dependence of the Bias Voltage for quadratic $dG/dV_{\rm b}$ and $dG/dT$}
\label{sec:app-VT}

For a quadratic dependence of  both $dG/dV_{\rm b}$ versus $T$ and $dG/dT$ versus $V_{\rm b}$, we have
\begin{equation}
\frac{ dG(V_{\rm b},~T)}{dT}= a + b\cdot V_{\rm b}+ e \cdot V_{\rm b}^2,
\label{eq:11}
\end{equation}  
\begin{equation}
\frac{dG(V_{\rm b},~T)}{dV_{\rm b}}= c + d\cdot T+ f\cdot T^2.
\label{eq:12}
\end{equation}
\noindent This introduces two new parameters $e$ and $f$, which multiply the quadratic terms, respectively. Plugging eqs.~\ref{eq:11} and ~\ref{eq:12} into eq.~\ref{eq:5} yields
\begin{equation}
\frac{dV_{\rm b}}{dT}=- \frac{a + b\cdot V_{\rm b}+ e \cdot V_{\rm b}^2}{c + d\cdot T+ f\cdot T^2}
\label{eq:8}
\end{equation}
The solution for $b\neq 0$, $d\neq 0$,  $e\neq 0$ and $f\neq 0$ is simply
\begin{equation}
V_{\rm b}(T) = \frac{-b+\sqrt{B } ~\tan{\frac{1}{2}  \Big[2 (\sqrt{B}/\sqrt{D}) \arctan{\Big( (d+2 f\cdot T)/\sqrt{D}\Big)}+\sqrt{B}~C1}  \Big]} {2 e}
\end{equation}
\noindent where $C1$ is another integration constant and 
\begin{eqnarray}
B&=&-b^2 +4 a\cdot e\\
D&=&-d^2+4c \cdot f.
\end{eqnarray} 
For $e=0$, we get
\begin{equation}
V_{\rm b}(T) = -\frac{a}{b}+C1~\exp{\Bigg[- \frac{2b \arctan{\Big((d+2f\cdot T)/\sqrt{D}\Big)}}{\sqrt{D}}\Bigg]},
\label{eq:14}
\end{equation}
while for $f=0$ and $e \ne 0$ the solution is
\begin{equation}
V_{\rm b}(T) = \frac{-b+\sqrt{B } ~\tan{\frac{\sqrt{B}}{2}  \Big(C1 -  \log{(c+d\cdot T)}/d \Big)}} {2 e}.
\label{eq:15}
\end{equation}
The solutions for $c=0$ and $d=0$ or $a=0$ and $b=0$ are
\begin{eqnarray}
V_{\rm b}(T) &=& \Big[-b+\sqrt{B}\tan \Big( \frac{1}{2} \Big[ \sqrt{B}   /(f \cdot T)+\sqrt{B} ~C1 \Big] \Big) \Big]/(2e)    \nonumber \\
V_{\rm b}(T) &=& -\sqrt{D}/\Big[\sqrt{D }~C1 -2e \arctan \Big({(d+2 f\cdot T)/\sqrt{D}\Big)}\Big], 
\label{eq:16}
\end{eqnarray}
respectively.

\section{Gain Stabilization  of SiPMs with the First Gain Fit Model}
\label{sec:app-gainstab}
\begin{figure}[H]
\centering 
\includegraphics[width=2.9in]{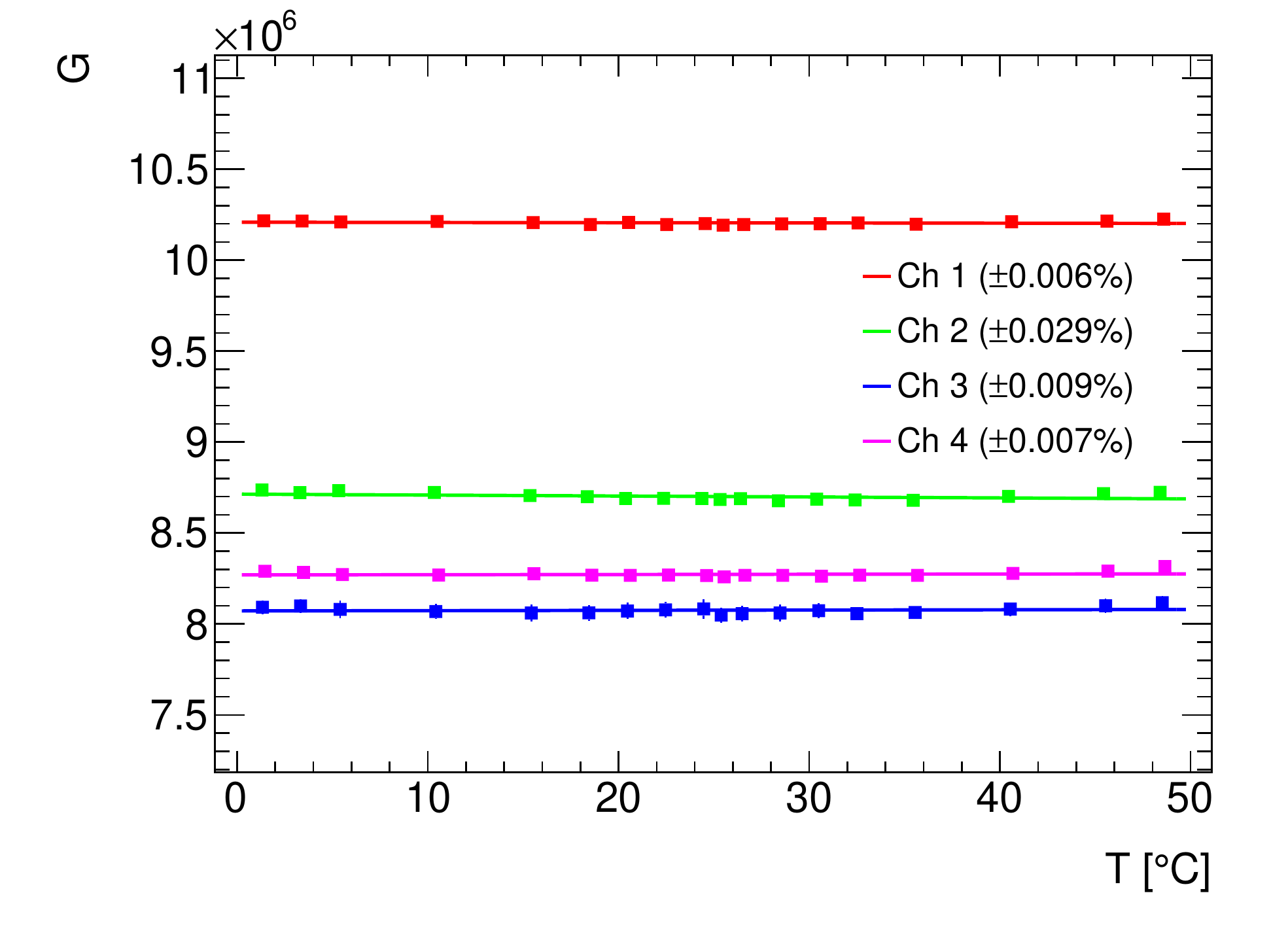}
\includegraphics[width=2.9in]{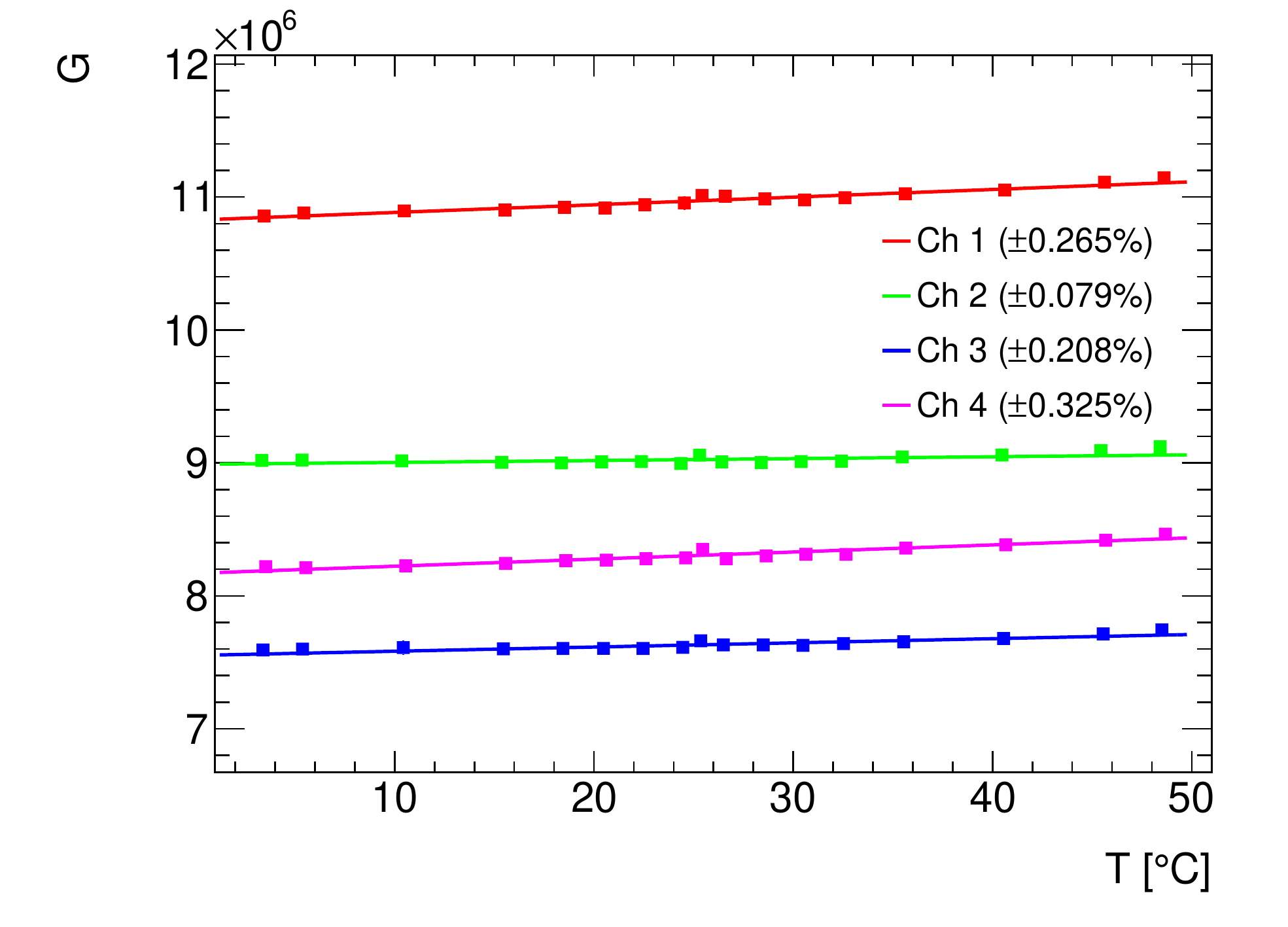}\\
\includegraphics[width=2.9in]{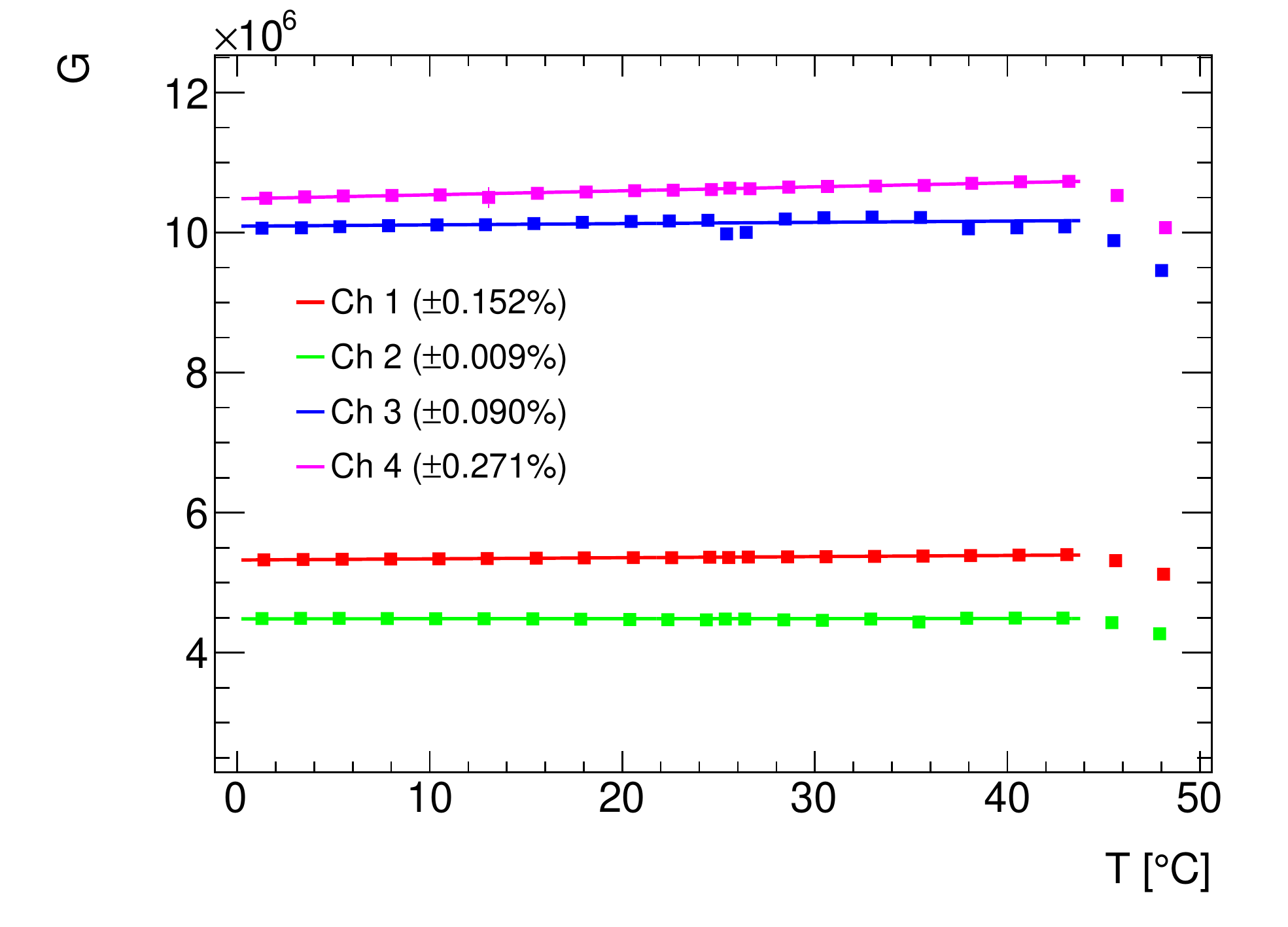}
\caption{ Measurements of stabilized gain versus temperature obtained with the first gain  fit model. Top left: all $\rm A-$type MPPCs; top right: all $\rm B-$type MPPCs; bottom; all S12571  MPPCs.}
\label{fig:Ham_all}
\end{figure}

\begin{figure}[H]
\centering 
\includegraphics[width=2.9in]{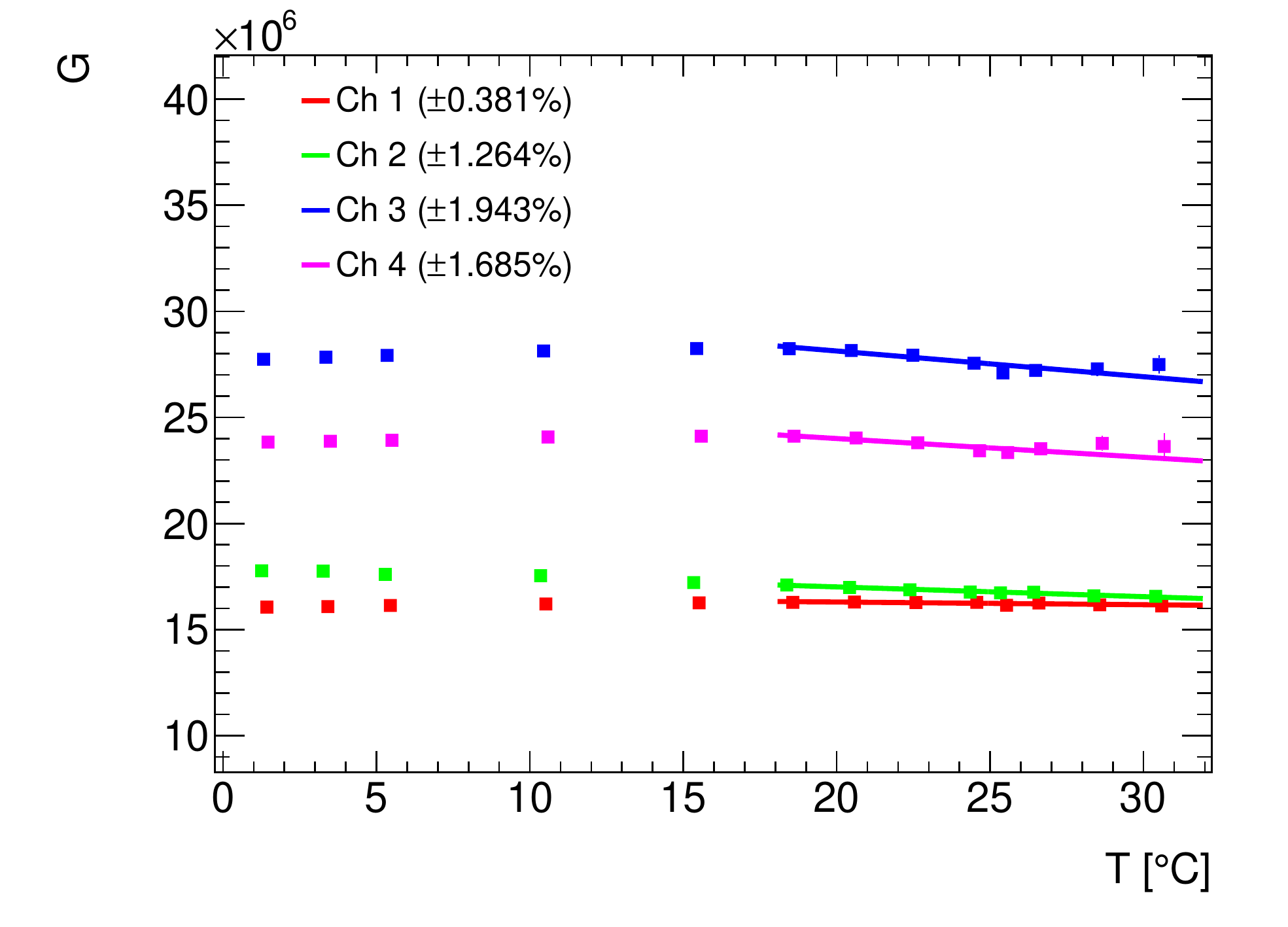}
\includegraphics[width=2.9in]{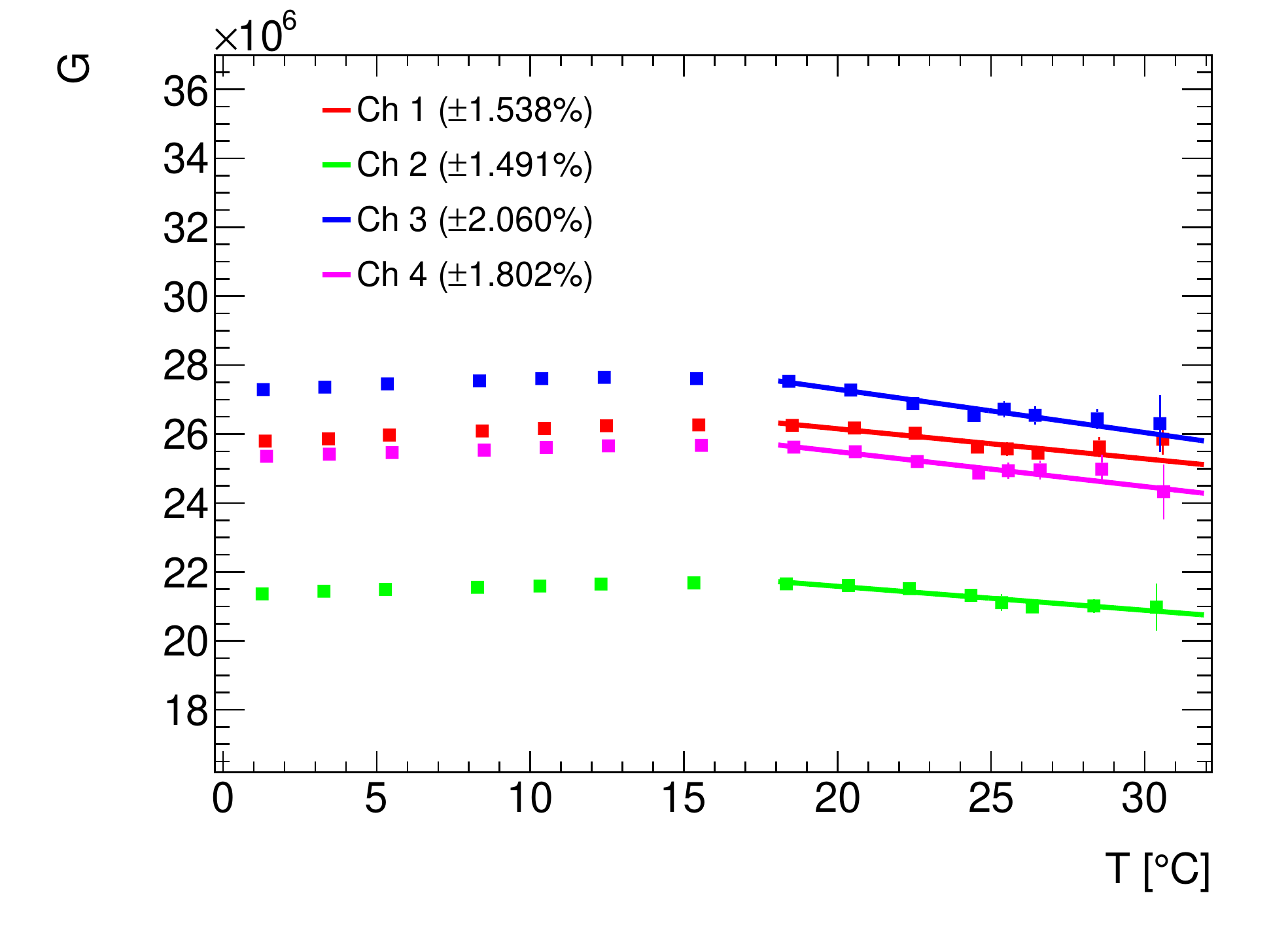}
\caption{\label{fig:4ch_PM} Measurements of stabilized gain versus temperature. Left: for W12A, W12B, PM3350\#1 and PM3350\#2 SiPMs; Right: for the other four PM3350 SiPMs (\#5 to \#8)  using the first gain fit model for the gain determination. }
\end{figure}

\begin{figure}[H]
\centering 
\includegraphics[width=2.9in]{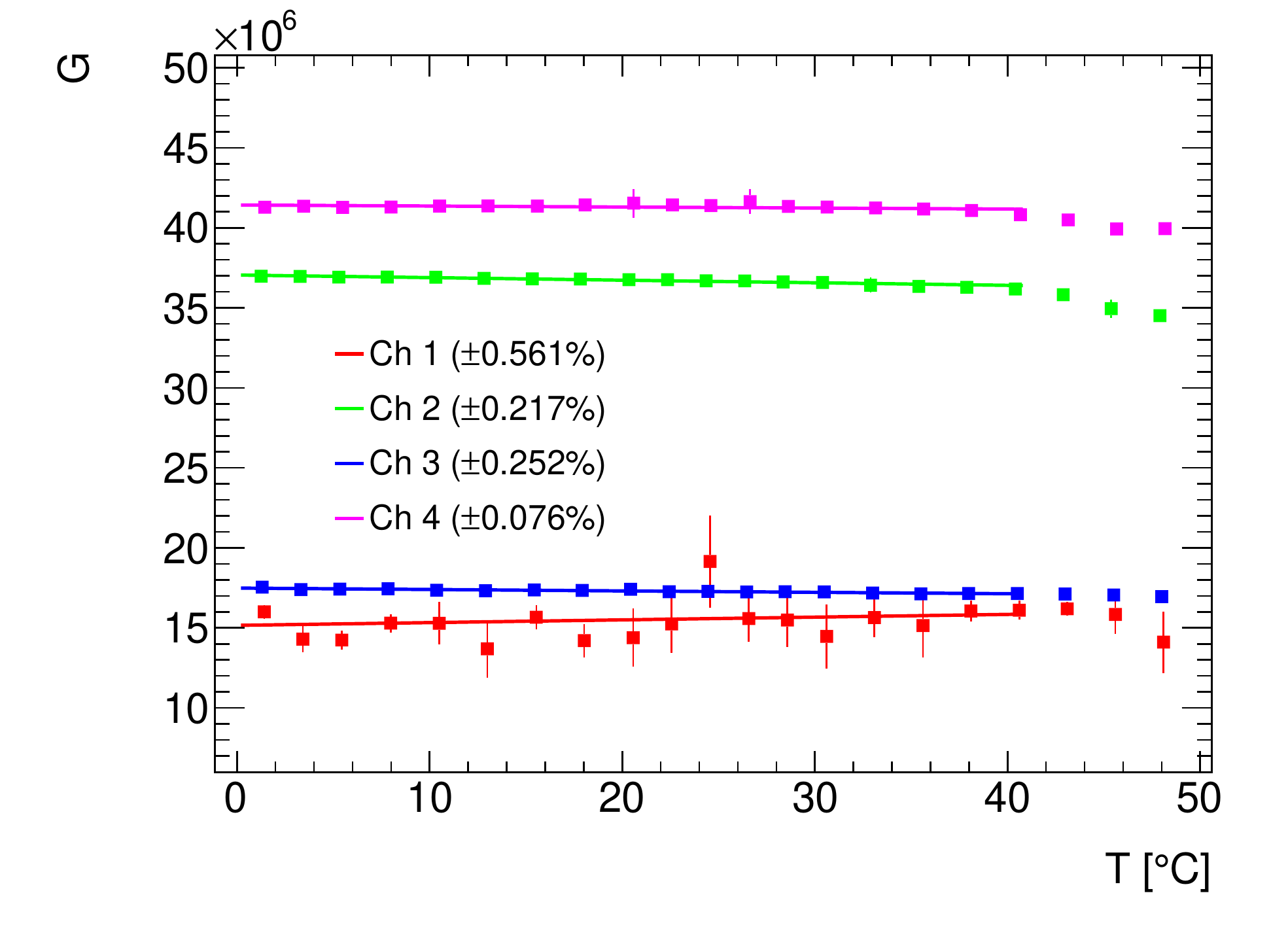}
\caption{\label{fig:CPTA} Measurements of stabilized gain versus temperature for CPTA SiPMs \#857, \#922, \#975 and \#1065 in which the gain was determined
with the first gain fit model. }
\end{figure}

Figure~\ref{fig:Ham_all} shows the gain stabilization fit results for Hamamatsu $\rm A-$type, $\rm B-$type and S12571 MPPCs using the first gain fit model while Fig.~\ref{fig:4ch_PM} and Fig.~\ref{fig:CPTA} show the corresponding results for the KETEK W12/PM3350 SiPMs and for CPTA SiPMs, respectively.
For the Hamamatsu SiPMs, these results agree well with those obtained with the second gain fit model. For KETEK and CPTA SiPMs the results are consistent with those obtained with the first gain fit model.


\end{document}